\documentclass[3p,twocolumn,article]{elsarticle} 

\usepackage{graphicx}
\usepackage{amsmath}
\usepackage{amssymb}
\usepackage{dcolumn}
\usepackage[euler]{textgreek}

\usepackage{xcolor}
\definecolor{myblue}{rgb}{0.0, 0.0, 0.6}
\usepackage{hyperref}
\hypersetup {
  colorlinks = true,
  citecolor  = myblue,
  linkcolor  = myblue,
  urlcolor   = myblue
}
\usepackage{lineno}
\modulolinenumbers[5]
\journal{ Nucl. Instrum. Methods A}
\begin{document}

\begin{frontmatter}
  
\title{
  Systematic error analysis in the absolute hydrogen gas jet polarimeter at RHIC
  \tnoteref{t1}}%
\tnotetext[t1]{This manuscript has been authored by employees of Brookhaven Science Associates, LLC under Contract No. DE-SC0012704 with the U.S. Department of Energy}

\author{A.~A.~Poblaguev}\corref{cor1}\ead{poblaguev@bnl.gov}\cortext[cor1]{Corresponding author}
\author{A.~Zelenski}
\author{G.~Atoian}
\author{Y.~Makdisi}
\author{J.~Ritter}

\address{Brookhaven National Laboratory, Upton, New York 11973, USA}

\date{\today}

\begin{abstract}
    The Polarized Atomic Hydrogen Gas Jet Target polarimeter (HJET) is operated at the Relativistic Heavy Ion Collider (RHIC) since 2004 to measure the absolute polarization of each colliding proton beam. Polarimeter detectors and data acquisition were upgraded in 2015 to increase the solid angle, energy range, and to improve the energy and time resolution. These upgrades along with an improved beam intensity and polarization allowed us to greatly reduce the statistical and systematic errors for the proton polarization measurements in RHIC Runs\,15 ($E_\text{beam}\!=\!100\,\text{GeV}$) and 17 (255\,GeV). For a typical 8 hour RHIC store, the measured proton beam average polarization was about $P_\text{beam}\!\sim\!55\pm2.0_\text{stat}\pm0.3_\text{syst}\%$. The elastic $\mathit{pp}$ analyzing power, $A_\text{N}(t)\!\sim\!0.04$, was determined with a precision of about $|\delta A_\text{N}(t)|\!\sim\!0.0002$ in the momentum transfer squared range $0.001\!<\!-t\!<\!0.020\,\text{GeV}^2$. In this paper we present a detailed systematic error analysis of the polarization measurements at HJET. Perspectives of using the HJET based absolute polarimeters in the future Electron Ion Collider (EIC) will be also discussed.

\end{abstract}

\end{frontmatter}

\section{Introduction}

A polarized hydrogen gas jet polarimeter is used to measure absolute proton beam polarization at the Relativistic Heavy Ion Collider (RHIC). This technique is based on concurrent measurements of the beam and target (jet) spin correlated asymmetries in the elastic proton--proton scattering in the Coulomb-Nuclear Interference (CNI) region\,\cite{bib:KL,bib:jetConcept,bib:Bunce}. Due to the identity of the beam and target particles, polarization of the accelerated proton beam is directly related to the proton target polarization, which can be determined precisely by a conventional Breit--Rabi polarimeter\cite{bib:BreitRabi}. The polarimeter target is a polarized atomic hydrogen beam (H-jet), which crosses the RHIC beam in the vertical direction. It is nearly a point-like target, which does not introduce any perturbation of recoil protons , due to the absence of any target walls or windows. With the high intensity circulating proton current ($\sim\!2\!\times\!10^{13}\,p$ at a $70\,\text{kHz}$ revolution frequency) in the storage rings and the jet thickness of $\sim\!10^{12}\,\text{atoms/cm}^2$, the event rate (for a reasonable detector acceptance) is sufficiently high to achieve $\sim5\%$ statistical error accuracy in an hour of measurement. At this target thickness, the polarimeter is non-destructive, i.e. it can be operated continuously, without any effect on the beams and with no background generation for the other experiments.

In the RHIC Spin Program\,\cite{bib:SpinProgram}, this technique was implemented in the Polarized Atomic Hydrogen Gas Jet Target\,\cite{bib:ABS} (HJET) which has been in operation since 2004. The original purpose of HJET was to measure the absolute proton beam polarization at $100\text{--}255\,\text{GeV}$ with accuracy better than $\delta P/P\lesssim5\%$\,\cite{bib:Bunce,bib:SpinProgram} and to calibrate the fast p-Carbon CNI polarimeter\,\cite{bib:pC} which was considered as the main tool to monitor the proton beam polarization. 

\begin{figure}[t]
  \begin{center}
    \includegraphics[width=0.6\columnwidth]{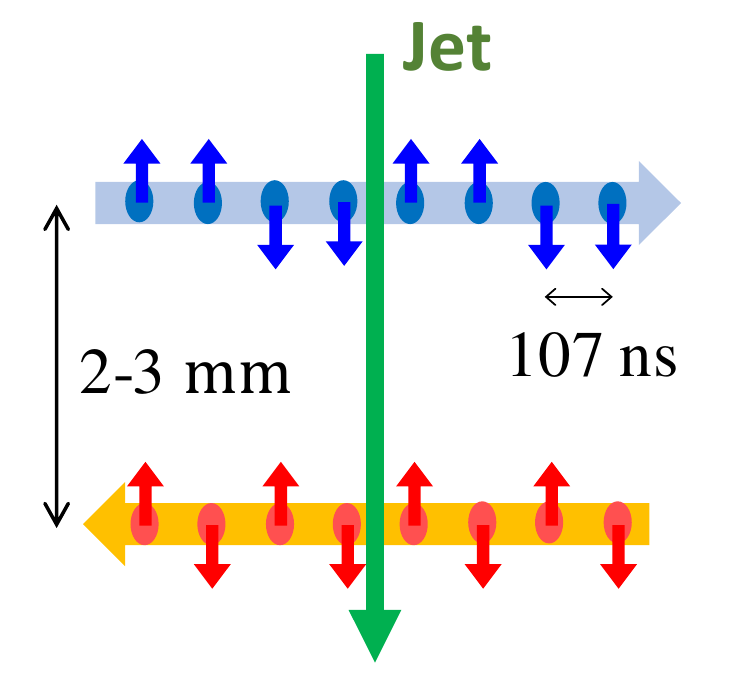}
  \end{center}
\caption{\label{fig:Beams} The RHIC polarized proton beams crossing the jet.}
\end{figure}
\begin{figure}[t]
\begin{center}
\includegraphics[width=0.95\columnwidth]{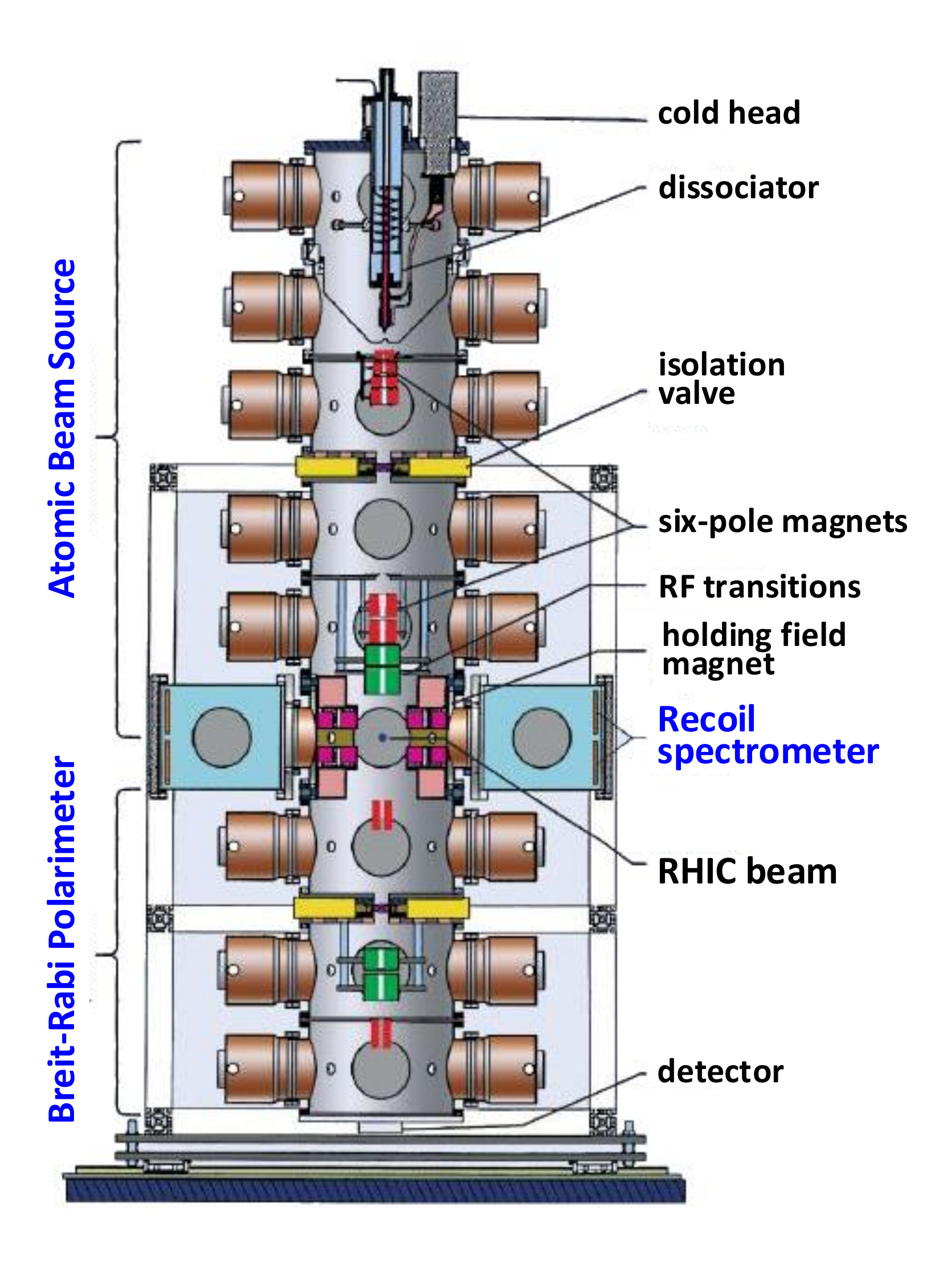}
\end{center}
\caption{The beam view of the Polarized Atomic Hydrogen Gas Jet Target polarimeter at RHIC.}
\label{fig:ABS}
\end{figure}

It was proven in the data analysis\,\cite{bib:HJET06, bib:HJET09}, that the HJET performance did satisfy the design goals.   The systematic error in absolute proton beam polarization measurement was dominated by uncertainty of the molecular hydrogen (or water vapor) contamination of the jet atomic beam and was evaluated to be $\delta P/P\!\sim\!3\%$. However, statistical accuracy of the measurements was relatively large, $\delta P\!\sim\!3\%$ for an 8 hour exposition.
Since the HJET operation stability appeared to be much better than that of the p-Carbon polarimeter, it was also realized that the continuous calibration of the p-Carbon is important to maintain the systematic errors under control.

The HJET detectors and data acquisition were upgraded in 2015\,\cite{bib:PSTP2015}.
The upgrades included the installation of new Si strip detectors, which provided better energy resolution and a significantly larger solid angle. The wave form digitizers were upgraded from CAMAC, 8 bit, 140\,MHz to VME, 12 bit, 250\,MHz.
Subsequently, the statistical error of the beam polarization measurements was greatly reduced to about $\delta P\!\sim\!2\%$ for an $8 \,\text{h}$ store. Large statistics of about $10^9$ elastic events per beam per Run acquired during RHIC Run\,15 ($E_\text{beam}=100\,\text{GeV}$)\,\cite{bib:Run15} and Run\,17 ($255\,\text{GeV}$)\,\cite{bib:Run17} allowed us to study in depth systematic uncertainties and to develop an efficient background subtraction method.  As a result, the systematic errors in the beam polarization measurements were reduced to $\sigma_P^\text{syst}/P\!\lesssim\!0.5\%$. Also, the elastic $\mathit{pp}$ single $A_\text{N}(t)\!\sim\!0.04$ and double $A_\text{NN}(t)\!\sim\!0.002$ spin analyzing powers have been precisely, $|\delta A_\text{N,NN}(t)|\!\sim\!2\!\times\!10^{-4}$, measured\,\cite{bib:HJET19} in the momentum transfer squared range of $0.001\!<\!-t\!<\!0.020\,\text{GeV}^2$.

In this paper we present the detailed description of the HJET data processing including the analysis of the systematic errors in  the RHIC proton beam polarization and elastic $\mathit{pp}$ analyzing powers measurements. Based on the obtained results, perspectives of using the HJET based absolute polarimeters at future Electron Ion Collider (EIC) will be also discussed.

\section{Polarized proton beams at RHIC}

RHIC is the world's first and only polarized proton--proton collider\,\cite{bib:RHIC}. The two independent storage rings, known as: {\em blue} and {\em yellow}, have six intersection points (IP's), where beam collisions are possible. At HJET location (IP12), both vertically polarized proton beams cross the jet as depicted in Fig\,\ref{fig:Beams}. The beam separation vertically was 3\,mm in RHIC Run\,15 (100\,GeV) and 2\,mm in Run\,17 (255 GeV).

Each beam consisted of 111 bunches with 106.5\,ns spacing. The beam intensity was $(2.0\text{--}2.3)\!\times\!10^{11}\,p/\text{bunch}$. The bunch spin pattern (see Fig.\,\ref{fig:Beams}) is produced in the source\,\cite{bib:OPPIS} fulfilling a requirement to minimize the spin correlated systematic uncertainties in the RHIC experiments. The proton beam polarization in RHIC Runs 15 \& 17, averaged over the beam intensity profile,  was about 55-60\% with a low $<\!1\%/h$ decay rate.

Since the RHIC transverse rms beam size of about 0.3--0.5\,mm is much smaller than the jet transverse profile, the average beam polarization and not its profile is measured by HJET.

\section{HJET polarimeter}

\begin{figure}[t]
  \begin{center}
    \includegraphics[width=0.8\columnwidth]{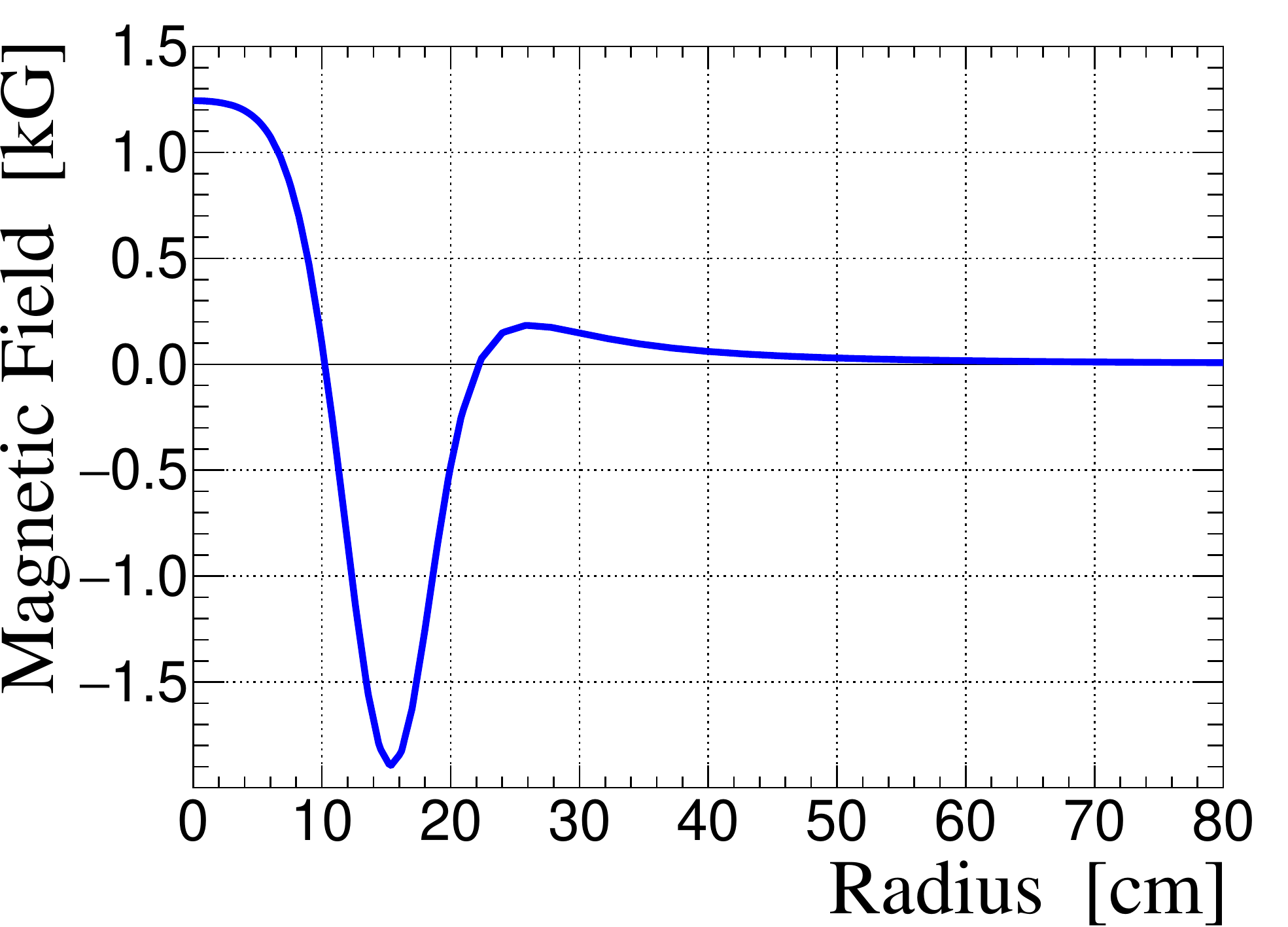}
  \end{center}
\caption{\label{fig:HoldingField} The holding magnetic field as a function of distance from the chamber center on the horizontal plane.}
\end{figure}

The HJET polarimeter\,\cite{bib:ABS} includes three major parts: the polarized Atomic Beam Source (ABS), a scattering chamber with a holding field magnet, and the Breit--Rabi polarimeter (BRP) (see Fig.\,\ref{fig:ABS}). The polarimeter axis is vertical and the recoil protons are detected in the horizontal plane in the direction perpendicular to the beam. The common vacuum system is assembled from nine identical vacuum chambers, which provide nine stages of differential pumping. The system building block is a cylindrical vacuum chamber 50\,cm in diameter and 32\,cm long. Each chamber is pumped by two turbomolecular pumps with 1000~l/s pumping speed each.

\subsection{The Atomic Beam Source}

The ABS part of HJET includes five vacuum chambers and five differential vacuum stages (see Fig.\,\ref{fig:ABS}). The distance from the nozzle to the collision point is 127 cm.
The dissociator design is described in detail in Ref.\,\cite{bib:ABS}. The combination of powerful vacuum system with a carefully designed separating magnet system allows for the delivery of a record atomic beam intensity of  $1.2\times10^{17}\,\text{atoms/s}$\,\cite{bib:ABS}. The atomic beam is focused precisely at the collision point, where the jet thickness is about $1.2\times10^{12}\,\text{atoms/cm}^2$ and the profile is well approximated by a Gaussian distribution with $\sigma\!\approx\!2.6\,\text{mm}$.

The proton spin direction  in HJET is controlled by switching of RF-transitions, the Strong Field Transition (SFT) and Weak Field Transition (WFT). Typically, the jet spin was reversed every 5 min.

The proton polarization is vertical in the scattering chamber and is determined by the holding field magnet strength. The magnetic field steering effect on the recoil protons is minimized by using compensation coils, whose fields are adjusted (see Fig.\,\ref{fig:HoldingField}) to keep the total vertical field integral along the proton paths close to zero.

\subsection{The Breit--Rabi polarimeter}
The RF-transitions are shielded from the longitudinal holding magnetic field by additional magnetic screens. The polarimeter magnet system was designed for high efficiency beam transport to the atomic beam detector. This allowed a very accurate measurement of the RF transition efficiencies, which appeared to be $99.9\pm0.1\%$ both for WFT and SFT. In this case the proton polarization in the jet hydrogen atoms is defined by the strength ($\sim\!1.2\,\text{kG}$) of the holding field magnet and, thus, is known with a high accuracy, e.g. in Run\,17 it was
\begin{equation}
  P_\text{jet}=0.957\pm0.001.
\end{equation}

\subsection{The recoil proton spectrometer}

\begin{figure}[t]
  \begin{center}
    \includegraphics[width=1.\columnwidth]{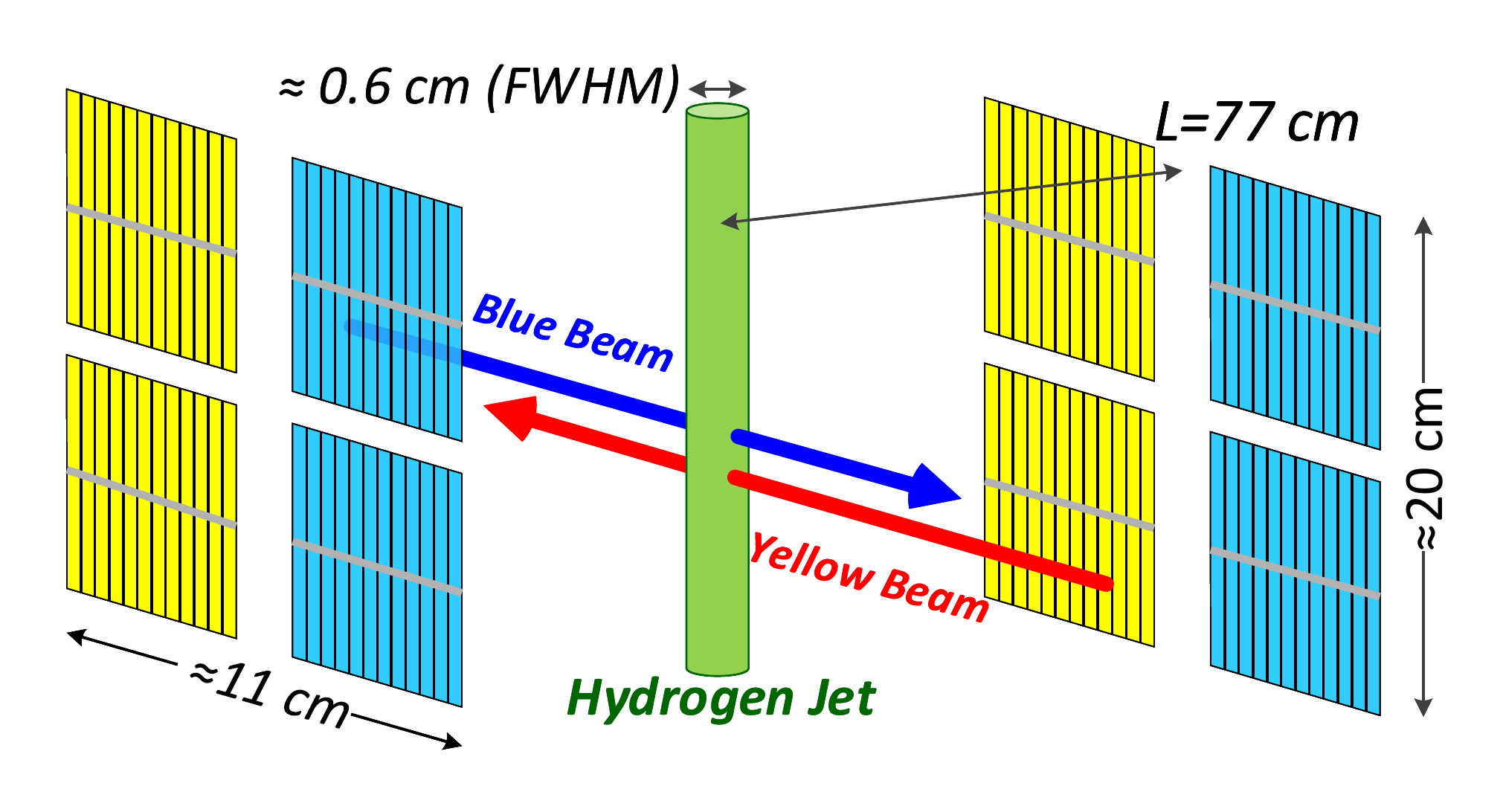}
  \end{center}
  \caption{ A schematic view of the HJET recoil spectrometer.
  }
  \label{fig:HjetView}
\end{figure}

The recoil protons from the RHIC beam scattering on the jet in the scattering chamber (Chamber\,6) are counted in the recoil spectrometer depicted in Fig.\,\ref{fig:HjetView}.

For the recoil proton measurements, we use 8 pairs of the Si strip detectors. Each detector consists of 12 vertically oriented strips of $3.75\!\times\!45\,\text{mm}^2$ area, $470\,\text{\textmu{}m}$ thickness, and a $\sim\!0.37\,\text{mg/cm}^2$ uniform dead-layer.

For elastic scattering, the spectrometer geometry  allows us to detect CNI recoil protons with kinetic energy up to $T_R\approx10\text{-}11\ \mathrm{MeV}$. Protons with an energy above $7.8\,\text{MeV}$ punch through the Si detector (only part of the proton kinetic energy is detected).

For signal readout we use a 12-bit $250\,\text{MHz}$ FADC250 wave-form digitizers\,\cite{bib:FADC250}. A full waveform (80 samples) is recorded for every triggered ($\sim\!0.5\,\text{MeV}$ threshold) event.

The recoil proton energy, time, and the signal shape parametrization determined in the waveform fit as well as the recoil angle tagged by the Si strip location provided an ability to reliably isolate the elastic $\mathit{pp}$ events.

\subsection{Special test measurements}

\begin{figure}[t]
  \begin{center}
    \includegraphics[width=0.8\columnwidth]{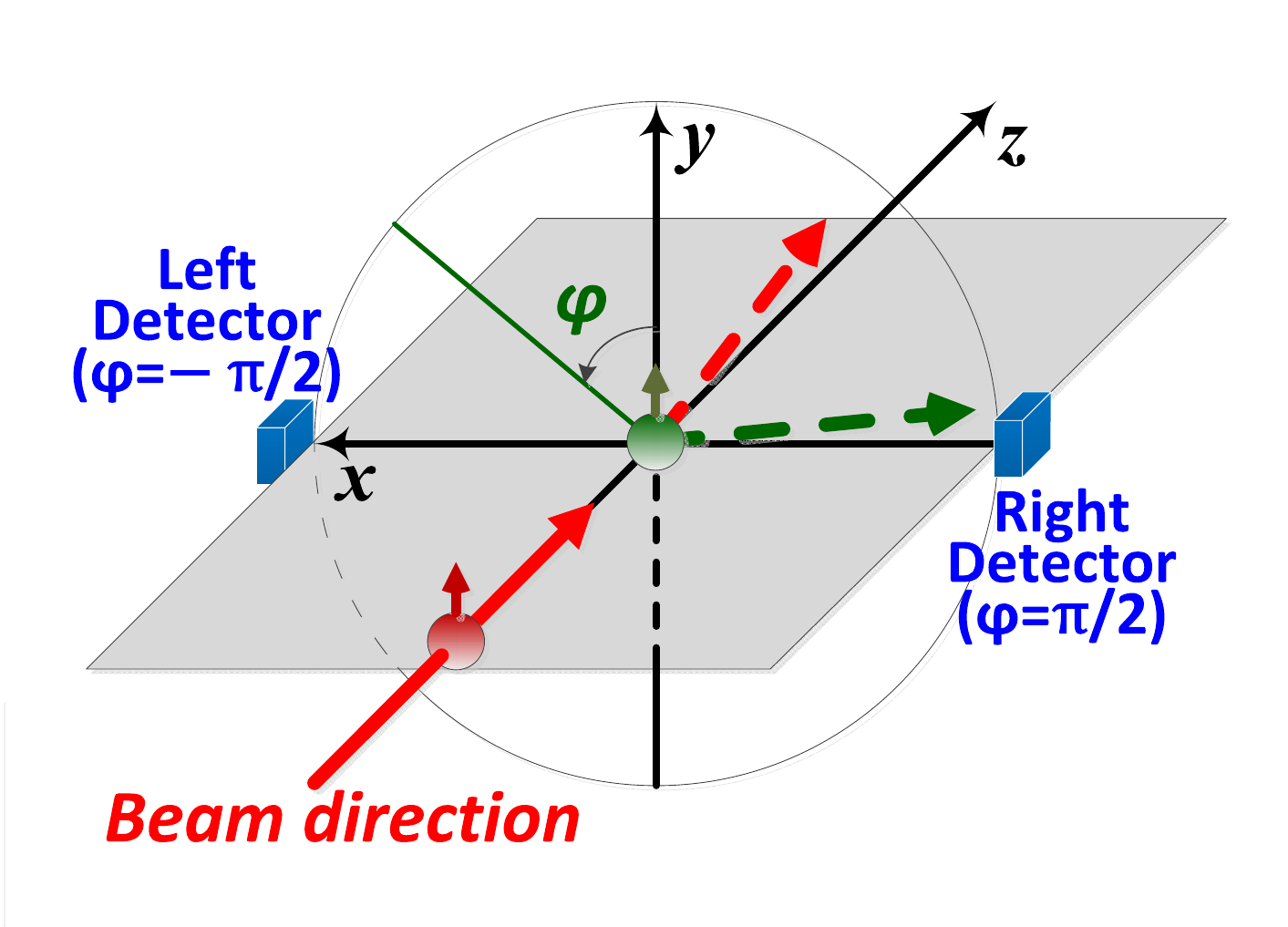}
    \caption{A schematic view of the $p^\uparrow p^\uparrow$ spin correlated asymmetry measurements at HJET. The recoil protons are counted in left/right symmetric detectors. The beam moves along the $z$-axis. The transverse polarization direction is along the $y$-axis.}
      \label{fig:Angles}
  \end{center}
\end{figure}

For the detailed study of the HJET performance including evaluation of the specific backgrounds, it was helpful to operate the HJET in special modes. For example, injecting $\text{H}_2$ (or $\text{O}_2$) gas (while the jet off) to Chamber\,7 allowed us to evaluate how the recoil proton tracking in the magnetic field affects the results of the measurements. This diffused $\text{H}_2$  to the scattering chamber emulated beam scattering off the beamline gas.

To evaluate the atomic beam contamination by $\text{H}_2$, we carried out the measurement with the dissociator RF-power off.

Special tests also included the empty target (jet off) and single RHIC beam measurements.

HJET was routinely operated in parasitic mode during Heavy Ion RHIC Runs. This not only allowed us to study the analyzing power of the $p^\uparrow\text{A}$ scattering in the CNI region for many nucleus species but to also conduct various tests of HJET.

\section{General analysis of the spin dependent measurements at HJET}

\begin{figure}[t]
  \begin{center}
    \includegraphics[width=0.8\columnwidth]{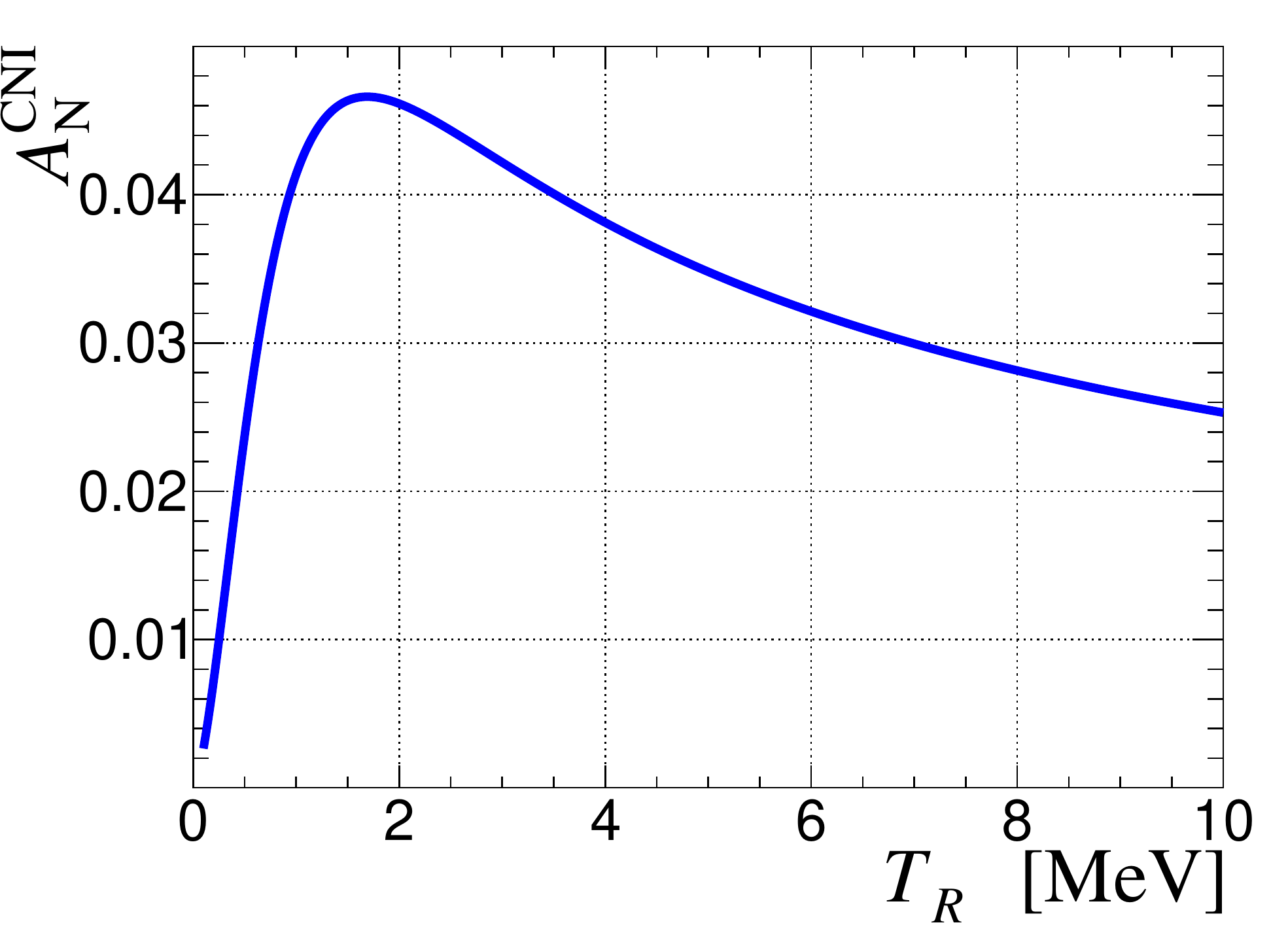}
    \caption{A basic theoretical prediction\,\cite{bib:KL} for the elastic $\mathit{pp}$ single spin analyzing power at RHIC energies.}
      \label{fig:AN0}
  \end{center}
\end{figure}

\subsection{Spin dependent correlations}

For the vertically polarized proton beam and target (the jet), the recoil proton azimuthal angle $\varphi$ distribution can be written\,\cite{bib:Convention} by
\begin{equation}
\begin{aligned}
\frac{d^2\sigma}{dtd\varphi} = \frac{1}{2\pi}\frac{d\sigma}{dt}\times%
\Big[1+\left(A_\text{N}^jP_j+A_\text{N}^bP_b\right)\sin{\varphi} \\ 
  +\left(A_\text{NN}\sin^2{\varphi}+A_\text{SS}\cos^2{\varphi}\right)P_bP_j  \Big].
\label{eq:dsdt}
\end{aligned}
\end{equation}
The positive signs of the beam $P_b$ and target $P_j$ polarizations correspond with the spin up direction. In Eq.\,(\ref{eq:dsdt}), $\varphi$ is defined in accordance with Fig.\,\ref{fig:Angles}. Since $\cos{\varphi}\approx0$ at HJET, the measurements are not sensitive to $A_\text{SS}$. The analyzing powers $A_\text{N}^{b,j}(t,s)$ and $A_\text{NN}(t,s)$ are, generally, functions of the central mass energy squared $s$ and 4-momentum transfer squared $t$, which for the fixed target proton--proton scattering can be expressed as
\begin{equation}
s=2m_p(m_p+E_\text{beam}), ~~~~ t=-2m_pT_R.
\end{equation} 
where $E_\text{beam}$ is the beam energy, $T_R$ is the recoil proton kinetic energy, and $m_p$ is a proton mass. 

For elastic $p^\uparrow p^\uparrow$ scattering, the target and beam single spin analyzing powers are the same, $A_\text{N}^j\!=\!A_\text{N}^b\!=\!A_\text{N}$. At the HJET, i.e. for small angle scattering at high energies, the analyzing power can be well approximated\,\cite{bib:KL} by the CNI term
\begin{equation}
  A_\text{N}^{\text{CNI}}(s,t) \approx \sqrt{\frac{2T_R}{m_p}}\:\frac{\varkappa\,T_c/T_R}{(T_c/T_R)^2+1}
  \label{eq:AN_KL}
\end{equation}
where $\varkappa\!=\!\mu_p\!-\!1\!=\!1.793$ is proton's anomalous magnetic moment, $T_c\!=\!4\pi\alpha/m_p\sigma_\text{tot}\!\approx\!1\,\text{MeV}$, and  $\sigma_\text{tot}(s)\!\sim\!40\,\text{mb}$ is proton--proton total cross section. $A_\text{N}^\text{CNI}(T_R)$, shown in Fig.\,\ref{fig:AN0}, has a maximum of about 0.047 at $T_R\!=\!\sqrt{3}T_c\!\approx\!1.7\,\text{MeV}$.

The HJET  measured\,\cite{bib:HJET19} single $A_\text{N}(T_R)$ and double $A_\text{NN}(T_R)$ spin analyzing powers are shown in Figs.\,\ref{fig:AN} and \ref{fig:ANN}, respectively.

\subsection{Single spin analyzing power}

Using the HJET, the beam polarization can be determined with no prior knowledge of $A_\text{N}(T_R)$.  However, since the analyzing power can be precisely measured at HJET, a theoretical understanding of $A_\text{N}(T_R)$ is important for the interpretation of the results of this study. A detailed theoretical analysis of $A_\text{N}(t,s)$ and $A_\text{NN}(t,s)$ was given in Ref.\,\cite{bib:BKLST}. In this approach, the single spin analyzing power can be written as
\begin{equation}
  A_\text{N}(T_R) = A_\text{N}^{(0)}(T_R)\times \alpha_\text{N}\,\left(1+\beta_\text{N}T_R/T_c\right)
  \label{eq:AN_norm}
\end{equation}
where $A_\text{N}^{(0)}(T_R)$ is the analyzing power if the hadronic spin--flip amplitude is disregarded. For the elastic $\mathit{pp}$ scattering, $A_\text{N}^{(0)}(T_R)$ can be calculated\,\cite{bib:BKLST} using QED helicity amplitudes in one photon approximation\,\cite{bib:BGL} and the experimental results\,\cite{bib:PDG} for the $\mathit{pp}$ differential cross-section and for the proton rms charge radius. Such defined $A_\text{N}^{(0)}$ is consistent with $A_\text{N}^\text{CNI}$ with the relative accuracy of about 5\%. Possible corrections to  $A_\text{N}^{(0)}$, e.g. due to the hadronic spin flip amplitudes\,\cite{bib:BKLST}, absorptive corrections\,\cite{bib:AbsorptiveCorr}, etc. can be approximated by a linear function of  $T_R$, parametrized by two factors\,\cite{bib:HJET19},
\begin{equation}
  \alpha_\text{N}=1+{\cal O}(10^{-2})\quad\text{and}\quad \beta_\text{N}={\cal O}(10^{-2}).
\end{equation}

In Ref.\,\cite{bib:BKLST}, the expression for $A_\text{N}(t,s)$ was given with some simplifications. However, for the experimental accuracy achieved at HJET, some of the omitted corrections to $A_\text{N}(t,s)$\,\cite{bib:AbsorptiveCorr,bib:AN_corr} should be taken into account.

\subsection{Experimental determination of the spin correlated asymmetries}

At the HJET, three spin-correlated asymmetries
\begin{equation}
  a_\text{N}^j\!=\!A_\text{N}^j|P_j|,\:\:
  a_\text{N}^b\!=\!A_\text{N}^b|P_b|,\:\:
  a_\text{NN}\!=\!A_\text{NN}|P_jP_b|
  \label{eq:asym}
\end{equation}
can be determined (see \ref{app:SqrtFormula}) from the elastic events discriminated by the left/right ($\text{LR}$) detector location and the beam ($\uparrow\downarrow$) and jet ($+-$) spin directions:
\begin{equation}
\begin{aligned}
  N^{(\uparrow\downarrow)(+-)}_\text{RL}\!\propto\!\left(1\!+\!\eta_\epsilon\eta_ja_\text{N}^\text{j}\!+\!\eta_\epsilon\eta_ba_\text{N}^\text{b}\!+\!\eta_j\eta_ba_\text{NN}\right)\\
  \times
  \left(1+\eta_j\lambda_\text{j}\right)
  \left(1+\eta_b\lambda_\text{b}\right)
  \left(1+\eta_\epsilon\epsilon\right)
  \label{eq:N}
\end{aligned}
\end{equation}
where $\lambda_\text{j,b}$ are the jet/beam intensity asymmetries and $\epsilon$ is the right/left detector acceptance asymmetry. The sign of the $\eta_j(+-)=\pm1$ is defined by the jet spin $(+-)$  and similarly for $\eta_b(\uparrow\downarrow)$ and $\eta_\epsilon(\text{RL})$.

\subsection{General description of the systematic errors' sources \label{sec:SystErr}}

In a basic approximation, i.e. if all asymmetries, $a_\text{n}^{j,b}$, $a_\text{NN}$, $\lambda^{j,b}$, and $\epsilon$, used in Eqs.\,(\ref{eq:N}) are mutually uncorrelated, the solution given by (\ref{eq:a_Nj})--(\ref{eq:epsilon}) is free of systematic errors.
However, more commonly, the possible correlations between the asymmetries might result in the systematic uncertainties in the measurements.

 Generally, the {\it effective} analyzing power $A_\text{N}^\text{(eff)}(T_R)$ in a measurement differs from the actual one, $A_\text{N}(T_R)$, due to the background and/or the errors in the energy calibration:
 \begin{equation}
   \begin{aligned}
     \delta A_N = & A_\text{N}^\text{(eff)} - A_\text{N}
     \\ = & \frac{b}{1+b}\left(A_N^\text{(bgr)}\!-\!A_N\right)\!-\!
     \frac{dA_N(T_R)}{dT_R}\delta T_R
  \end{aligned}
\end{equation}
where $b$ is the background to the signal ratio and $A_N^\text{(bgr)}$ is the effective analyzing power for background events. $\delta A_N$ may be interpreted as a systematic error in the definition of the analyzing power. Commonly, it is not the same for the left/right detectors and for the beam/jet spin asymmetries. For example, for the proton beam scattering on the beam gas unpolarized hydrogen, $A_N^\text{(bgr)}=0$ for the jet spin, while $A_\text{N}^\text{(bgr)}=A_\text{N}$ for the beam spin.

A possible spin correlated asymmetry of the detector acceptance $\omega$,
\begin{equation}
  \delta\omega = \frac{\omega^\uparrow - \omega^\downarrow}%
  {\omega^\uparrow + \omega^\downarrow},
\label{eq:omega}
\end{equation}
can lead to a systematic error in the measured $a_\text{N}$.

In Run\,15, a significant electronic noise correlation with the WFT (i.e. with the jet spin) was observed in two Si detectors. Subsequently, the event selection efficiency in these detectors appeared to be the jet spin dependent giving non-zero $\delta\omega$

It should also be kept in mind, that the absolute values of the polarization for the up and down spins are not necessarily the same. In this paper, except Eq.\,(\ref{eq:dsdt}), we assume 
\begin{equation}
   P = |P| = \frac{|P^\uparrow|\!+\!|P^\downarrow|}{2},\quad%
  \delta P = \frac{|P^\uparrow|\!-\!|P^\downarrow|}%
  {|P^\uparrow|\!+\!|P^\downarrow|}.
\end{equation}
For the HJET data analysis, $\delta P_\text{jet}\!=\!\delta P_\text{beam}\!=\!0$ was found to be an acceptable approach. 

In summary, the leading order systematic errors in the measured jet (beam) asymmetries can be given by  
\begin{align}
  \delta a_N^\text{syst} &= P\,%
  \frac{\delta A_N^{(R)}\!+\!\delta A_N^{(L)}}{2} +
  \frac{\delta\omega_R\!-\!\delta\omega_L}{2}
  \label{eq:aN_syst}
  \\
  \delta\lambda^\text{syst} &= P\,%
  \frac{\delta A_N^{(R)}\!-\!\delta A_N^{(L)}}{2} +
  \frac{\delta\omega_R\!+\!\delta\omega_L}{2}
  \label{eq:Lambda_syst}
  \\
  \delta\epsilon^\text{syst} &= \delta P\,A_\text{N}
  \label{eq:accept_syst}
\end{align}
where $\delta A_N^{(L,R)}$ and $\delta\omega_{L,R}$ are the average corrections in left/right detectors for the jet (beam) analyzing power and for the acceptance dependence on the jet (beam) spin direction, respectively.

\subsubsection{Routine control of systematic errors}

For elastic $\mathit{pp}$ scattering and a systematic error free measurement, the measured ratio $a_\text{N}^b/a_\text{N}^j$ as well as the measured beam and jet intensity asymmetries $\lambda^{b,j}$ must not depend on the recoil proton energy $T_R$:
\begin{equation}
  \frac{d\left(a_\text{N}^b/a_\text{N}^j\right)}{dT_R} =
  \frac{d\lambda^b}{dT_R} =
  \frac{d\lambda^j}{dT_R} =
  b_\text{NN}(T_R) =
  0.  \label{eq:a(T_R)}
\end{equation}
Also, both the jet and beam spin {\em normalized} asymmetries
\begin{equation}
  a_n^{j,b} = a_\text{N}^{j,b}/A_\text{N}^{(0)} =
  P_{j,b}\,\alpha_\text{N}\,\left(1+\beta_\text{N}T_R/T_c\right)
  \label{eq:a_n(T_R)}
\end{equation}
[see Eq.\,\ref{eq:AN_norm})] are expected to be linear functions of $T_R$ with the same factors $\alpha_\text{N}$ and  $\beta_\text{N}$. 

Since the possible $\delta A_\text{N}$ and $\delta\omega$ are expected to be significantly dependent on $T_R$, the experimental tests of Eqs. (\ref{eq:a(T_R)}) and (\ref{eq:a_n(T_R)}) are critically important for detecting systematic errors in the HJET measurements.

\section{Recoil proton kinematics}

\subsection{Elastic $\mathit{pp}$ scattering \label{sec:kappa}}
For elastic proton--proton scattering, there is a strict correlation between the $z$ coordinate of the recoil proton in the detectors and its kinetic energy
\begin{equation}
  z\!-\!z_\text{jet}=L\sqrt{\frac{T_R}{2m_p}%
    \frac{E_\text{beam}\!+\!m_p}%
         {E_\text{beam}\!-\!m_p\!+\!T_R}}=\kappa_p\sqrt{T_R}.
  \label{eq:kappa}
\end{equation}
Here, $z_\text{jet}$ is the coordinate of the scattering point (in the jet). At HJET, $\kappa_p$ has only a weak dependence on the beam energy, $\kappa_p\!=\!17.92\,\text{mm/MeV}^{1/2}$ (100 GeV) and  $\kappa_p\!=\!17.82\,\text{mm/MeV}^{1/2}$ (255 GeV).  

For the vertically oriented Si strips with fixed 3.75 mm $z$ coordinate spacing, it is convenient to define the strip-corresponding energy
\begin{equation}
  \sqrt{T_\text{strip}} = \frac
       {z_\text{strip} -\left\langle z_\text{jet}\right\rangle}
       {\kappa_p}.
       \label{eq:Tstrip}
\end{equation}
The average recoil proton energy shift between two consecutive strips is
\begin{equation}
  \Delta\sqrt{T_\text{strip}}\approx 0.210\,\text{MeV}^{1/2}
  \label{eq:DeltaStrip}
\end{equation}

Eq.\,(\ref{eq:kappa}) allows one to relate the recoil proton energy distribution in a Si strip 
\begin{equation}
  \frac{dN_\text{strip}}{d\sqrt{T_R}} \propto \frac{d\sigma}{dt}\sqrt{T_R} \times
  f\left(\kappa_p\sqrt{T_R}-\kappa_p\sqrt{T_\text{strip}}\right).
  \label{eq:jetProfile}
\end{equation}
to the jet target density profile $f(z-\langle z_\text{jet}\rangle)$. Such an evaluated jet profile must be the same for all Si strips. However, due to the actual strip width of 3.75\,mm, the measured profile rms will be overestimated by about 9\%.

\subsection{Inelastic $\mathit{pp}$ scattering}

For inelastic scattering, $p\!+\!p\to X\!+\!p$, we should replace in Eq.\.(\ref{eq:kappa}) with 
\begin{equation}
  \begin{aligned}
    \kappa_p\to\kappa'_p=\kappa_p\!\times
  \left(1+\frac{m_p\Delta}{T_RE_\text{beam}}\right),\\
  \Delta = M_X-m_p\ge m_\pi.
  \end{aligned}
  \label{eq:kappa'}
\end{equation}
For the 100\,GeV proton beam, the inelastic events are mostly beyond the detector's acceptance. However, for the $255\,\text{GeV}$ beam, the acquired $\mathit{pp}$ data contamination by the inelastic events should be properly treated in the data analysis.

\subsection{Beam ion scattering on the jet proton}

We also operated HJET in the Heavy Ion RHIC Runs. The main goal was to measure the $p^\uparrow A$ analyzing powers for several nuclei, Au, Ru, Zr, Al, and d. Since the beam energy $E_\text{beam}$ is reported at RHIC in GeV/nucleon units, the equivalent jet proton energy in the ion frame is equal to
\begin{equation}
  E_p = E_\text{beam}\,\frac{m_pA}{M}\approx E_\text{beam}
\end{equation}
where $A$ is the ion's mass number and $M$ is its mass. For the elastic scattering of the ion beam on the jet proton, one substitutes in Eq.\,(\ref{eq:kappa}) with
\begin{equation}
  \kappa_A \approx \frac{L}{\sqrt{2m_p}}\,\sqrt{\frac{E_p\!+\!m_p^2/M}{E_p\!-\!m_p^2/M}}.
  \label{eq:kappaA}
\end{equation}
For heavy ions, e.g. for Au, $\kappa_A\!\approx\!L/\sqrt{2m_p}\!=\!17.76\,\text{mm/MeV}^{1/2}$ is beam energy independent. In case of an inelastic scattering of the ion, $M_X\!=\!M\!+\!\Delta$, the correction factor to $\kappa_A$ is the same as given in Eq.\,(\ref{eq:kappa'}), if we substitute $E_\text{beam}\!\to\!E_p$.

\subsection{Tracking in the holding field magnet \label{sec:mfTracking} }

\begin{figure}[t]
  \begin{center}
    \includegraphics[width=0.8\columnwidth]{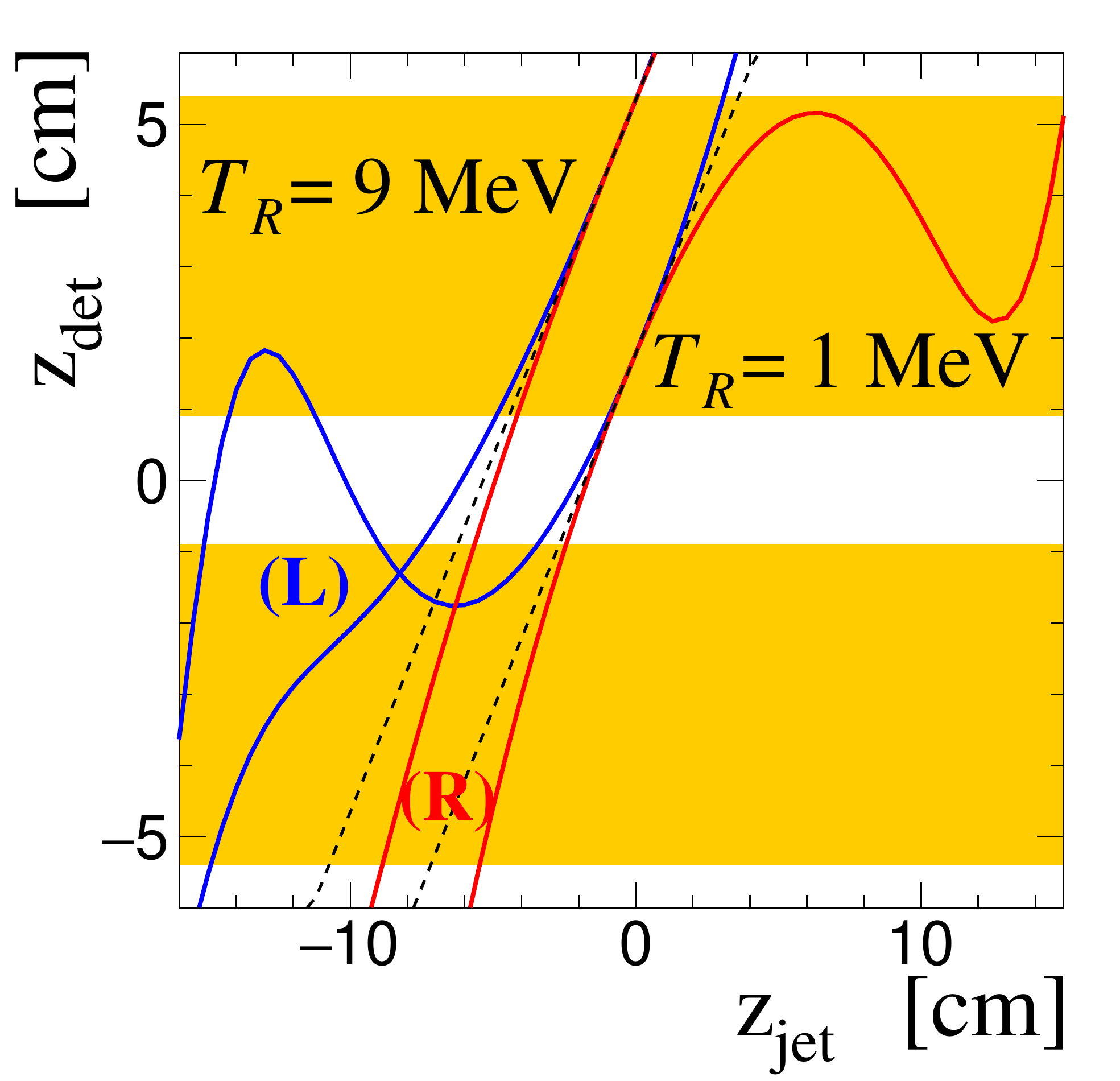}
  \end{center}
  \caption{\label{fig:TracksMF} 
    The correlation between the $z$-coordinates in the detectors (filled areas) $z_\text{det}$ and in a vertex $z_\text{jet}$ for $1\,\text{MeV}$ and $9\,\text{MeV}$ recoil protons. The results for left and right detectors are labeled by (L) and (R), respectively. Dashed lines show the dependencies if there is no magnetic field.}
\end{figure}

The magnetic field (see Fig.\,\ref{fig:HoldingField}) may significantly alter the correlation (\ref{eq:kappa}) between the recoil proton's energy and coordinate in the Si detector
\begin{equation}
  z \to z + b_\text{MF}/\sqrt{T_R}.
  \label{eq:dz}
\end{equation}
  The displacement factor $b_\text{MF}$ is defined by the field integral 
\begin{equation}
 |b_\text{MF}| = 
\frac{qL}{c}\!\times\!%
\int_0^L{\left(1\!-\!\frac{r}{L}\right)\,\frac{H(r)dr}{\sqrt{2m_p}}}.
\label{eq:bMF}
\end{equation}
where $qL/c\!=\!2.3\times10^{-2}\,\text{MeV/G}$ and $H(r)$ is vertical component of the holding magnetic field shown in Fig.\,\ref{fig:HoldingField}. The sign of $b_\text{MF}$ is different for the left and right detectors. To minimize the effect, the magnet coils currents are adjusted to keep the displacement factor close to zero.

A horizontal beam shift $\Delta x$  can also contribute to (\ref{eq:dz})
\begin{equation}
  |b'_\text{MF}| = \frac{qL}{c}\!\times\!%
  \frac{H(0)\Delta x}{\sqrt{2m_p}}.
    \label{eq:b'MF}
\end{equation}

\begin{figure*}[t]
  \begin{minipage}[t]{\columnwidth}
  \begin{center}
    \includegraphics[width=0.8\columnwidth]{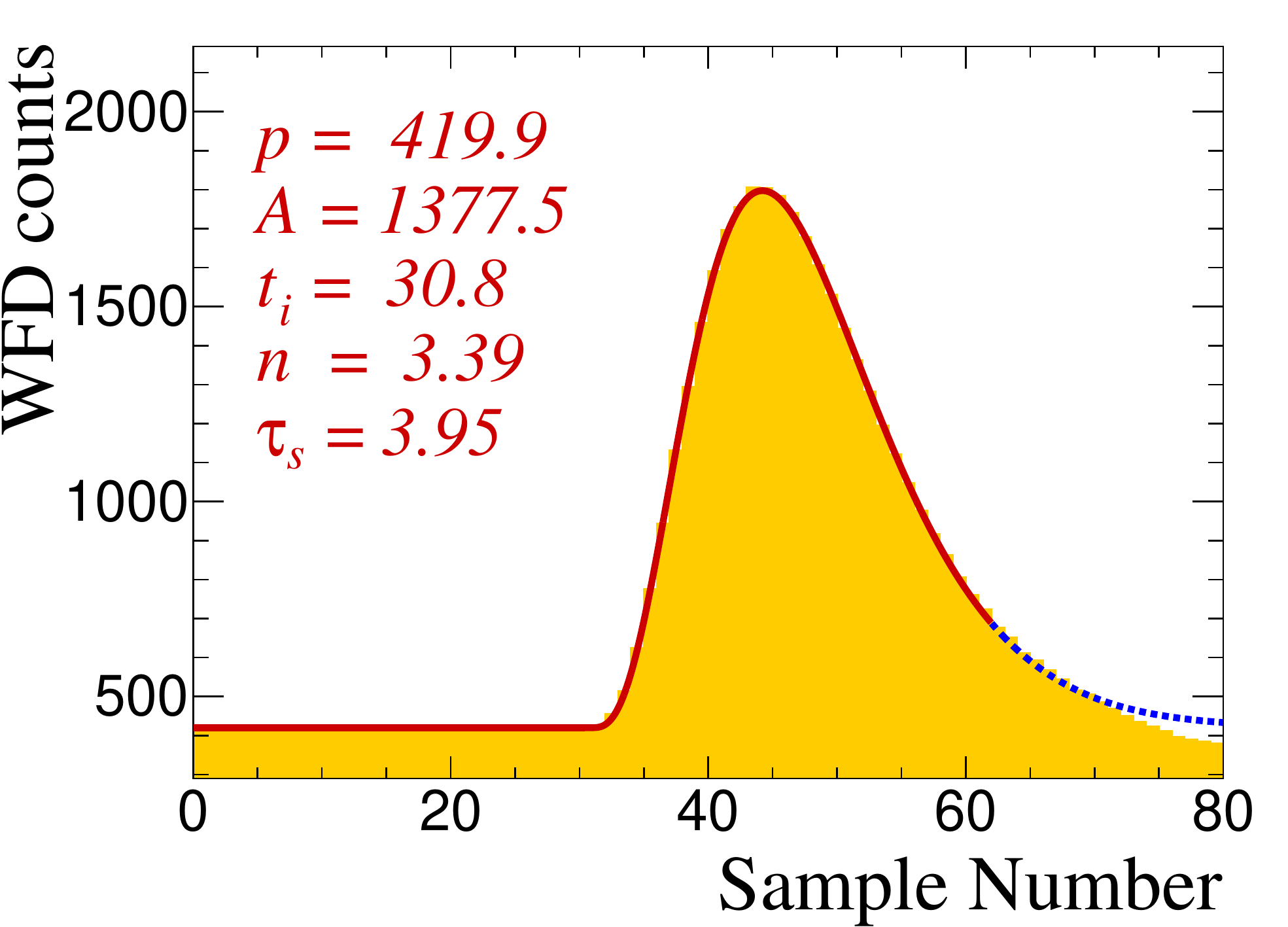}
  \end{center}
  \caption{\label{fig:Waveform} A signal waveform in HJET (the filled histogram). The solid red line indicates the time interval which is used in the waveform fit. The dashed blue line is the waveform function $W(t)$ beyond this interval. The sampling time is about $4.1\,\text{ns}$.}
  \end{minipage} \hfill \begin{minipage}[t]{\columnwidth}
  \begin{center}
    \includegraphics[width=0.8\columnwidth]{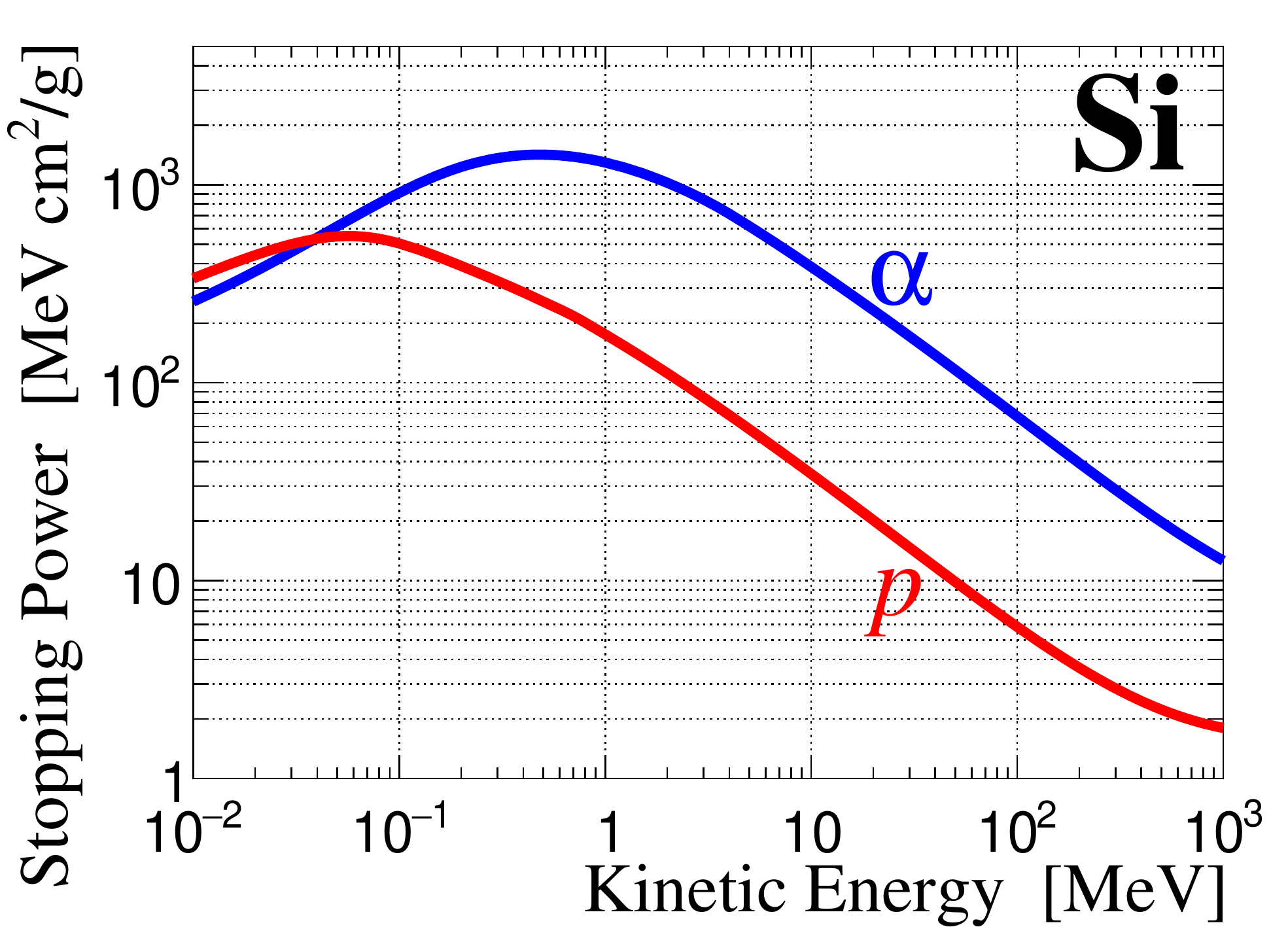}
  \end{center}
  \caption{\label{fig:StoppingPower} Stopping power in silicon material for \textalpha-particle and proton as a function of the injection energy of each particle.}
\end{minipage}
\end{figure*}

\begin{figure*}[h]
  \begin{center}
    \includegraphics[width=0.95\textwidth]{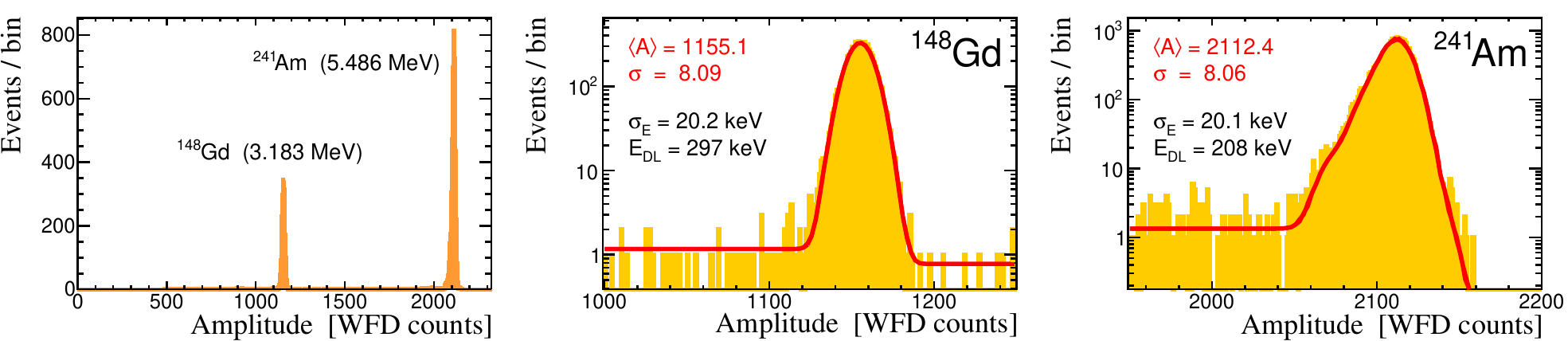}
    \caption{\label{fig:AlphaCalib} Signal amplitude distribution in the \textalpha-source calibration. For ${}^{241}\mathrm{Am}$, multiple peaks around $5.486\,\text{MeV}$ are taken into account in the fit.}
  \end{center}
\end{figure*}

Factors $b_\text{MF}$ and $b'_\text{MF}$ effectively alter $\Delta\sqrt{T_\text{strip}}$ (\ref{eq:DeltaStrip}) depending on the strip location and, thus, can be determined in the data analysis. In RHIC Run\,15, it was
$|b_\text{MF}|\!\sim\!0.7\,\text{mm}\!\cdot\!\text{MeV}^{1/2}$
and
$|b'_\text{MF}|\!\sim\!2\,\text{mm}\!\cdot\!\text{MeV}^{1/2}$.

The magnetic field correction (\ref{eq:dz}) was evaluated assuming beam scattering at $z\!=\!\langle z_\text{jet}\rangle$. It will be shown below that the beam scattering on the beam gas hydrogen in a wide range of $z$ is important for evaluation of the background contribution. For such events, the dependence (\ref{eq:kappa}) can also be strongly violated. The calculated dependencies $z_\text{det}=f(z_\text{jet},T_R)$ for two recoil proton energies, $1\,\text{MeV}$ and $9\,\text{MeV}$ are shown in Fig.\,\ref{fig:TracksMF}.

According to (\ref{eq:kappa}), for any given recoil proton energy $T_R$, this background event rate is expected to be the same in all Si strips (neglecting the small variation of the strips acceptance). For small $T_R$, the strip-to-strip rate alteration may be substantial due to the magnetic field.

\section{Data Processing}

\subsection{Signal waveform parametrization}

 In the data analysis, the signal shape (Fig.\,\ref{fig:Waveform}) was parametrized by the following function 
\begin{equation}
W(t)=p+A\,\left(\frac{t-t_i}{n\tau_s}\right)^n\exp\left(-\frac{t-t_i}{\tau_s}+n\right)
\label{eq:Waveform}
\end{equation}
if $t\!>\!t_i$ and $W(t)\!=\!0$ if ($t\!<\!t_i$). Here, $p$ is the pedestal, $A$ is signal amplitude, $t_i$ is signal start time, $n$ and $\tau_s$ are signal shape parameters.  The waveform has maximum $W(t_m)\!=\!p\!+\!A$  at $t_m\!=\!t_i\!+\!n\tau_s$.

Due to energy losses in the Si detector entrance window, the reconstructed kinetic energy of the recoil proton is expressed via the measured signal amplitude as
\begin{equation}
  T_R = gA+E_\text{loss}(gA,x_\text{DL})
  \label{eq:DL}
\end{equation}
where $g$ is the ADC gain and $x_\text{DL}$ is the dead layer thickness. Energy losses $E_\text{loss}$ were evaluated using stopping power dependence on $T_R$\,\cite{bib:StoppingPower} shown in Fig.\,\ref{fig:StoppingPower}.

\subsection{Energy calibration of the Si detectors}

\begin{figure*}[t]
  \begin{minipage}[t]{\columnwidth}
    \begin{center}
      \includegraphics[width=0.8\columnwidth]{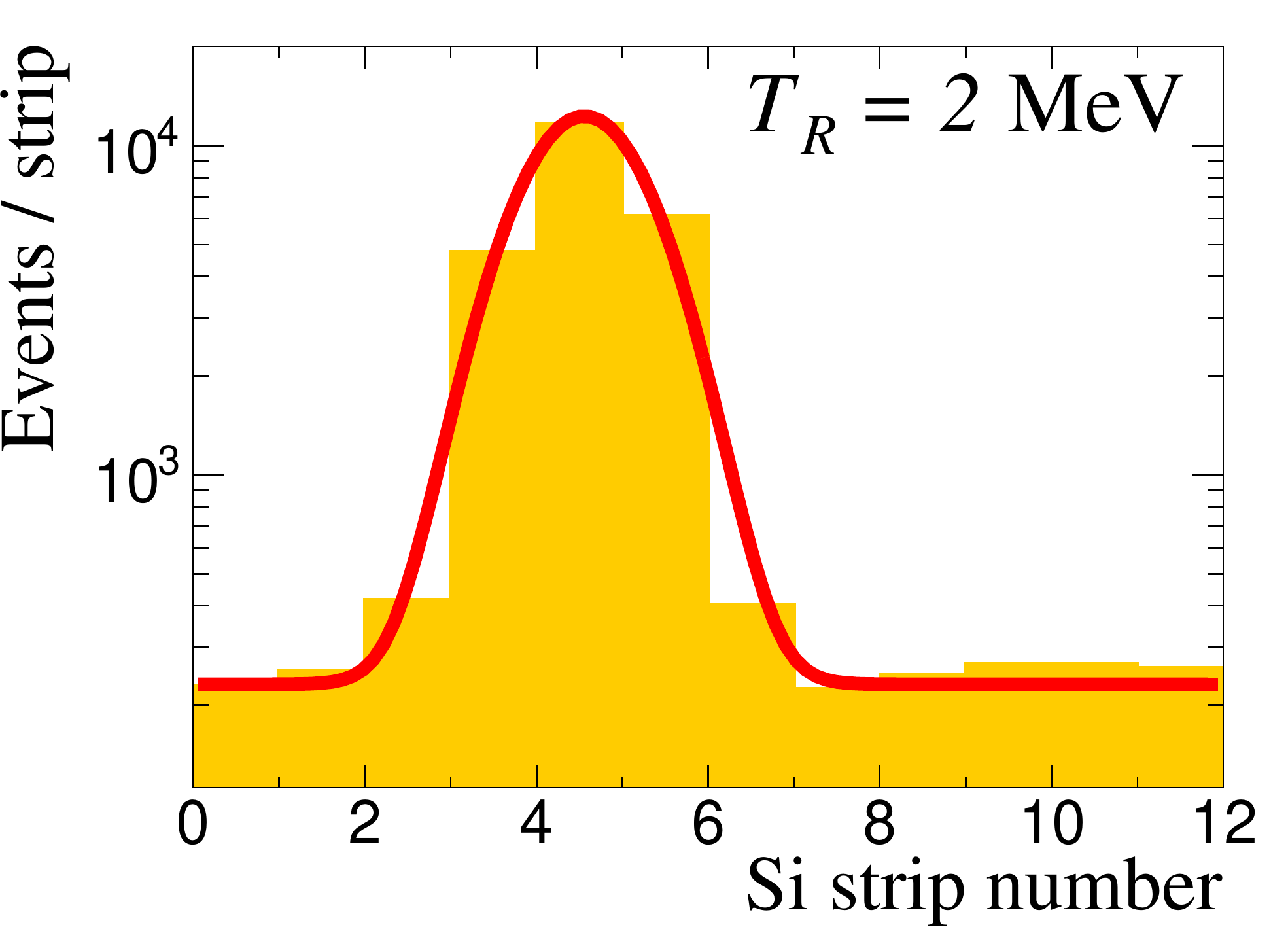}
      \caption{\label{fig:z2MeV}
        Elastic event rates in a Si detector strips for fixed recoil proton energy. The coordinate $z_0(T_R)$ corresponding to the jet center can be found in a Gaussian (plus flat background) fit of the distribution.}
    \end{center}
  \end{minipage}
  \hfill
  \begin{minipage}[t]{\columnwidth}
    \begin{center}
      \includegraphics[width=0.8\columnwidth]{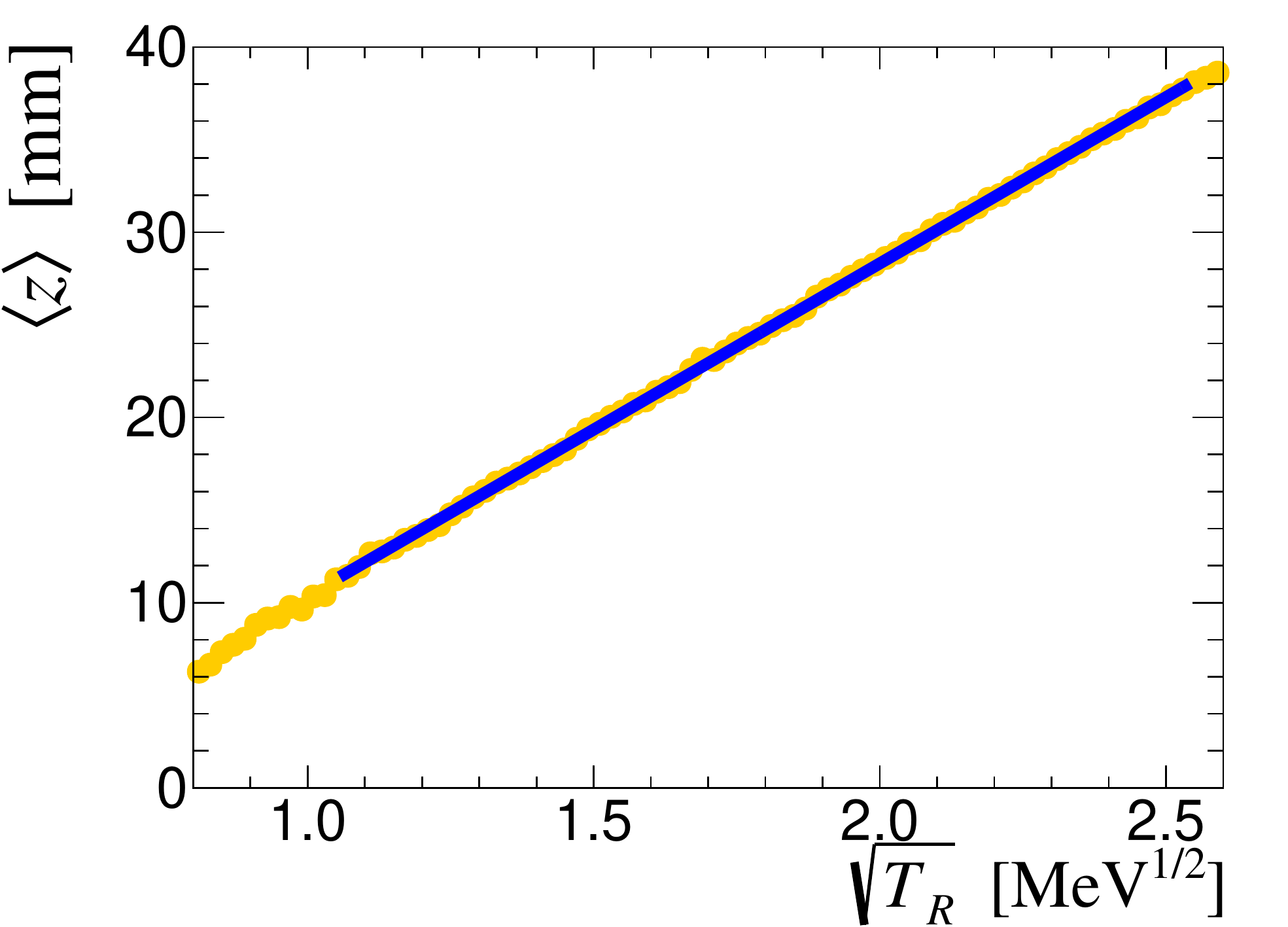}
      \caption{\label{fig:zSlope}
        The average (all HJET detectors) jet center coordinate $\langle{z_0}\rangle$ dependence on the recoil proton energy $T_R$. The solid line indicates the result of linear fit in the $1.1\,\text{--}\,6.5\,\text{GeV}$ energy range.}  
    \end{center}
  \end{minipage}
\end{figure*}

For energy calibration, all Si strips are exposed to \textalpha-particles from two alpha-sources, ${}^\mathrm{148}\mathrm{Gd}$ (3.183 MeV) and  ${}^\mathrm{241}\mathrm{Am}$ (5.486 MeV). A typical signal amplitude distribution in a Si strip is shown in Fig.\,\ref{fig:AlphaCalib}.
The ${}^{241}$Am spectrum was approximated by five Gaussian distributions
\begin{equation}
  dN/dA \propto \sum_{i=1}^5w_i\exp\frac{-\left(A-\langle A\rangle E_i/E_3\right)^2}{2\sigma^2}
\end{equation}
where
$E_i\!=\!($5.388, 5.443, 5.486, 5.511, 5.544$)\,\text{MeV}$ are the ${}^{241}$Am alpha energies and  $w_i\!=\!(1.66, 13.1, 84.8, 0.225, 0.37)$ are the respective relative intensities\,\cite{bib:Am141}. For ${}^{148}$Gd, a simple Gaussian function was used. 

The two different energies of \textalpha-particles allow us to determine both the gain $g\!\sim\!2.5\,\text{keV/cnt}$ and dead-layer thickness $x_\text{DL}\!\sim\!0.37\,\text{mg/cm}^2$ in every Si strip. The energy resolution, $\sigma_E\!\sim\!20\,\text{keV}$, is dominated by electronic noise.

The described method appeared to be a reliable instrument to calibrate the HJET detectors, which gives well reproducible results. However, it does not contain an intrinsic mechanism to control for systematic errors, e.g. due to possible uncertainties of the dead-layer description.
Extrapolation of the \textalpha-particle measurements to the proton detection is also model dependent. Thus, we cannot exclude the calibration related systematic error 
\begin{equation}
  \Delta_\text{calib}T = \delta_T^0 + \delta_T^1T_R.
  \label{eq:systCalDef}
\end{equation}
in the recoil proton energy $T_R$ measurement.
Even though such an uncertainty must not affect the result of the beam polarization measurement, it may be essential for the experimental determination of $A_\text{N}(t)$.

A simple way to evaluate the calibration quality is to test how it satisfies Eq.\,(\ref{eq:kappa}). For that, $T_R$ dependence of the recoil proton mean $z(T_R)$ coordinate in a Si detector was being determined as shown in Fig.\,\ref{fig:z2MeV}. On average (over all Si detectors)  the value $\langle z\rangle$, the magnetic field corrections (\ref{eq:bMF}) and (\ref{eq:b'MF}) are canceled and the value of $\kappa_p$ can be determined in a linear fit of $\langle{z(T_R)}\rangle$ (see Fig.\,\ref{fig:zSlope}). For the energy range $1.1\,\text{--}\,6.5\,\text{MeV}$, the values of $\kappa_p$ found, $17.94(2)\,\text{mm/MeV}^{1/2}$ (100 GeV) and $17.80(2)\,\text{mm/MeV}^{1/2}$ (255 GeV), are in good agreement with those calculated in section\,\ref{sec:kappa}.    

However, it should be noted that such evaluated values of $\kappa_p$ are fit range dependent. A more detail study of systematic uncertainties (\ref{eq:systCalDef}) should include a combined analysis of the measured jet profiles (\ref{eq:jetProfile}) and the measured time dependence on the kinetic energy\,\cite{bib:GeomCalib}. Also, calibration systematic errors may lead to the non-linearity of $A_\text{N}/A_\text{N}^{(0)}$ (\ref{eq:AN_norm}), which also may be used to constrain $\delta_T^0$ and $\delta_T^1$. Admitting, that the result of this study appeared to be unstable due to the large number of free parameters used, we established the following upper limit on the systematic errors in the energy calibration
\begin{equation}
  \delta_T^0=0\pm15\,\text{keV},\qquad\delta_T^1=0\pm0.01,
  \label{eq:systCalib}
\end{equation}
where $\delta_T^0$ and $\delta_T^1$ [see Eq.\,(\ref{eq:systCalDef})] variations are uncorrelated.

\subsection{Overview of the signal time-amplitude distribution at HJET}
\begin{figure}
  \begin{center}
    \includegraphics[width=1.0\columnwidth]{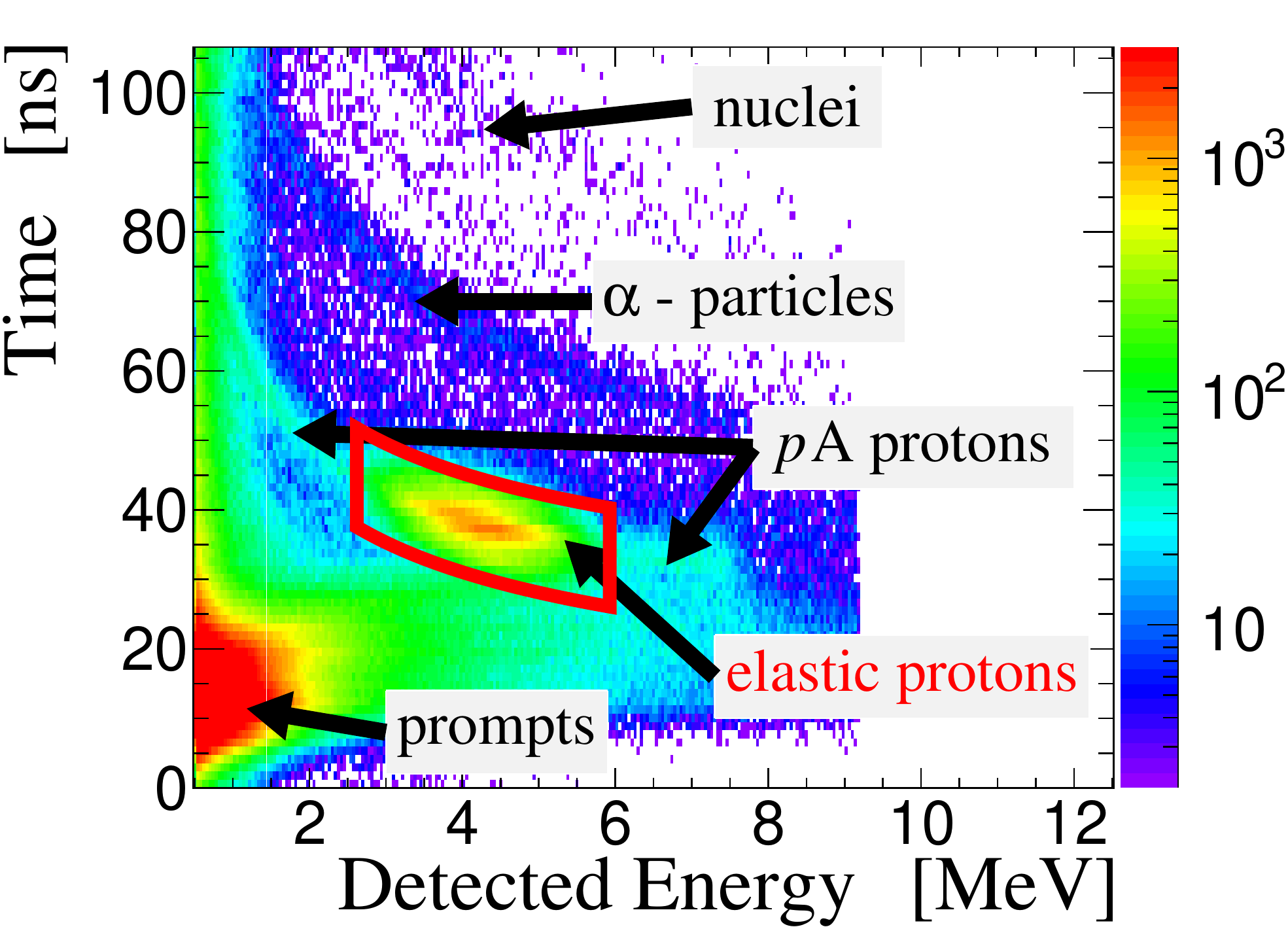}
    \caption{\label{fig:ATraw}
      A typical time-amplitude distribution in a Si strip. The number of events per bin is limited to 3000 to clearly isolate elastic protons in the plot.
    }
  \end{center}
\end{figure}

\begin{figure*}[t]
\begin{center}
\includegraphics[width=0.3\textwidth]{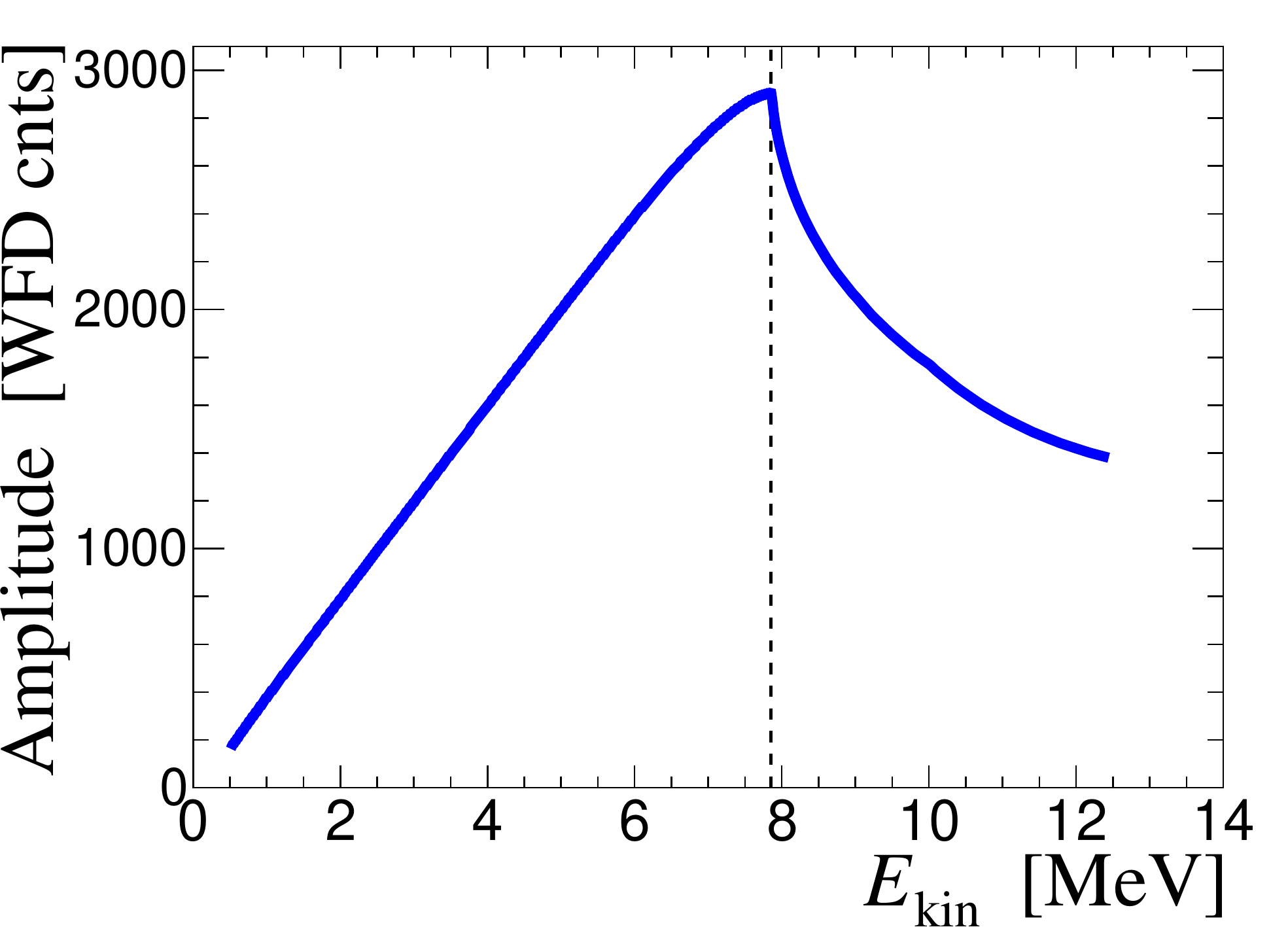}\hfill
\includegraphics[width=0.3\textwidth]{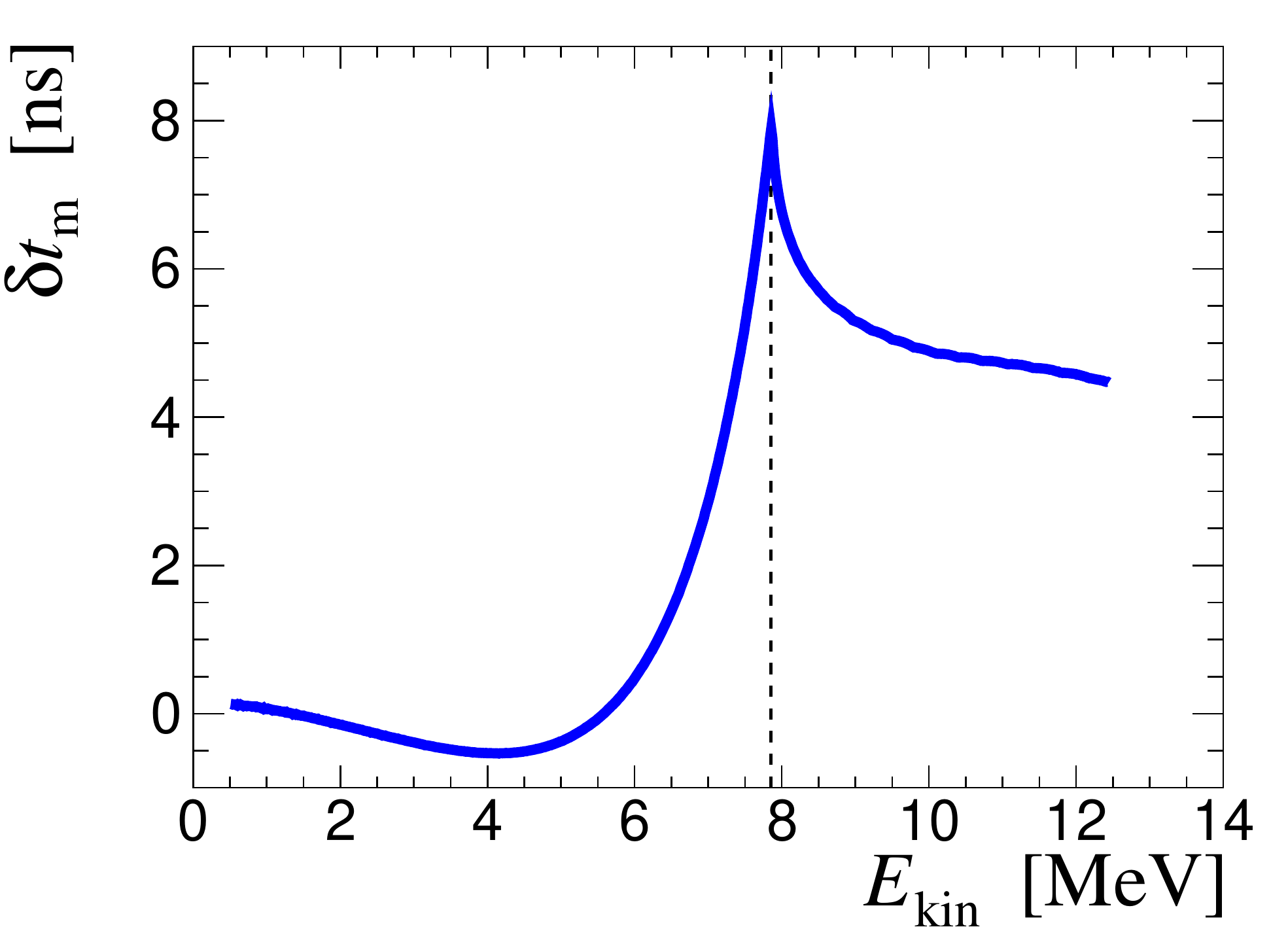}\hfill
\includegraphics[width=0.3\textwidth]{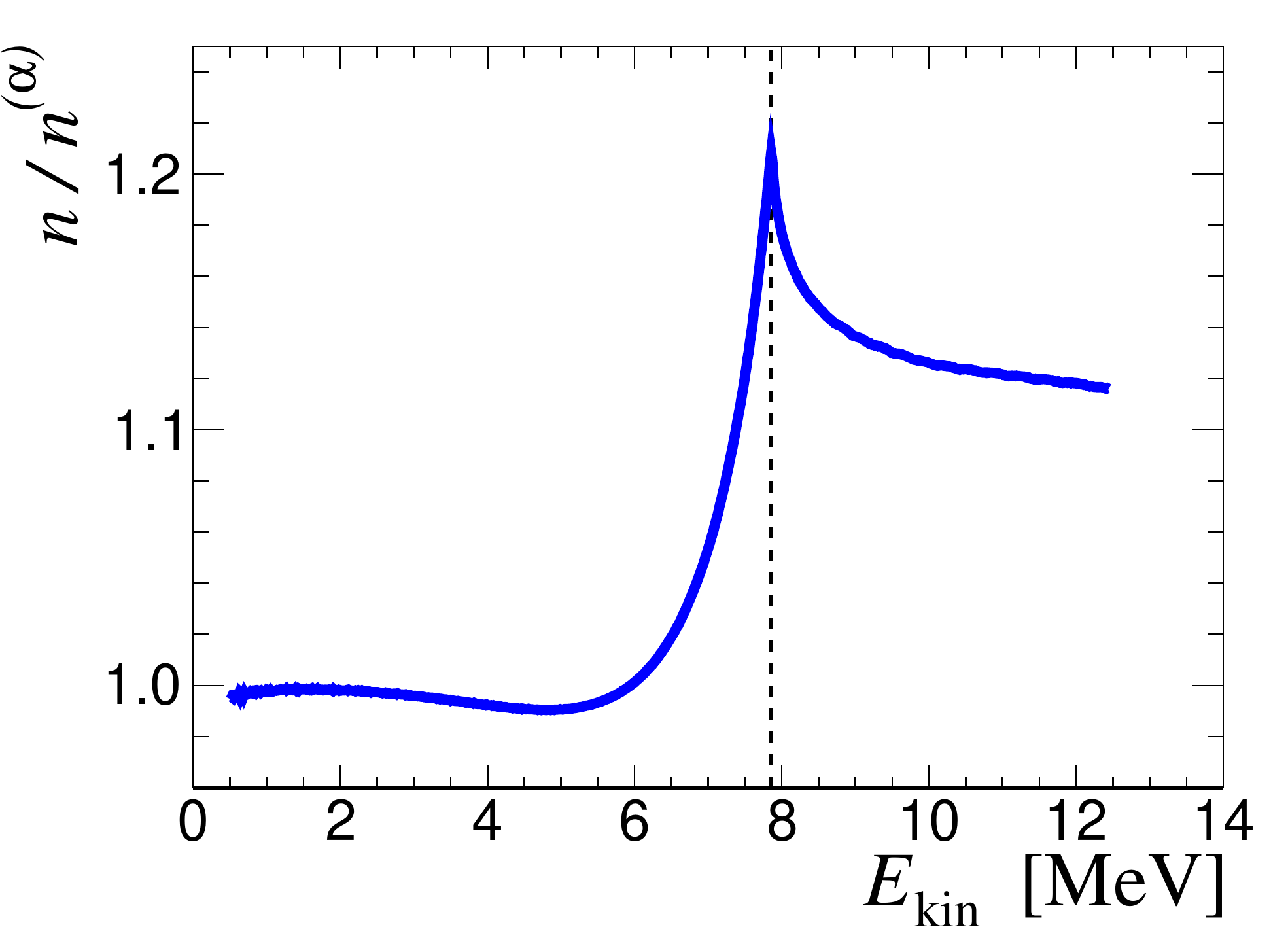} 
\caption{Simulation of waveform shape parameters dependence on recoil proton energy. $n^{(\alpha)}$ denotes mean waveform shape parameter $n$ determined in the \textalpha-calibration. Vertical dashed lines indicate the threshold energy of 7.8 MeV for punch-through protons.
  \label{fig:EkinWF}}
\end{center}
\end{figure*}

A typical time-amplitude distribution of the events detected in a Si strip is displayed in Fig.\,\ref{fig:ATraw}. The detected particles can be readily identified by the correlation between measured energy and time of flight. It should be noted that a detected particle with energy above some threshold ($7.8\,\text{MeV}$ for a proton) punch through the Si strip and, thus, only part of the kinetic energy is measured. For such events, the larger kinetic energy the smaller signal amplitude is determined.

Elastic $\mathit{pp}$ events can be readily recognized by the measured energy peak around the strip location dependent value of  $T_\text{strip}$ (\ref{eq:Tstrip}). In addition to the proton line containing the elastic events, one can also identify \textalpha-particles and heavier nuclei. These species which cannot be produced in the $\mathit{pp}$ scattering, indicate a noticeable contribution from inelastic $p\text{A}$ scattering where A is a nucleus (Nitrogen, Oxygen, \dots) in the beam gas or in the jet.

The event statistics is strongly dominated by fast particles, so-called {\em prompts}, which are supposed to be $\pi$, $p$, $\alpha$, \dots, emitted in the inelastic $\mathit{pp}$ and/or $p\text{A}$ scattering. Switching off the jet reduces {\em prompts} rate by factor 3--4.

\begin{figure}[t]
  \begin{center}
    \includegraphics[width=0.85\columnwidth]{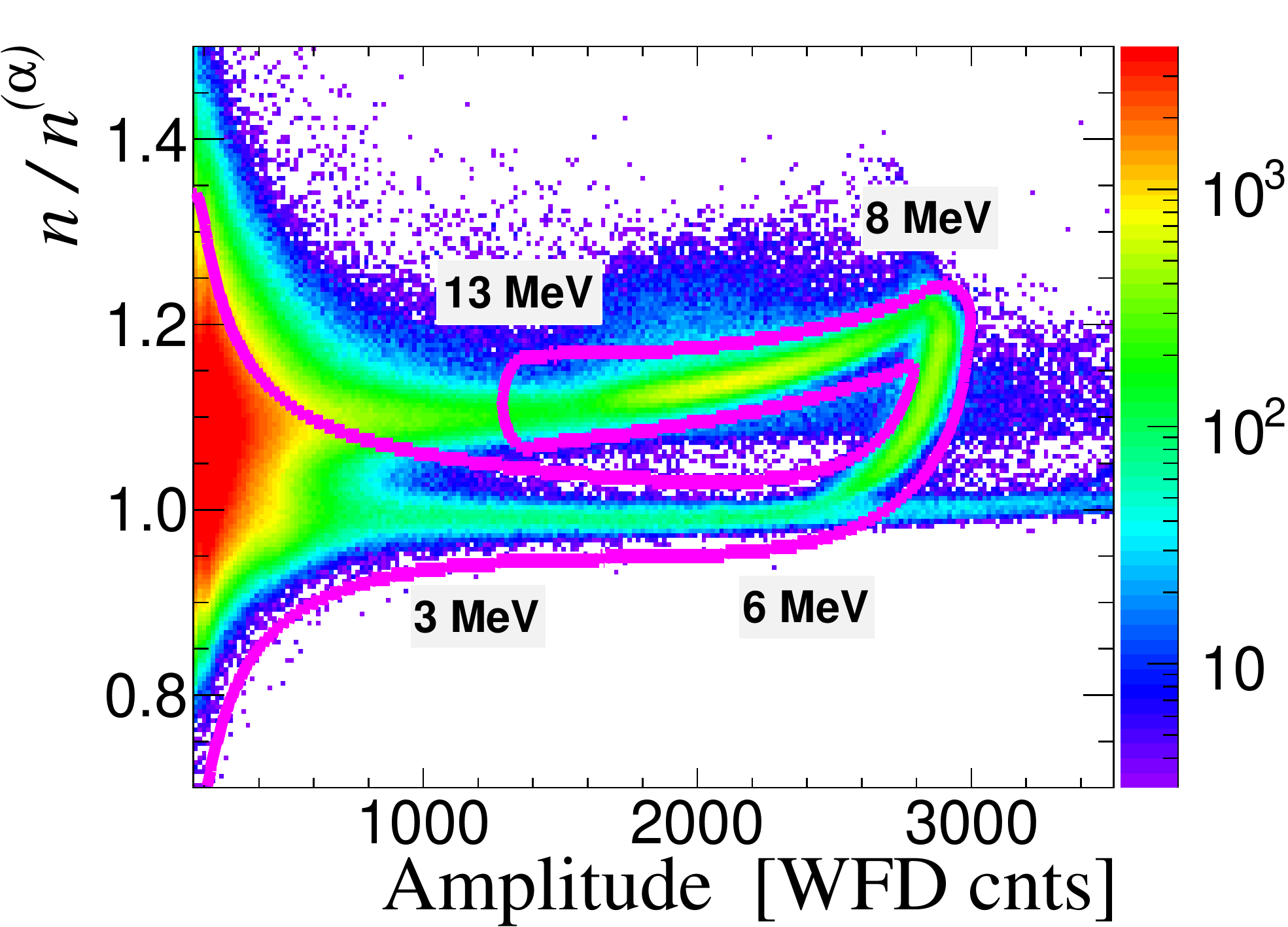}
  \end{center}
  \caption{\label{fig:nA_Cut} Event selection cut (solid magenta line) to separate punch-through and stopped protons for the $0.5\!<\!T_R\!<\!13\,\text{MeV}$ energy range. No other event selection cuts had been applied in this plot.}
\end{figure}

\subsection{Reconstruction of the punch-through protons}

The recoil spectrometer geometry allows detection of the recoil protons with kinetic energy up to $10\text{--}11\,\text{MeV}$.
For the detected energy above $5\,\text{MeV}$ the stopped and punch-through protons cannot be efficiently separated by comparing the measured signal time and the deposited energy only.

\begin{figure}[t]
  \begin{center}
    \includegraphics[width=0.85\columnwidth]{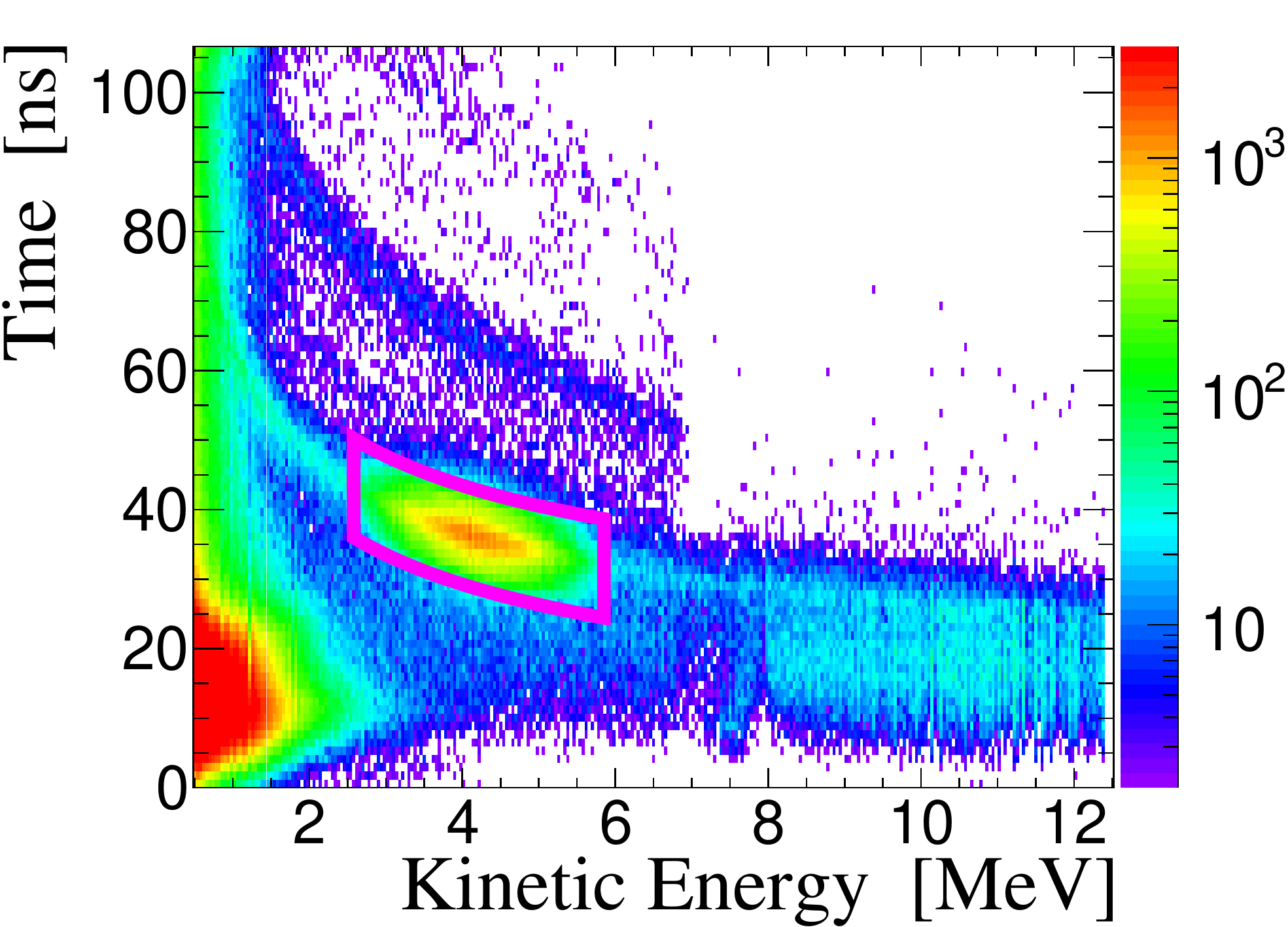}\\
  \end{center}
  \caption{\label{fig:AT}
    The same data as in Fig.\,\ref{fig:ATraw}, but after reconstruction of the recoil proton kinetic energy and time $t_m$.}
  \end{figure}

It was found in the data analysis that the signal waveform shape (\ref{eq:Waveform}) is significantly dependent on the kinetic energy $E_\text{kin}$ of an incident recoil proton. With parameter $\tau_s\!=\!\tau_s^{(\alpha)}$ fixed to the value determined in the \textalpha-calibration, the simulated dependence of $A$, $\delta t_m$ (the correction to the measured time), and $n/n^{(\alpha)}$ on $E_\text{kin}$ is shown in Fig.\,\ref{fig:EkinWF}. The simulation\,\cite{bib:SiSimulation} is based on the evaluation of charge collection in the Si strip and it was adjusted using the \textalpha-calibration data.

To separate stopped and punch-through protons, for every pair of measured $A$ and $n$ within the event selection cut shown in Fig.\,\ref{fig:nA_Cut} the corresponding recoil proton kinetic energy was determined in the simulation. The resulting time-amplitude distribution (including the signal time correction) is displayed in Fig.\,\ref{fig:AT}. 

\subsection{Event selection \label{sec:EventSelection}}

\begin{figure*}[t]
  \begin{minipage}{\columnwidth}
  \begin{center}
    \includegraphics[width=0.85\columnwidth]{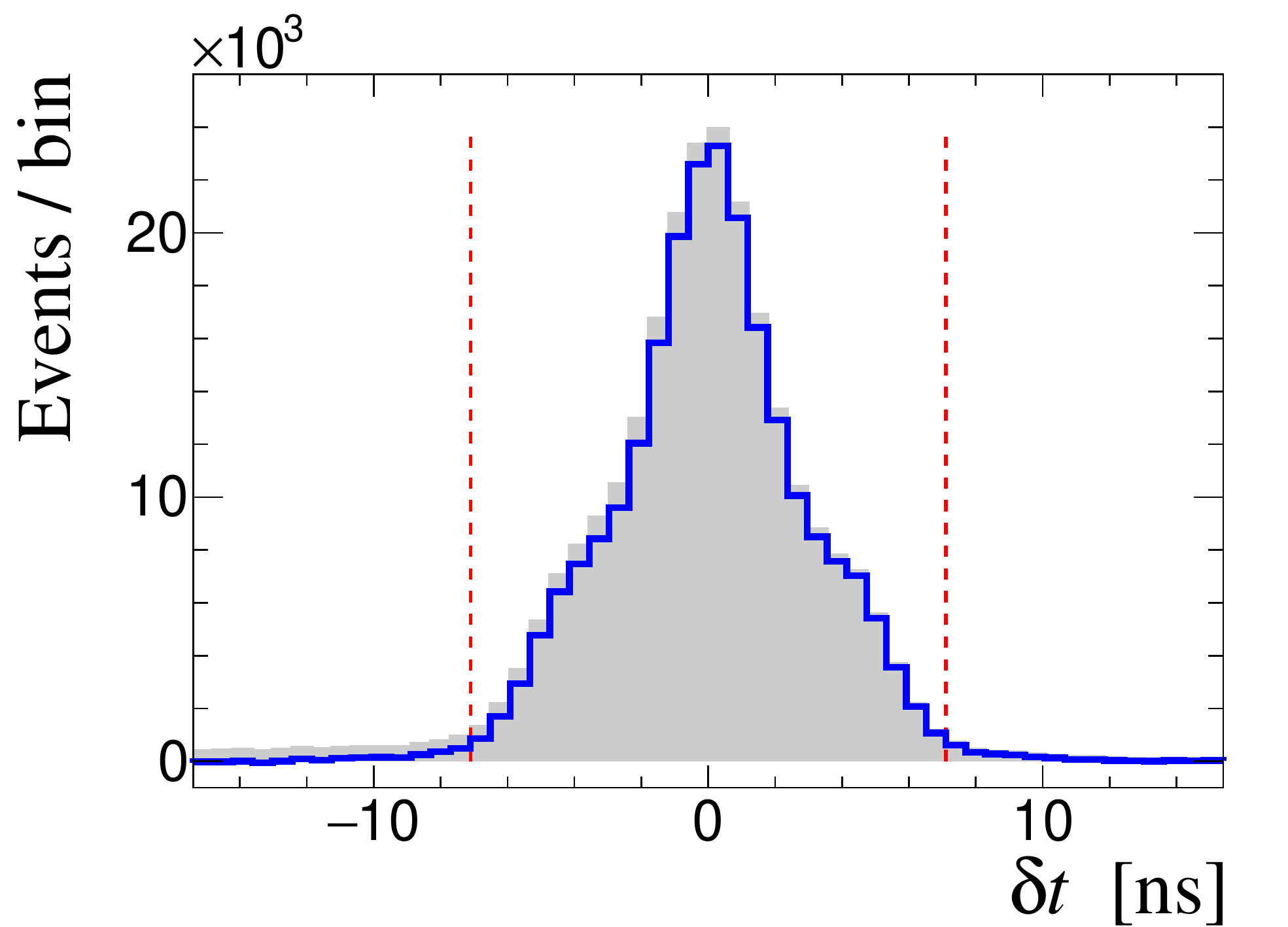}
    \end{center}\end{minipage} \begin{minipage}{\columnwidth}\begin{center}
    \includegraphics[width=0.85\columnwidth]{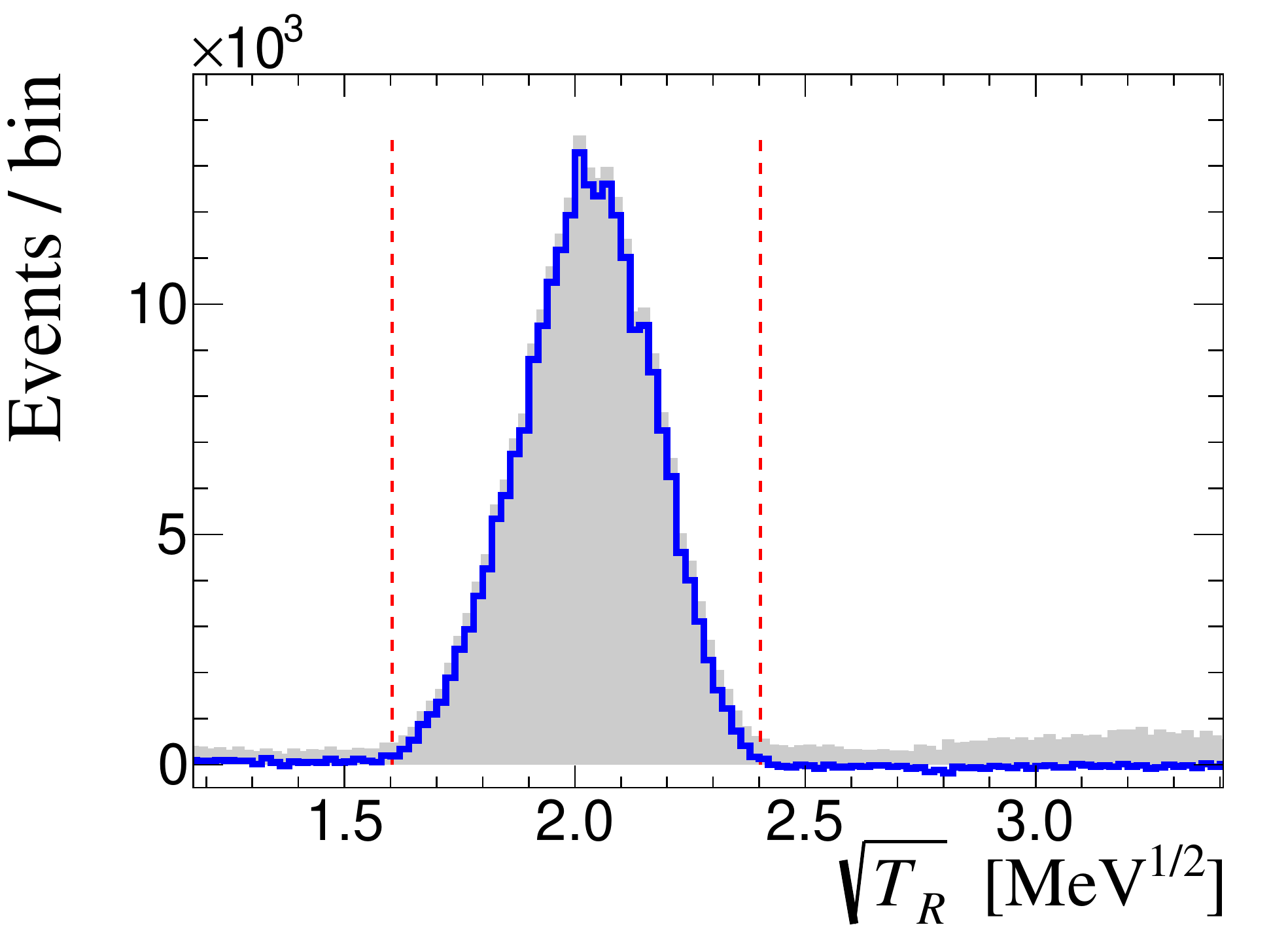}
    \end{center}
  \end{minipage}
  \caption{\label{fig:EventSelection}
    Elastic $\mathit{pp}$ events isolation. Event selection cuts are shown by the dashed red lines. The $\delta\sqrt{T}$ cut is applied for events in the $\delta t$ histogram, and the $\delta t$ cut is applied in the $T_R$ histogram. Filled histograms show the distributions before the background subtraction.
}
\end{figure*}

\begin{figure*}[t]
\begin{center}
\includegraphics[width=0.32\textwidth]{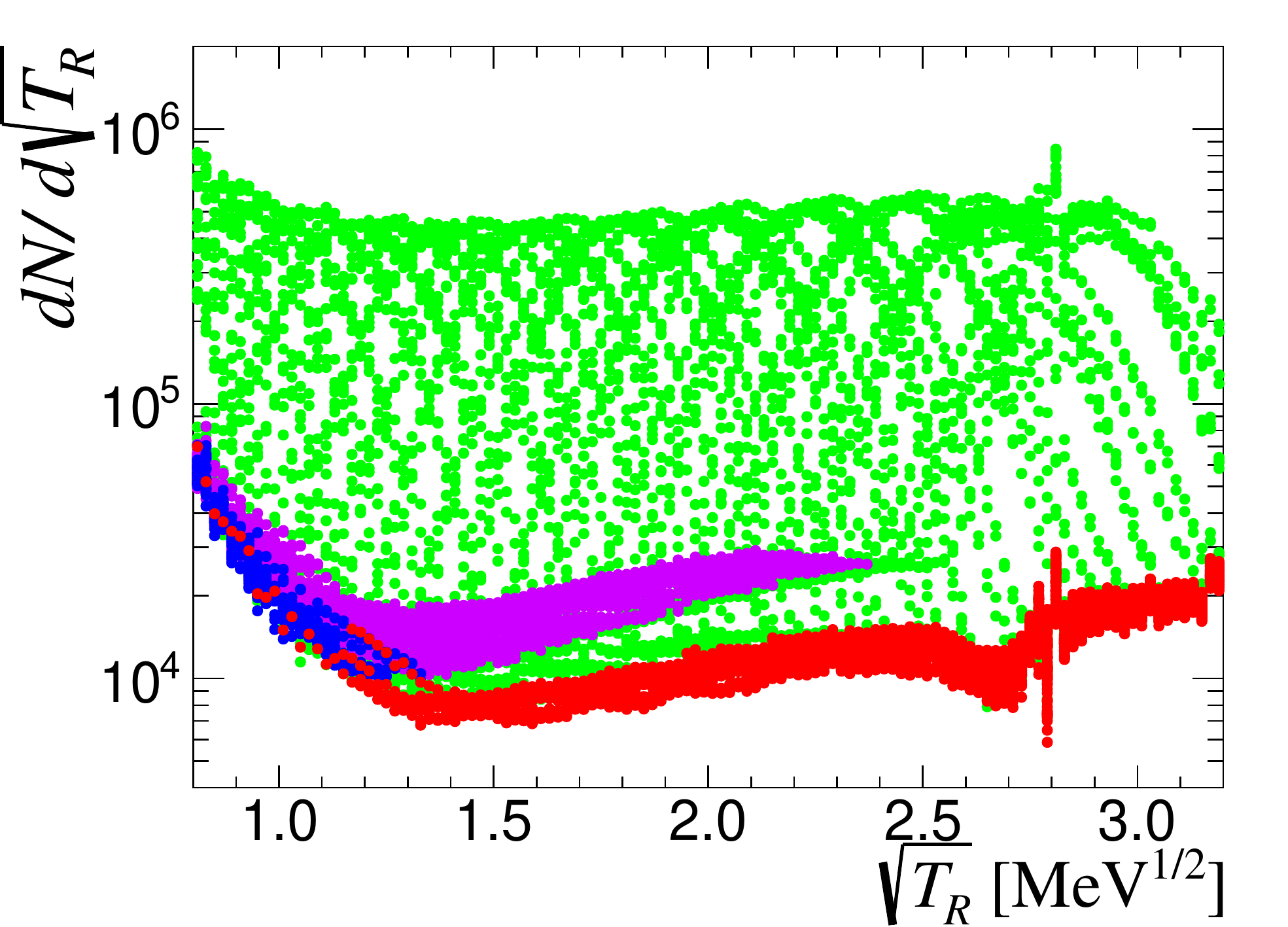}\hfill
\includegraphics[width=0.32\textwidth]{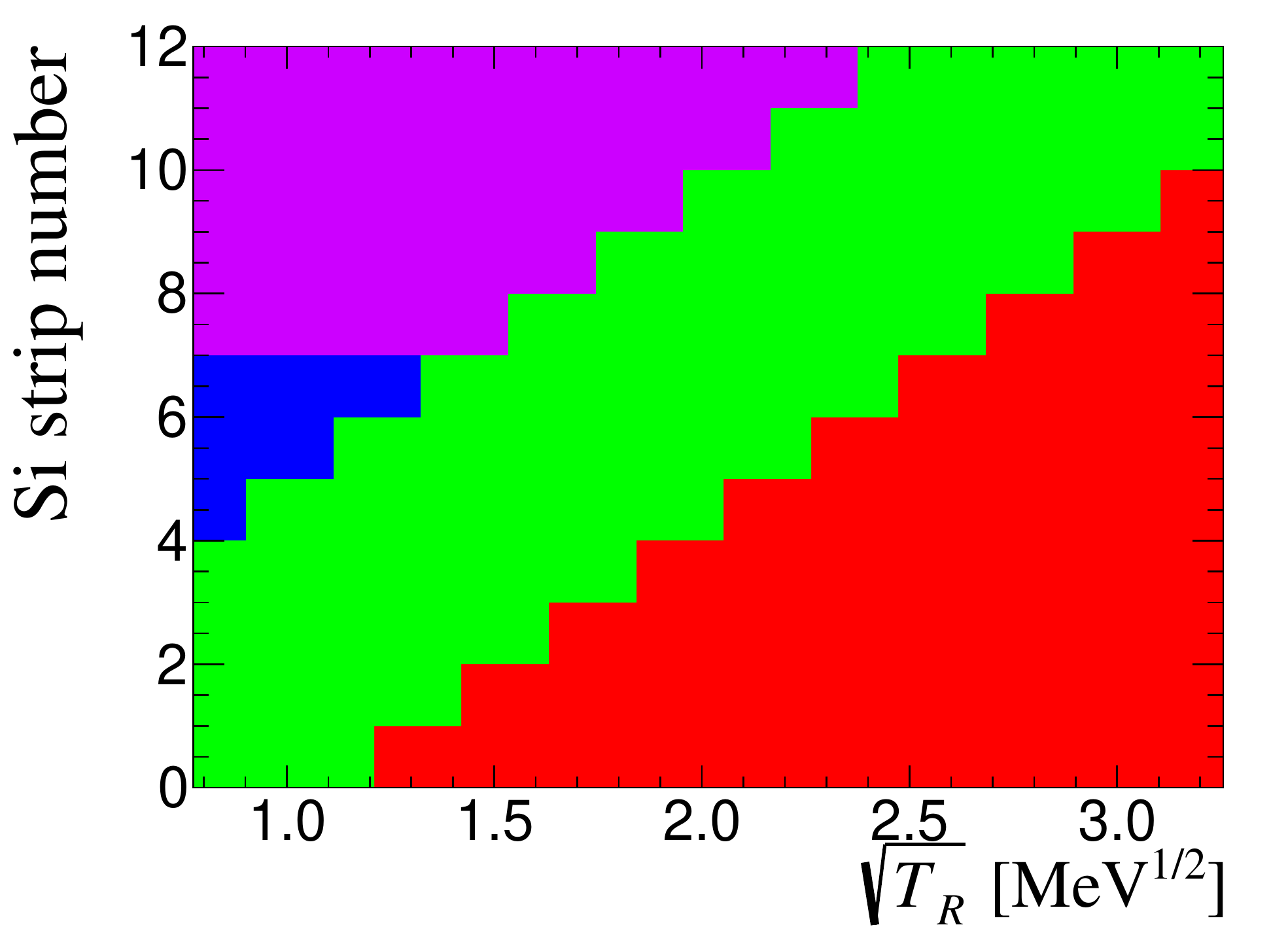}\hfill 
\includegraphics[width=0.32\textwidth]{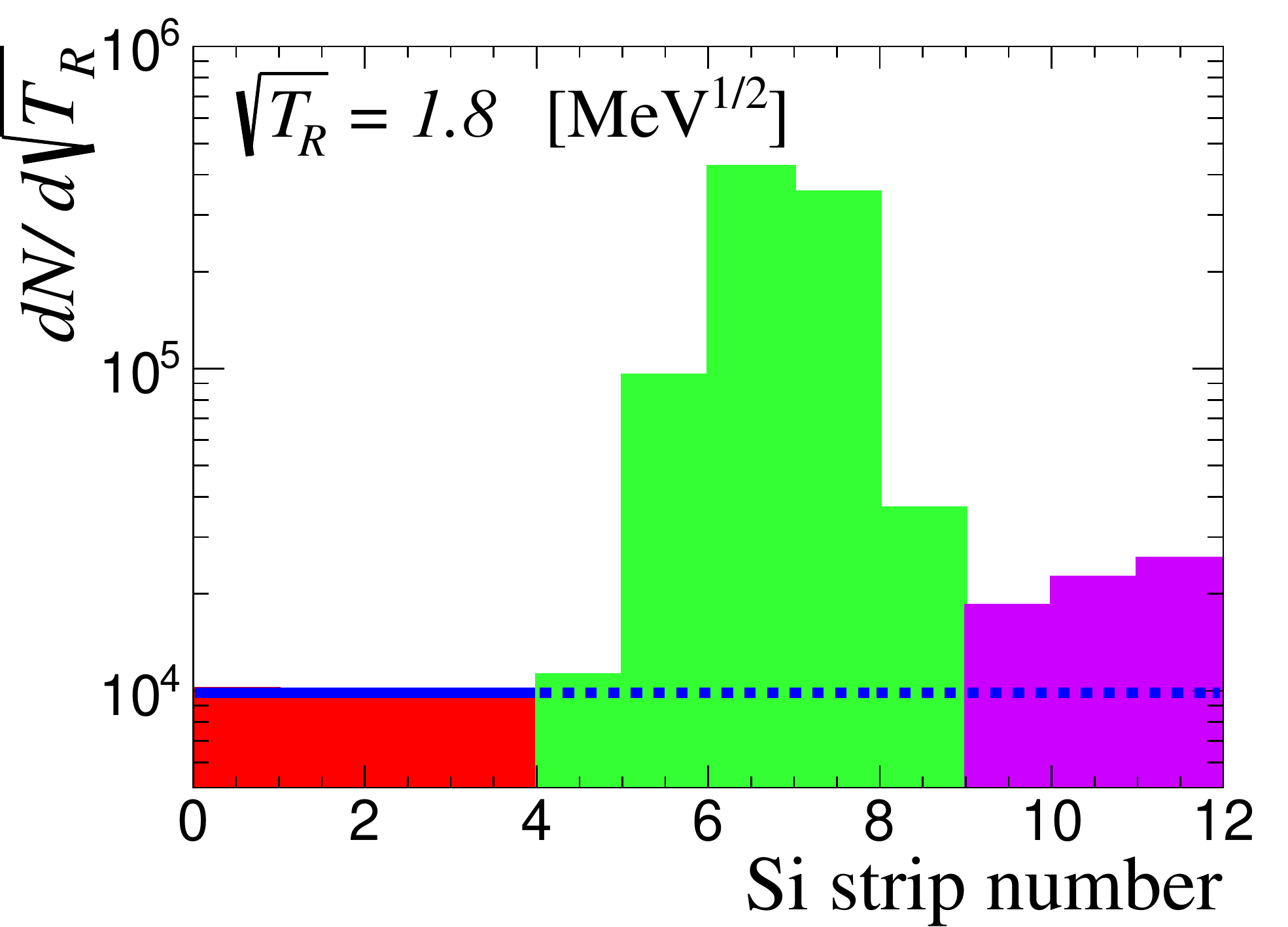}
\end{center}
\caption{\label{fig:Background} The superposition of $dN/s\sqrt{T_R}$ distributions for all Si strips (left). The markers colors depending on the strip location in a detector and recoil proton energy  are explained in center histogram. Green color is used for elastic events, red and blue for backgrounds, and violet specifies the area contaminated by the inelastic events. An example of background subtraction for fixed, $T_R\!=\!3.6\,\text{MeV}$ recoil proton energy is shown in the right histogram. The background determined in the red (and/or blue, depending on $T_R$) area is extrapolated to the signal (green) area.}  
\end{figure*}

To study the spin correlated asymmetries, we have to isolate the elastic events.

First, we should verify that the detected particle is a proton. For that, we compare the measured signal time $t$ with the time corresponding to the measured recoil proton kinetic energy $T_R(A,n)$
\begin{equation}
  \delta t = t\!-\!t_0\!-\!\text{ToF}
  = t\!-\!t_0\!-\!\frac{L}{c}\sqrt{\frac{m_p}{2T_R}} \approx 0
  \label{eq:dt}
\end{equation}
Here, $L\!=\!769\,\text{mm}$ is the distance to detector, $c$ is speed of light, and $t_0$ is the time offset. Since the $\delta t$ distribution is dominated by the beam bunch longitudinal profile, it must be the same for all Si strips.

Second, we should check that the missing mass $M_X$, defined as 
\begin{equation}
  M_X^2=\left(p_\text{beam}+p_\text{jet}-p_\text{recoil}\right)^2,
\end{equation}
is equal to a proton mass, $M_X\!=\!m_p$. This condition leads to Eq.\,(\ref{eq:kappa}). Therefore, using definition (\ref{eq:Tstrip}), it may be presented as 
\begin{equation}
  \delta\sqrt{T} = \sqrt{T_R}-\sqrt{T_\text{strip}} \approx 0
\end{equation}
Since $\delta\sqrt{T}$ distribution is dominated by the jet density profile; it has to be the same in all of the Si strips. 

An elastic $\mathit{pp}$ event isolation based on the $\delta t$ and $\delta\sqrt{T}$ cuts is illustrated in Fig.\,\ref{fig:EventSelection}.

\subsection{Background subtraction}

\begin{figure*}[t]
\begin{center}
\includegraphics[width=0.32\textwidth]{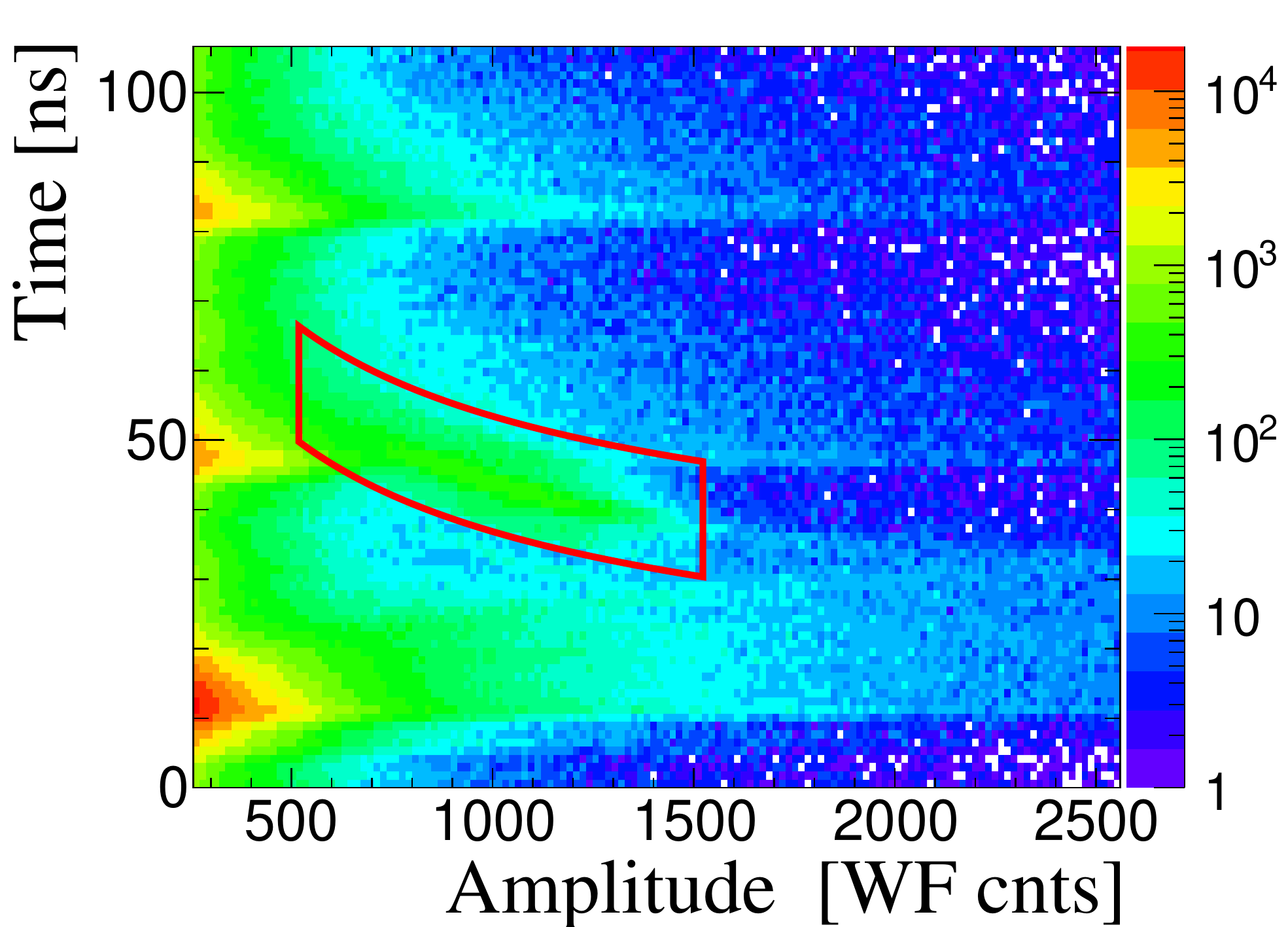}\hfill 
\includegraphics[width=0.32\textwidth]{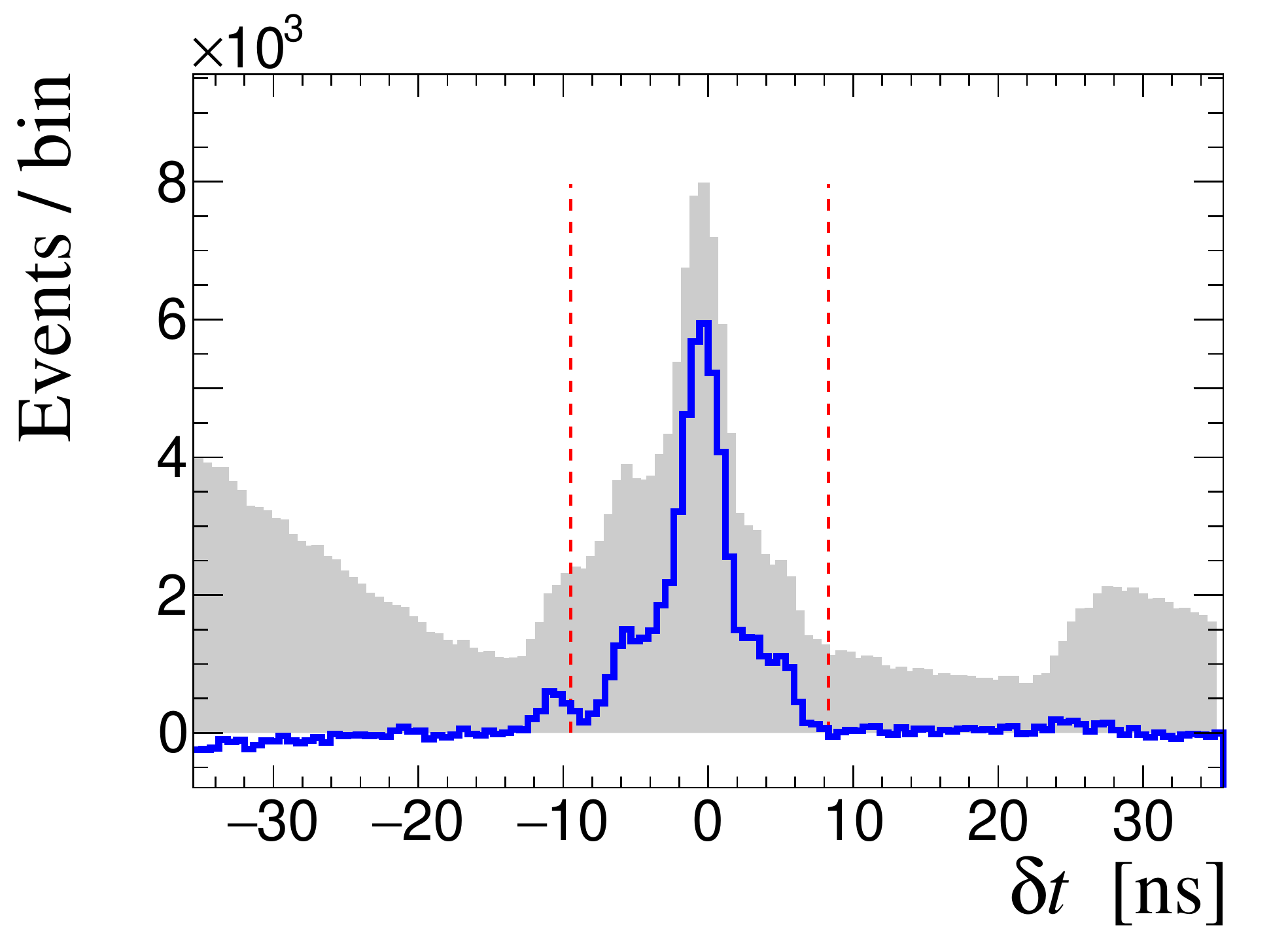}\hfill
\includegraphics[width=0.32\textwidth]{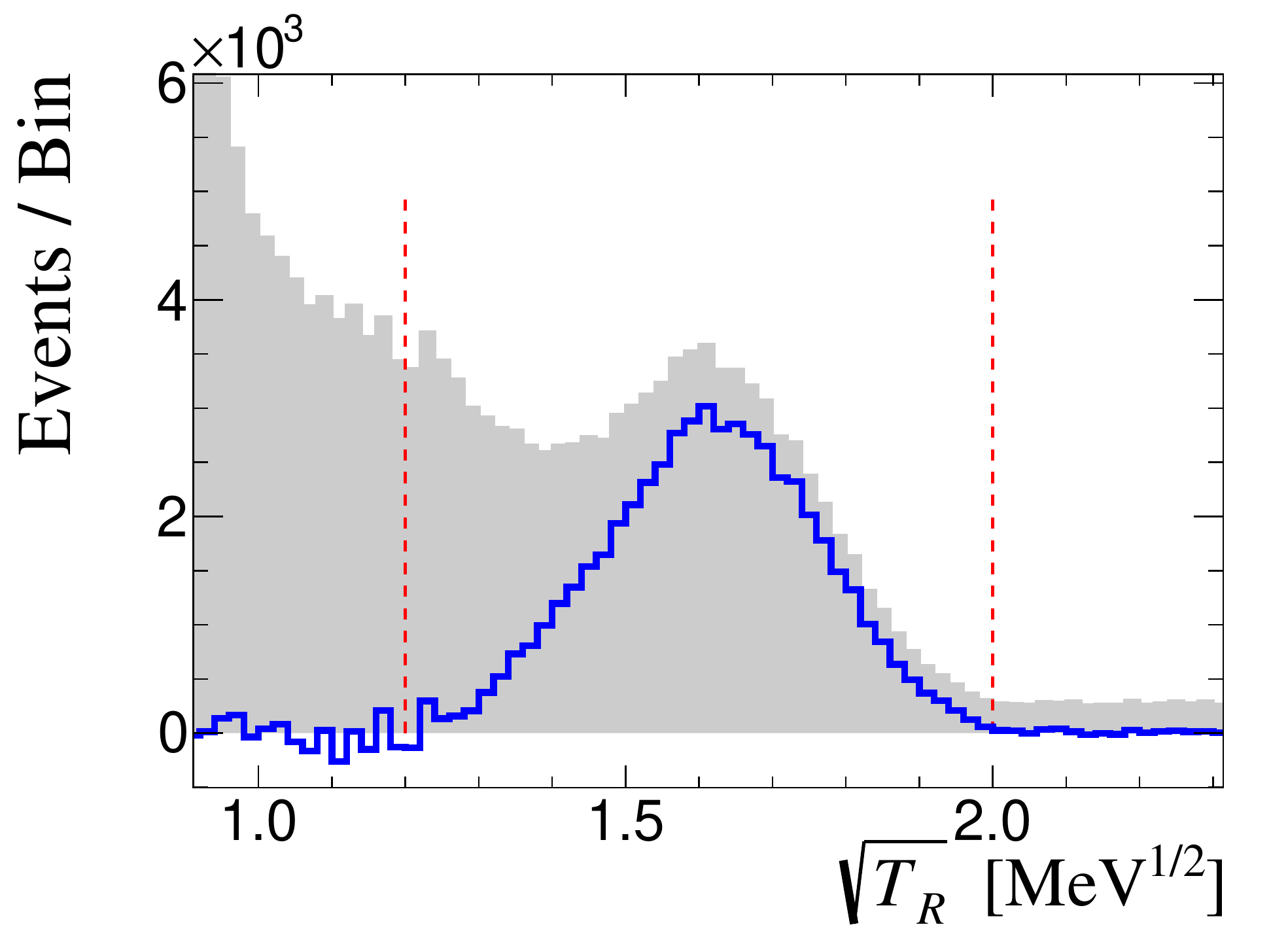}
\end{center}
\caption{\label{fig:dAuBgr}
  An example of the background subtraction in case of very high background rate. Gray filled and blue solid line histograms show event distributions respectively before and after background subtraction. Dashed red lines indicate event selection cuts used.}
\end{figure*}

The molecular hydrogen ($\text{H}_2$) in the beam gas (including water vapor and  $\text{H}_2$ from the jet atomic hydrogen recombination on the walls) effectively dilute the target polarization. The $\text{H}_2$ contamination in the polarized jet was measured by using a quadrupole mass-analyzer\,\cite{bib:H2}. Only an upper limit  of $\sim3\%$ on possible polarization dilution was obtained in these earlier studies and the molecular hydrogen was considered to be the main source of systematic errors of about $\sigma^\text{syst}_P/P\sim2\text{--}3\%$\,\cite{bib:HJET06,bib:HJET09}.

To improve, a background subtraction method was developed and implemented. The method is based on an assumption, that, for any $T_R$, the background rate is the same in all strips of a Si detector (see, for example, Fig.\,\ref{fig:z2MeV}).  Thus, the measured event rates for $T_R$ outside the  elastic peak $|\delta\sqrt{T}|\!>\!0.6\,\text{MeV}$ can be used to evaluate (and subtract) the background under the elastic peak in the other Si strips. To illustrate, in Run\,17 ($255\,\text{GeV}$) the measured $dN/d\sqrt{T_R}$ distributions in all 96 Si strips are displayed in Fig.\,\ref{fig:Background}. The 255 GeV data is noticeably contaminated by inelastic $\mathit{pp}$ events and complicates the basic assumption of the background subtraction. This is why the background selection criteria were modified to cut off the inelastic events.

In the data analysis, the background rate was determined, separately for each Si detector, as a function of $\delta t$ (\ref{eq:dt}) and $\sqrt{T_R}$. To properly account for the possible spin correlated effects associated with the background, every combination of the beam and jet spins was analyzed independently. A typical result of the background subtraction, seen in Fig.\,\ref{fig:EventSelection}, suggests that the residual background contribution to the elastic events is well below 1\%. 

The efficiency of background subtraction is demonstrated in Fig.\,\ref{fig:dAuBgr}. For the illustration, we used uncommon data, obtained in RHIC\,16 deuteron--Gold Run, with very high background level. Even in this case, the method works sufficiently well.

The background subtraction is routinely used in the data analysis which allows us to dilute the background to a sub-percent (on average) level.

The following speculations were used to justify the method. For the $p\text{A}$ background, which is dominant in the HJET measurements, the rate should be strip independent due to a relatively small size of the detectors and absence of a strict correlation between the recoil angle and kinetic energy.

For the beam gas $\text{H}_2$ background, one can expect almost uniform hydrogen density in the scattering chamber. Furthermore, according to Eq.\,(\ref{eq:kappa}), the background rate $dN_{\text{H}_2}/dT_R$  should be the same in all Si strips. However, the recoil proton tracking in the magnetic field (see Fig.\,\ref{fig:TracksMF}) results in strip-to-strip alteration of this background rate. That effect requires special data processing which will be discussed in section \ref{sec:H2BeamGas}.

Other possible errors in the background subtraction  will be analyzed in sections \ref{sec:H2Jet} (the atomic hydrogen jet contamination by $\text{H}_2$) and \ref{sec:Inelastic} (the inelastic $\mathit{pp}$ background).

\section{Absolute proton beam polarization measurements at HJET}

To determine the RHIC proton beam polarization, we compare  the measured beam and jet single spin asymmetries
\begin{equation}
  P_\text{beam} =
   \left({a_\text{N}^\text{beam}}/{a_\text{N}^\text{jet}}\right)\times P_\text{jet}.
   \label{eq:Pbeam0}
\end{equation}
Here, $a_\text{N}^\text{beam}$ and $a_\text{N}^\text{jet}$ are the measurement averaged values, depending on the event selection cuts used. Since both asymmetries are determined using the same events, the result must be independent of the analyzing power $A_\text{N}(t)$ details.

To discuss experimental uncertainties of the measurements, we will use $255\,\text{GeV}$ polarized proton beam data acquired during a 14-week RHIC Run\,17.

Two main sets of event selection cuts were used. The first one accepts as many elastic events as reasonably possible to minimize the statistical uncertainty of the measurements:
\begin{equation}
  \begin{aligned}
    \text{Cuts\,I:}\qquad
    & 0.6<T_R<10.6\,\text{MeV},\\
    & |\delta t|<7\,\text{ns},\\
    & \left|\delta\sqrt{T}\right|<0.40\,\text{MeV}^{1/2}.
  \end{aligned}
\label{eq:CutsI}
\end{equation} 
The second set minimizes uncontrollable systematic corrections:
\begin{equation}
  \begin{aligned}
    \text{Cuts\,II:}\qquad
    & 2.0<T_R<9.5\,\text{MeV},\\
    & |\delta t|<7\,\text{ns},\\
    &  -0.18<\delta\sqrt{T}<0.30\,\text{MeV}^{1/2}.
  \end{aligned}
\label{eq:CutsII}
\end{equation}
Explanation of these cuts will be given below.

\begin{figure}[t] \begin{center}
\includegraphics[width=0.95\columnwidth]{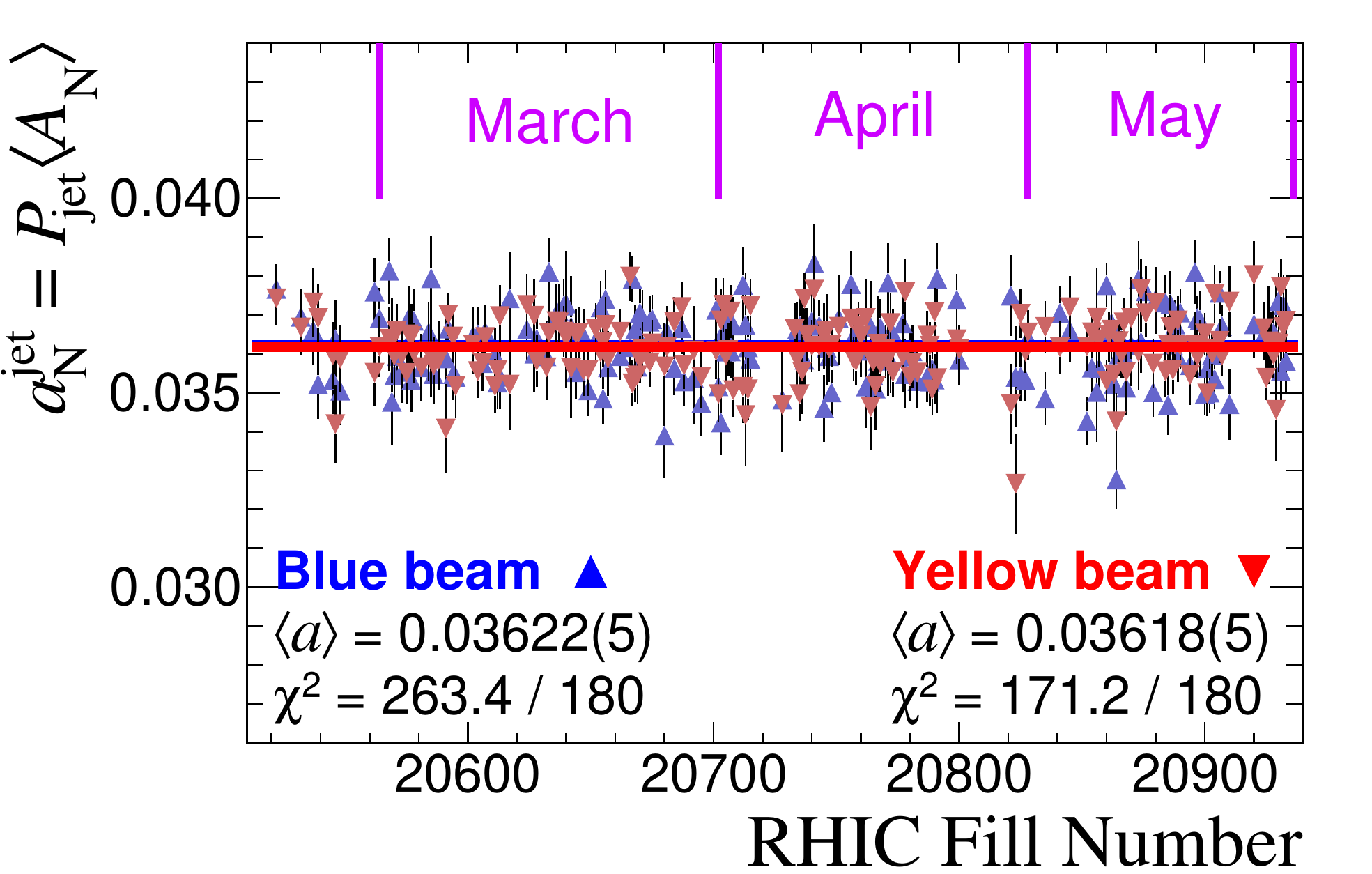}
\caption{\label{fig:RawAsymF} 
 The RHIC store average jet spin asymmetries measured with {\em blue} and {\em yellow} beams in Run\,17.}
\end{center} \end{figure}
\begin{figure}[t] \begin{center}
\includegraphics[width=0.95\columnwidth]{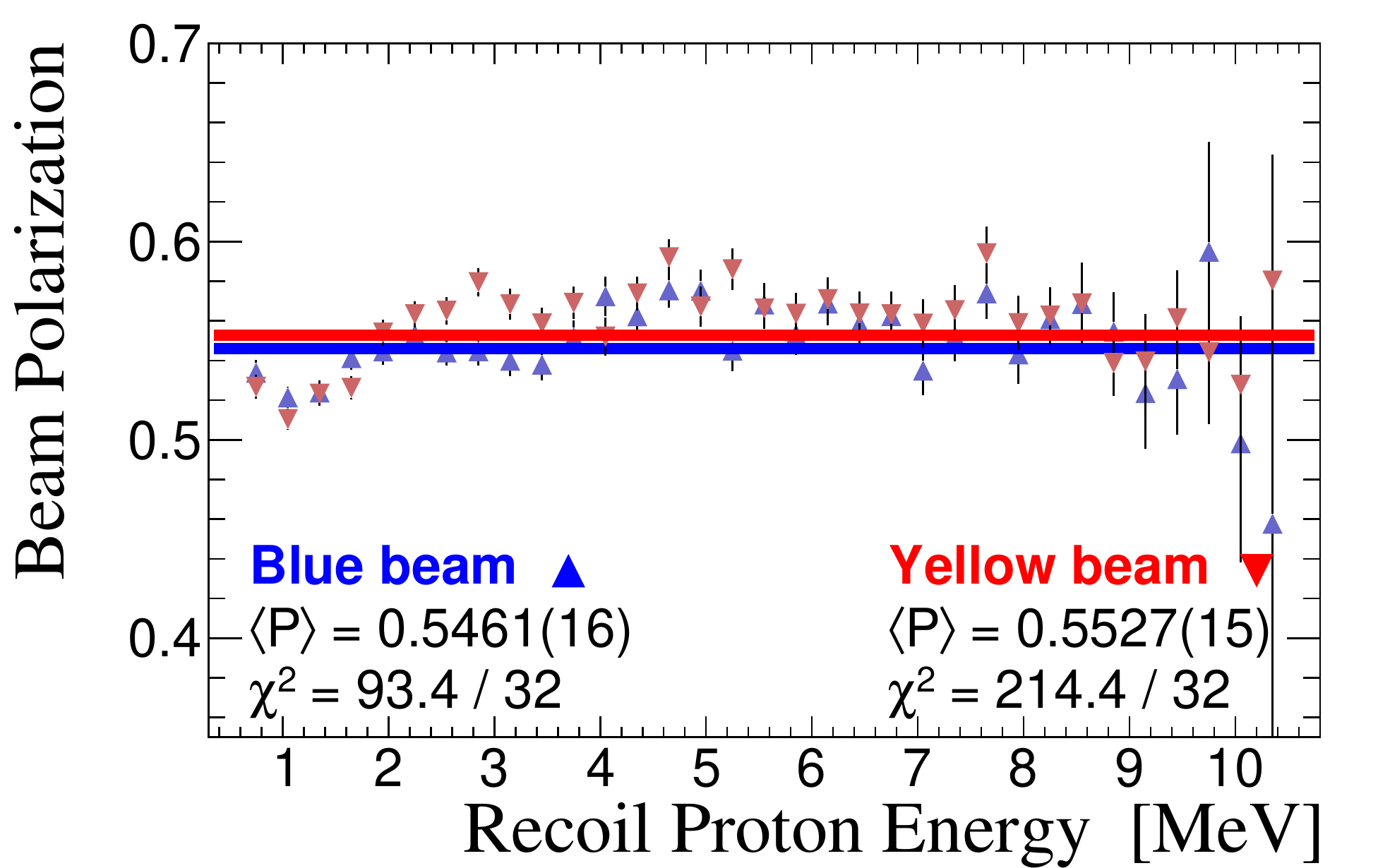}       
\caption{\label{fig:RawPol}
  The RHIC Run\,17 averaged {\em blue} and {\em yellow} beam polarization dependence on the recoil proton energy.}
\end{center} \end{figure}

\begin{figure}[t]
\begin{center}
  \includegraphics[width=0.8\columnwidth]{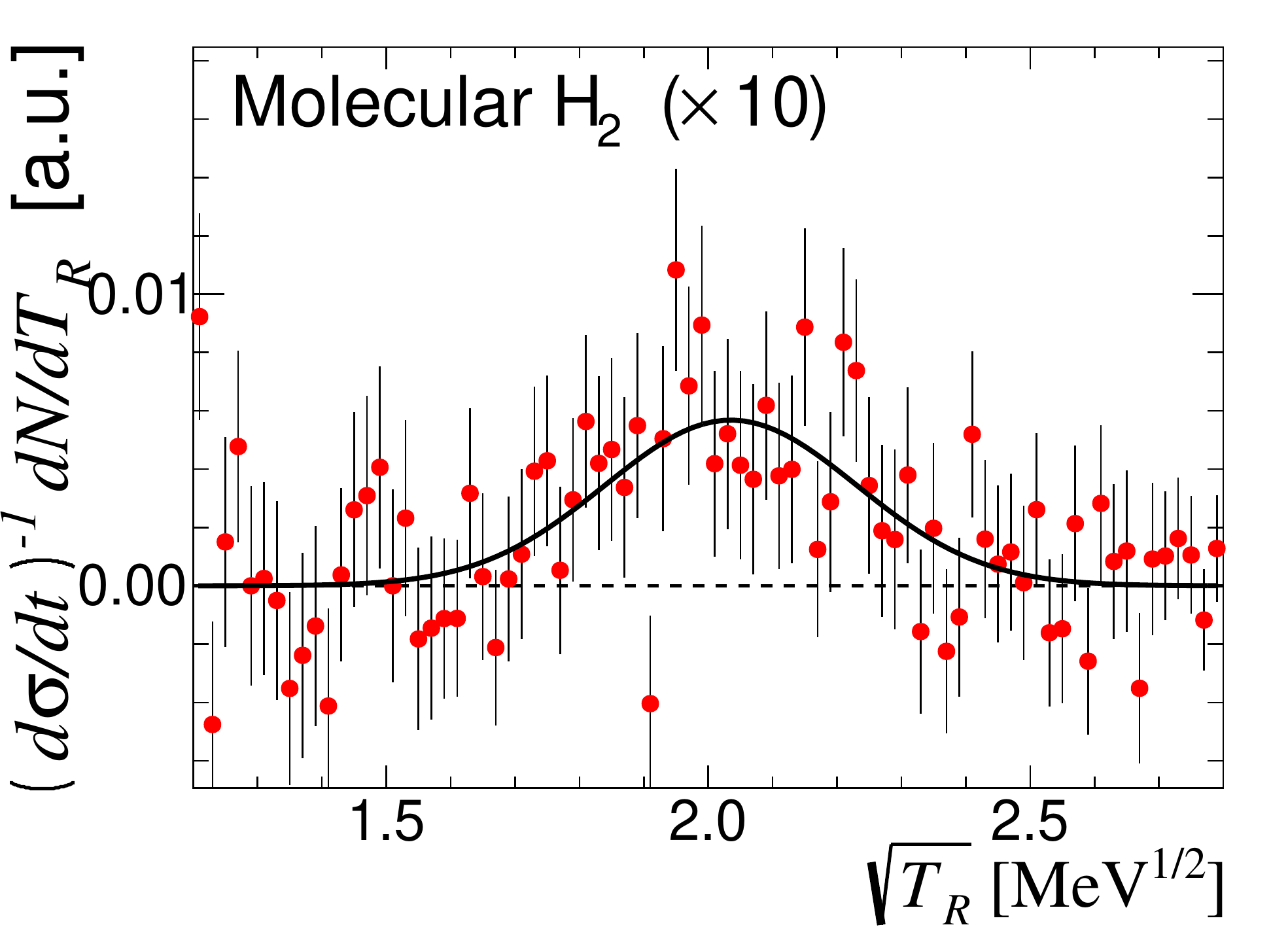}
  \includegraphics[width=0.8\columnwidth]{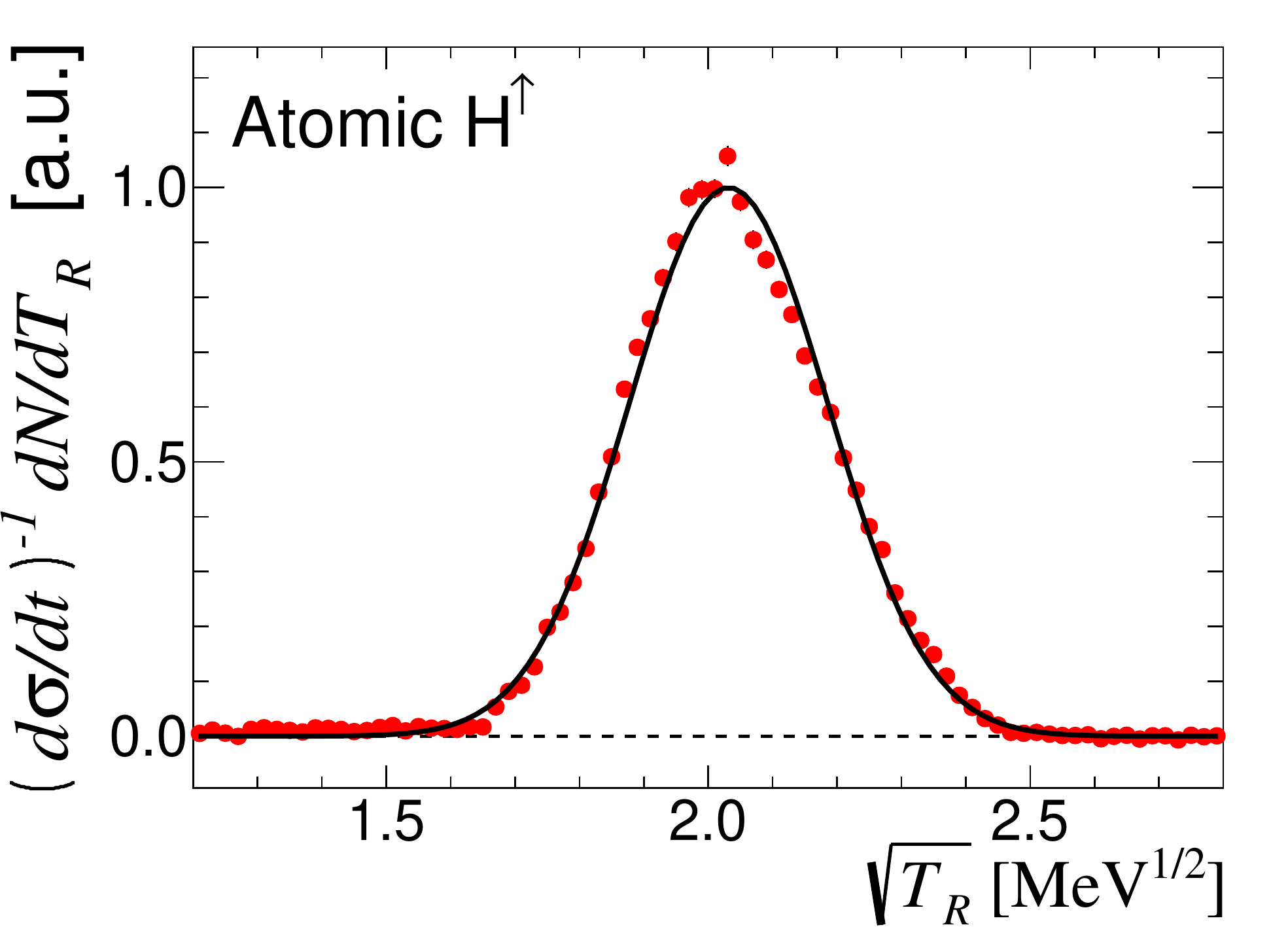}
\end{center}
\caption{\label{fig:jetH2}
  The jet density profile evaluation using the measured $dN/dT_R$ distribution in a Si strip. Upper: the molecular hydrogen component isolated in a run with the dissociator RF-discharge off. Lower: the polarized atomic hydrogen profile in a regular measurement. The graphs have the same normalization per integral beam intensity in the measurements.}
\end{figure}

For Cuts\,I, the RHIC fill number dependence of $a_N^\text{jet}$ is displayed Fig.\,\ref{fig:RawAsymF}. One can see good agreement between asymmetries measured with {\em blue} and {\em yellow} beams. Some excess of the $\chi^2$ over NDF, observed for {\em blue} beam only, can be attributed to possible indication of the fill-by-fill fluctuations of the systematic corrections. However, these fluctuations are smaller than the statistical errors of the measurements. For the RHIC Run\,17 average estimates, the long-term instability of the jet asymmetry measurement must not exceed  
\begin{equation}
\left|\frac{\delta a_N}{a_N}\right|_\text{long-term} \lesssim0.1\%
\end{equation}
This value includes the possible fluctuations of the jet polarization, instability of the energy calibration, fluctuations of the signal contamination by backgrounds, etc.

In a short term beam polarization measurement, $a_N^\text{jet}$ in Eq.\,(\ref{eq:Pbeam0}) can be substituted by the Run average $\left\langle a_N^\text{jet} \right\rangle$ value
\begin{equation}
  P_\mathrm{beam} = 
  \frac{a_N^\mathrm{beam}}{\langle a_N^\mathrm{jet}\rangle} P_\mathrm{jet}\left(1+\delta_\mathrm{corr}\right) =
   \frac{a_N^\mathrm{beam}}{A_N^\mathrm{eff}}
\end{equation}
where $\delta_\text{corr}$ denotes correction, which has to be applied to compensate the systematic error. The significance of this correction for Cuts\,I is followed from the measured beam polarization dependence on the recoil proton energy (Fig. \ref{fig:RawPol}).

Excellent stability of the measured jet spin asymmetry during the Run allows us to use the RHIC Run average effective analyzing power $A_N^\text{eff}$. This dilutes statistical uncertainties in short-term measurements. To incorporate the Run average systematic corrections to  $A_N^\text{eff}$, we may process the data using  a special set of event selection cuts (Cuts\,II) for which $\tilde{\delta}_\text{corr}$ can be precisely evaluated:
\begin{equation}
A_N^\mathrm{eff} = \frac%
{\left\langle\tilde{a}_N^\mathrm{jet}\right\rangle}
{P_\mathrm{jet}\left(1+\tilde{\delta}_\mathrm{corr}\right)}
\,\frac%
{\left\langle{a_N^\mathrm{beam}}\right\rangle}
{\left\langle\tilde{a}_N^\mathrm{beam}\right\rangle}.
\label{eq:ANeff}
\end{equation}
To distinguish between Cuts\,I and Cuts\,II values, the latter are marked by tilde.

\begin{figure}[t]
\begin{center}
  \includegraphics[width=0.5\columnwidth]{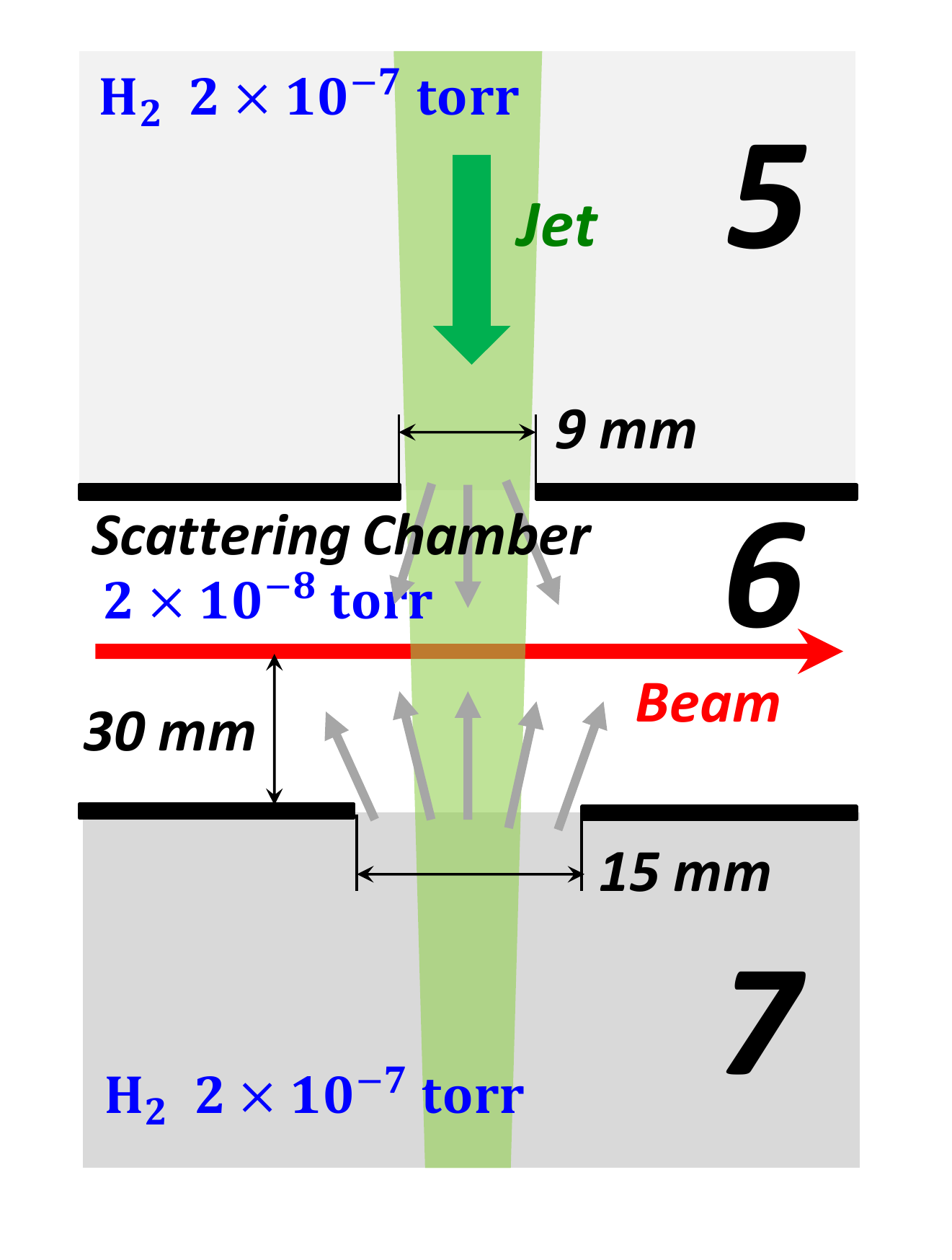}
\end{center}
\caption{\label{fig:MolecularHydrogen} The molecular hydrogen flow in HJET. The pressures in the chambers shown are given for a regular HJET run.}
\end{figure}

\begin{figure}[t]
  \begin{center}
    \includegraphics[width=0.8\columnwidth]{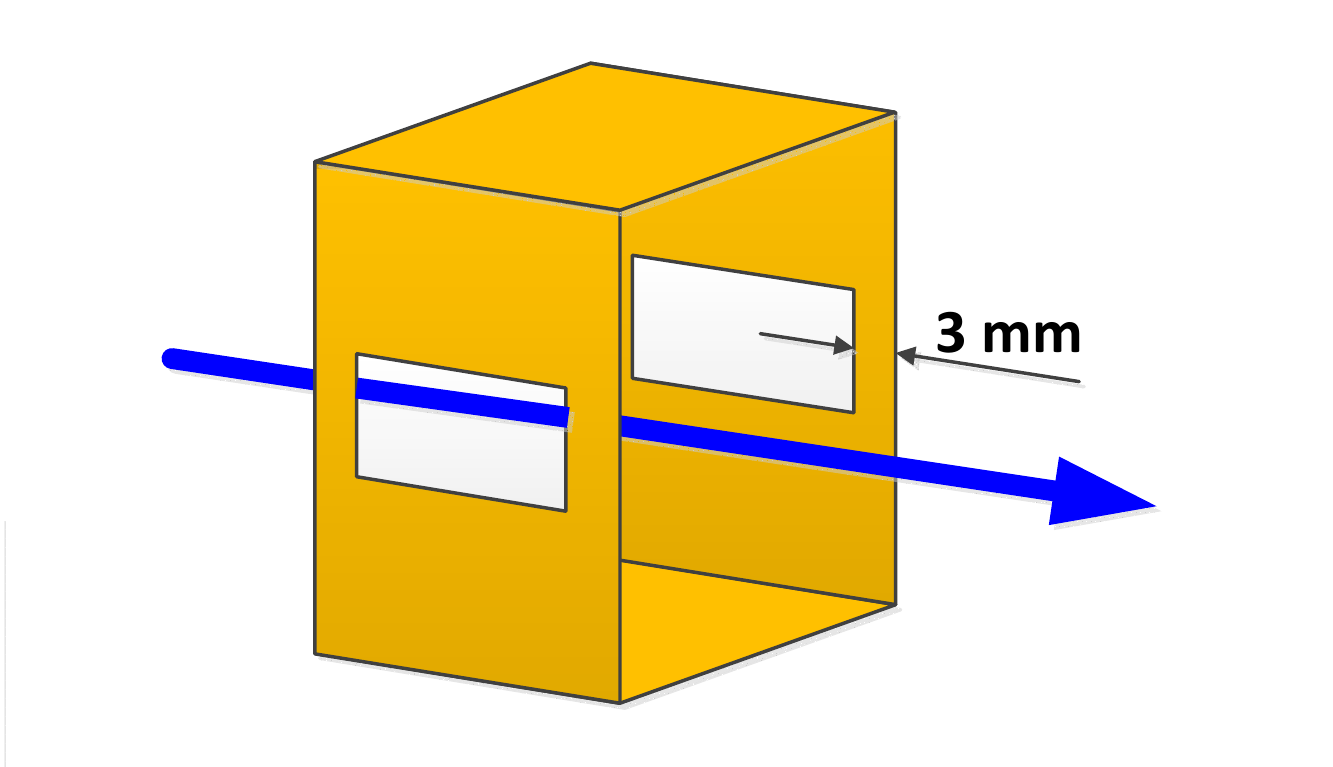} 
  \end{center}
  \caption{\label{fig:Collimators}
    HJET construction element which partially shadows recoil protons from the beam scattering on the beam gas H$_2$.}
\end{figure}

\begin{figure}[h]
  \begin{center}
    \includegraphics[width=0.8\columnwidth]{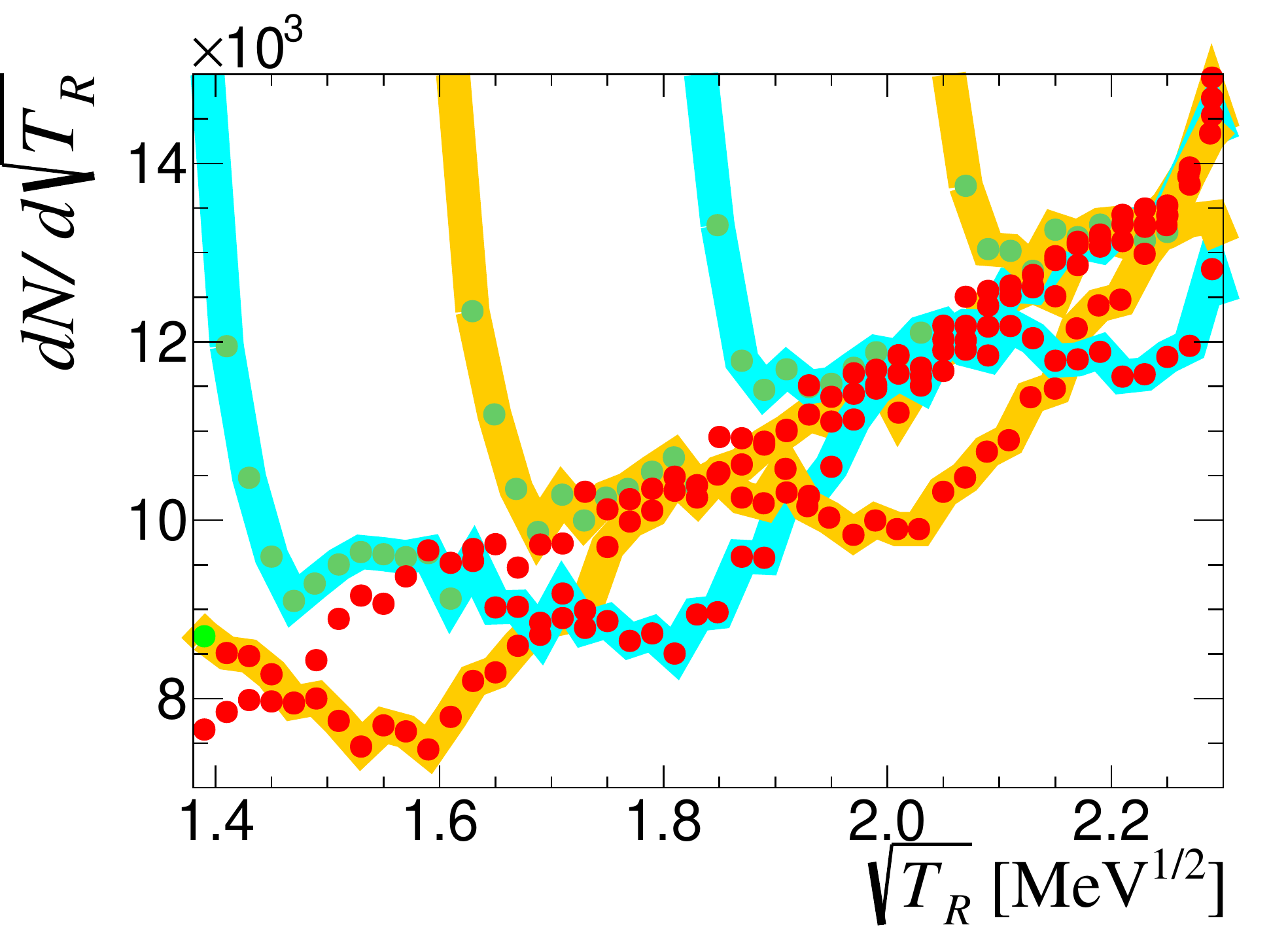} 
  \end{center}
  \caption{\label{fig:MHnorm} 
  Equidistant dips in the $dN/d\sqrt{T_R}$ distributions in the consecutive Si strips used to calibrate the beam gas $\text{H}_2$ background.
} 
\end{figure}

\subsection{Molecular hydrogen in the jet \label{sec:H2Jet}}

Typically, mass fraction of the molecular hydrogen flow from the dissociator is about 10\%. Since the $\text{H}_2$ component is not focused by the sextupole magnets, only a small fraction of these molecules can directly reach the collision point, 127\,cm away from the dissociator nozzle. By turning off the dissociator RF-discharge, the atomic component in the jet is eliminated while the $\text{H}_2$ density is increased by factor $\sim10$. In a such measurement (see Fig.\,\ref{fig:jetH2}), we have evaluated the molecular hydrogen fraction in the polarized atomic jet and, subsequently, the corresponding systematic correction to the measured beam polarization
\begin{equation}
  \left(\frac{\delta P}{P}\right)^\text{syst}_\text{jet H$_2$} =
  +0.06\pm0.06\,\%,
\end{equation}
where the specified error reflects the possible (upper limit) uncertainty in the dissociation degree.

\subsection{Molecular hydrogen in the beam gas \label{sec:H2BeamGas}}

As shown in Fig.\,\ref{fig:MolecularHydrogen}, the beam gas hydrogen excess in the scattering Chamber\,6 is caused by $\text{H}_2$ diffusion from Chamber\,7 (the scattered and recombined jet atomic hydrogen) and Chamber\,5 (unfocused hydrogen atoms recombined to molecules).
The diffusion $\text{H}_2$ density profile (seen by the beam)  is much wider than that of the atomic hydrogen profile in the jet.  It was confirmed in the special tests by injecting $\text{H}_2$ gas to Chamber\,7 or to Chamber\,5.
  
To evaluate the beam gas $\text{H}_2$ density, we noticed that the recoil protons from the beam scattering on H$_2$ at $z\!\approx\!\pm12\,\text{mm}$ are shadowed by $d\!=\!3\,\text{mm}$ wide wall in the HJET construction (depicted in Fig.\,\ref{fig:Collimators}). The efficiency of such shadowing is up to $d/d_\text{str}=80\%$ where $d_\text{str}\!=\!3.7\,\text{mm}$ is the Si strip active width. In the $dN/d\sqrt{T_R}$ distributions (see Fig.\,\ref{fig:MHnorm}), the effect manifests itself by the equidistant dips, relative to the strips common background. Comparing the event rates in the dips with the jet center (see, {\em for example}, Fig.\,\ref{fig:Background}) one can estimate the atomic mass fraction of the beam gas H$_2$ in the jet center
\begin{equation}  
  b_\text{H$_2$}^{(0)} = 0.41\pm0.04\,\%.
\end{equation}
which leads to the effective, Cuts\,II average, 
\begin{equation}
  b_\text{H$_2$} = 0.56\pm0.06\,\%.
\end{equation}
background rate in the Run\,17 measurements. Such evaluated  $b_\text{H$_2$}$ is provided by one ({\em forward}) beam only and gives about 20\% to the total background rate for the $2\text{--}5\,\text{MeV}$ recoil protons.

\begin{figure}[t]
\begin{center}
\includegraphics[width=0.8\columnwidth]{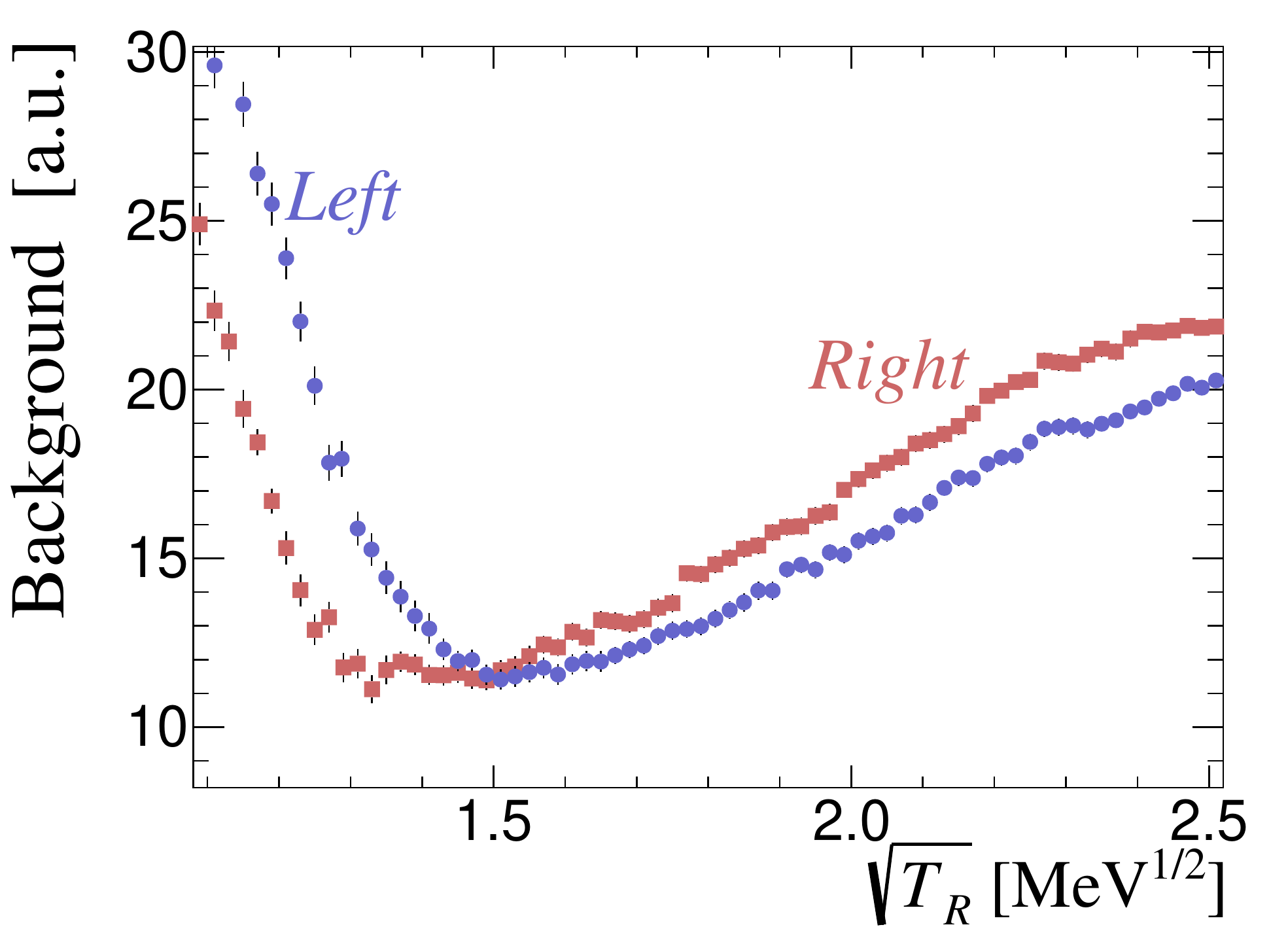} 
\end{center}
\caption{\label{fig:bgrLR}
  Comparison of the background $dN/d\sqrt{T_R}$ rate in left and right side detectors.
}
\end{figure}

\begin{figure}[t]
\begin{center}
\includegraphics[width=0.8\columnwidth]{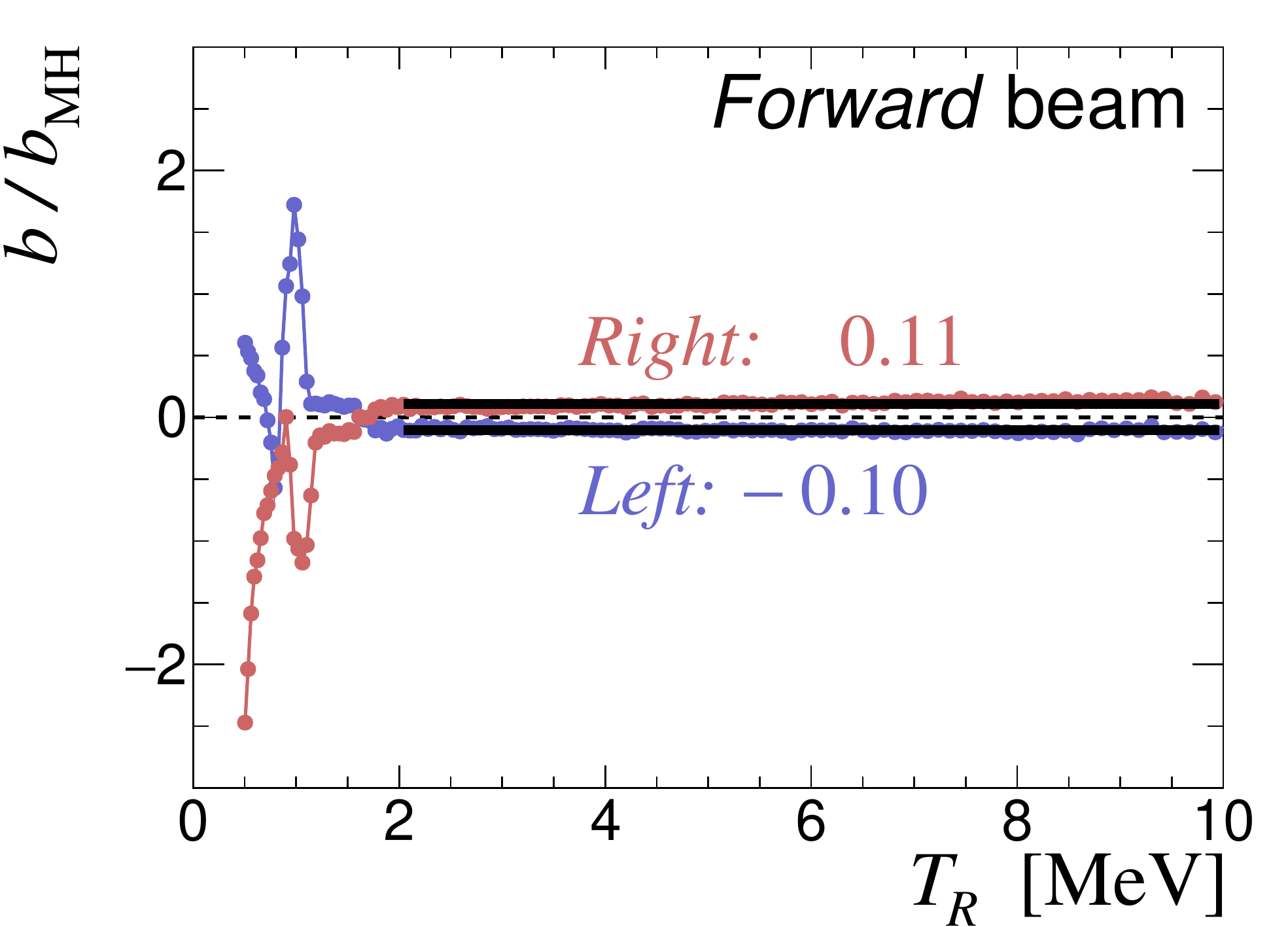} 
\includegraphics[width=0.8\columnwidth]{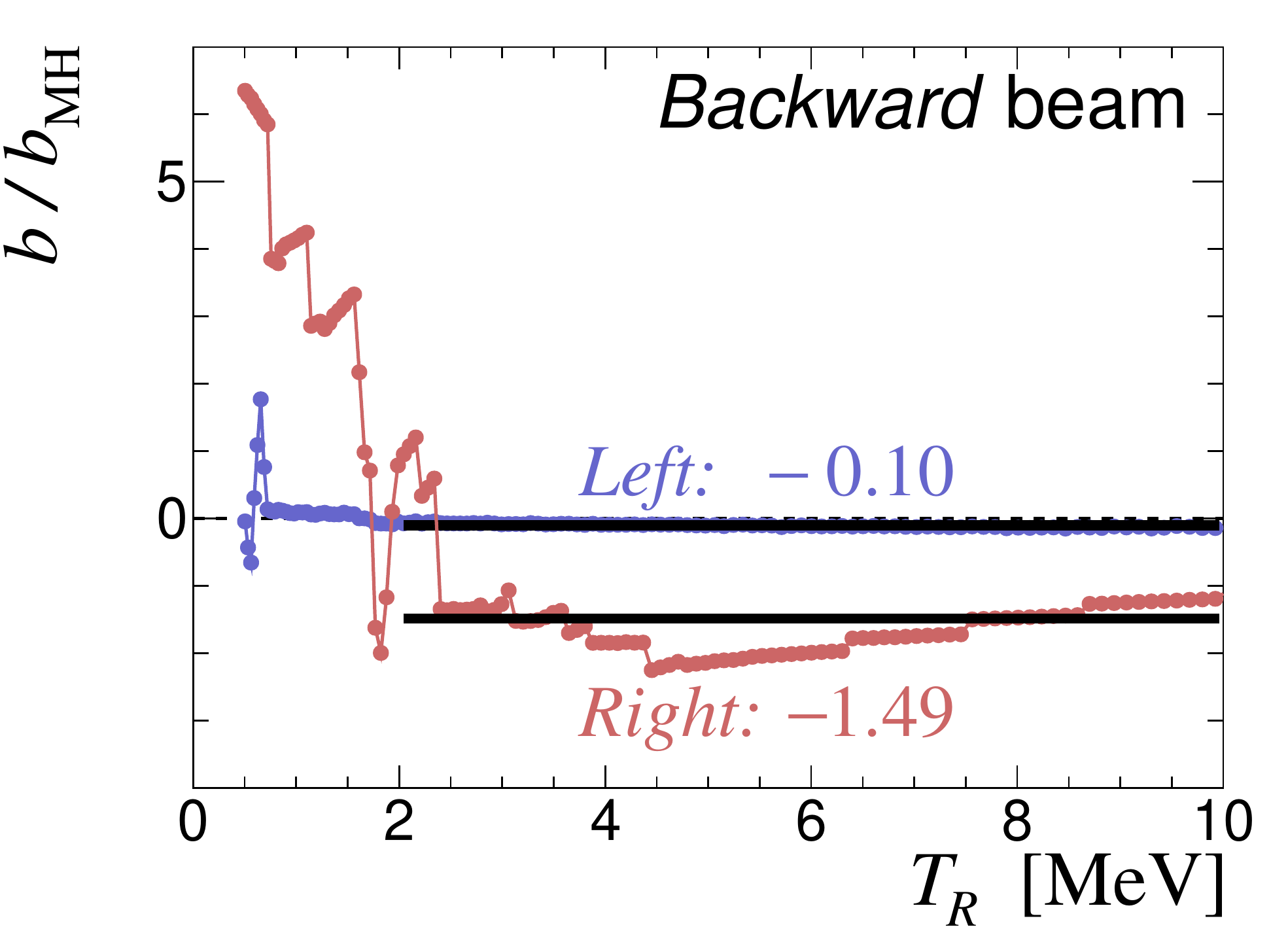}
\end{center}
\caption{\label{fig:mhBgr} 
  Simulation of the residual H$_2$ backgrounds for {\em forward} and {\em backward} beams. Backgrounds in left and right (relative to the {\em forward} beam) detectors are shown separately. The average values are given for the $2\!<\!T_R\!<\!10\,\text{MeV}$ energy range.}
\end{figure}
\begin{figure}[t]
  \begin{center}
    \includegraphics[width=0.95\columnwidth]{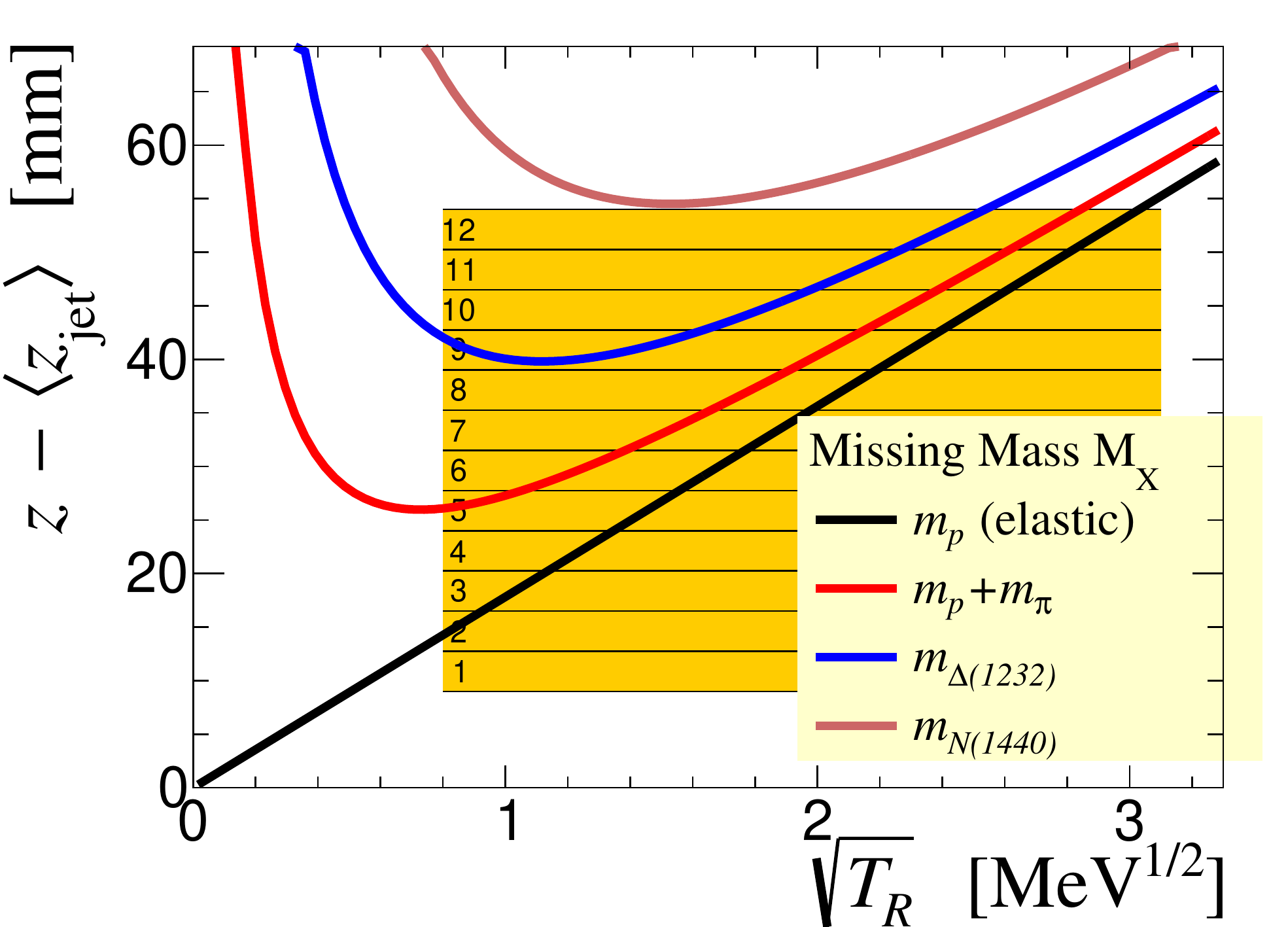} 
  \end{center}
  \caption{\label{fig:pp_pX} The recoil proton $z$-coordinate dependence on kinetic energy $T_R$ and missing mass $M_X$. The shown lines should be smeared with $\sigma_z\!\sim\!2.6\,\text{mm}$ due to the jet density profile.}
\end{figure}

The recoil proton  tracking in the holding field magnet (see Fig.\,\ref{fig:TracksMF}) strongly violates the dependence (\ref{eq:kappa}) and, thus, affects the beam gas $\text{H}_2$ background subtraction. The effect can be readily seen by comparing the experimentally evaluated background in left and right detectors (Fig.\,\ref{fig:bgrLR}).

Residual, i.e. after subtraction, backgrounds calculated separately for the {\em forward} (the beam which polarization is measured by the considered detector) and {\em backward} (the opposite direction beam) beams and for the left and right detectors are shown in Fig.\,\ref{fig:mhBgr}.

For the spin pattern used in RHIC, the polarizations of {\em forward} and {\em backward} beams are uncorrelated on average. Therefore, for the {\em backward} beam, the residual $\text{H}_2$ background does not alter the measured beam polarization (the ratio $a_N^\text{beam}/a_N^\text{jet}$). Subsequently, to evaluate respective systematic errors in the beam polarization measurement, one should only use the {\em forward} beam residual background. According to Fig.\,\ref{fig:mhBgr}, for the Cuts\,II recoil proton energy cut $T_R\!>\!2\,\text{MeV}$  the resulting  systematic corrections to the measured beam polarization is negligible. Possible variations in the simulation model do not change this conclusion.
  
However, there is some uncertainty in determining the $\text{H}_2$ background caused by non-uniform shadowing of a Si detector. This uncertainty may be accounted  as a related systematic error:
\begin{equation}
  \left(\frac{\sigma_P}{P} \right)^\text{syst}_\text{beam $\text{H}_2$}\lesssim0.1\%
\end{equation}

In the analyzing power measurements, we corrected the residual $\text{H}_2$ background using the simulation for the left/right detectors and {\em forward}/{\em backward} beams.

\subsection{Inelastic pp scattering \label{sec:Inelastic}}

\begin{figure}[t]
  \begin{center}
    \includegraphics[width=0.8\columnwidth]{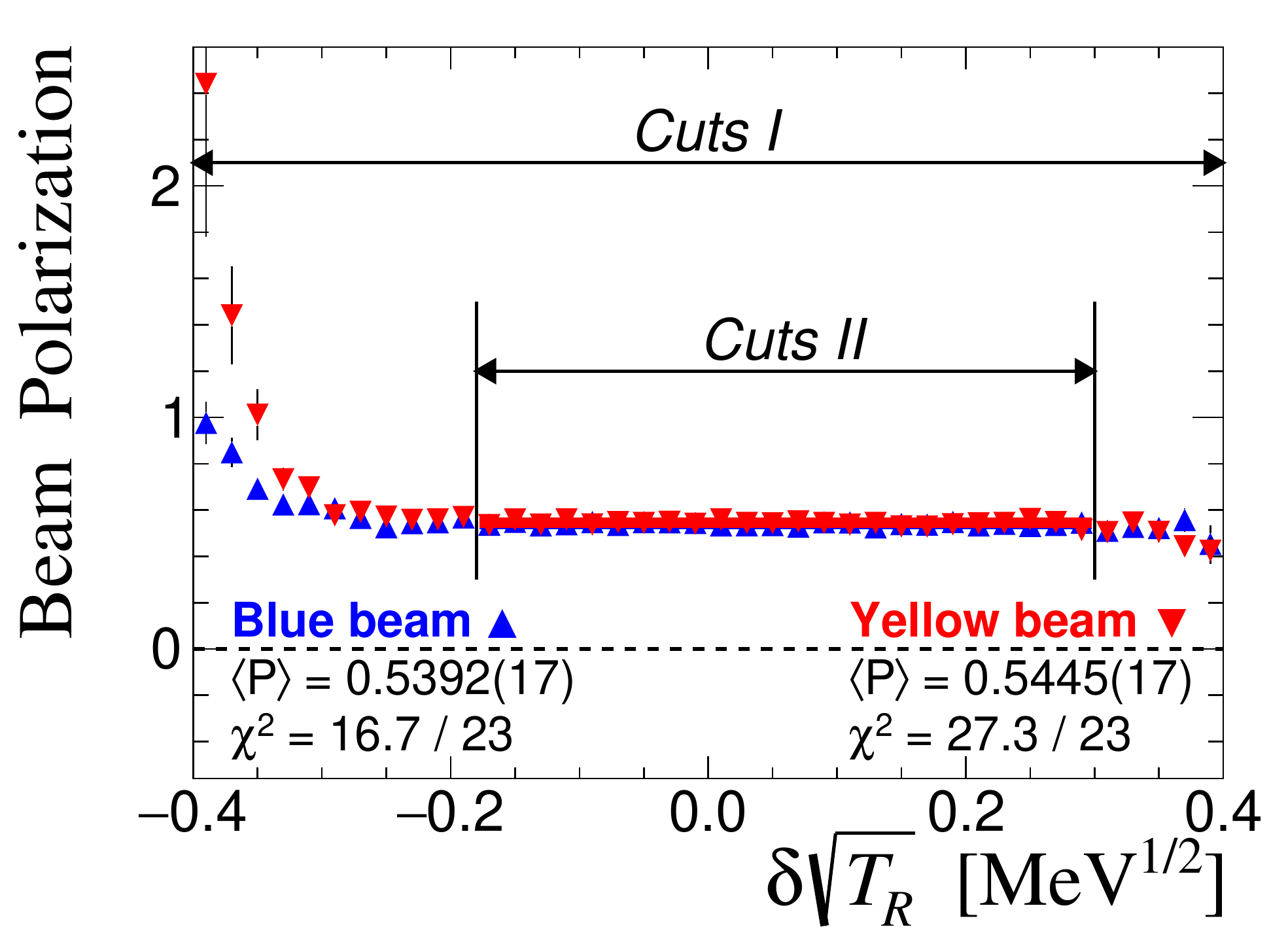} 
  \end{center}
  \caption{\label{fig:MMcut} The measured beam polarization dependence on $\delta\sqrt{T}$ event selection cuts. The fit results are given for Cuts\,II.}
\end{figure}
\begin{figure}[t] \begin{center}
\includegraphics[width=0.95\columnwidth]{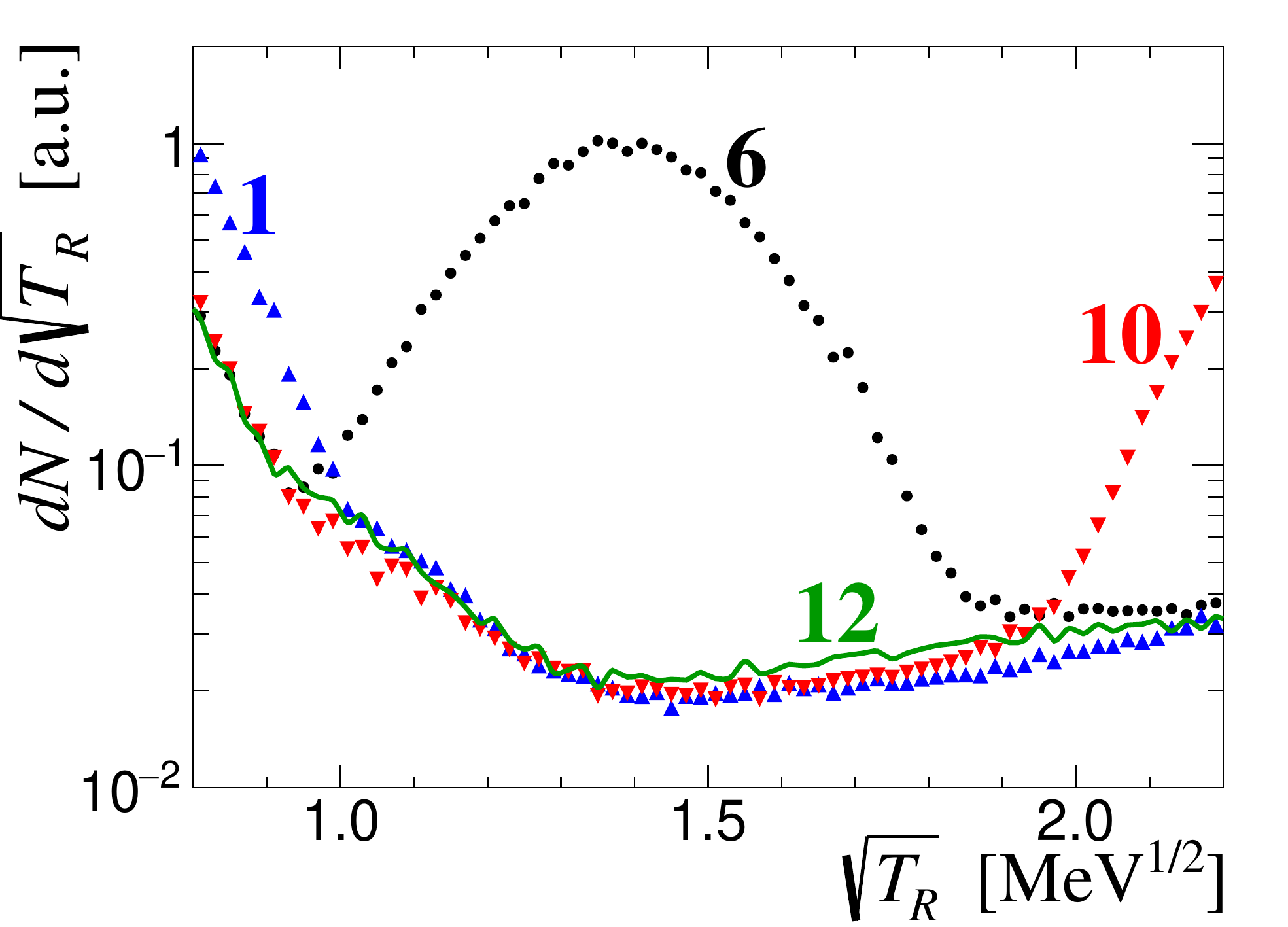}       
\caption{\label{fig:pA_bgr}
  Comparison of the $dN/d\sqrt{T_R}$ (background) distributions for Si strips 1 and 10. Elastic peak (strip 6) is shown for normalization. Solid line (strip 12) gives an estimate of the inelastic $\mathit{pp}$ background in this strip.
  }
\end{center} \end{figure}

For inelastic proton beam scattering on a proton target (\ref{eq:kappa'}), the recoil proton $z$-coordinate dependences on $T_R$ and $\Delta$ [see Eq.\,(\ref{eq:kappa'})] is shown in Fig.\,\ref{fig:pp_pX}. For the 255 GeV beam, about half of the Si strips are exposed to inelastic protons. It was observed (see Fig.\,27 in Ref.\,\cite{bib:PSTP2017}) that the inelastic beam spin asymmetry is larger than the elastic one while the jet spin asymmetry is smaller. Therefore, the inelastic background increases the measured beam polarization in HJET. To separate elastic and inelastic $\mathit{pp}$-scattering events we altered the $\delta\sqrt{T}$ cut as shown in Fig.\,\ref{fig:MMcut}. By varying the cut, the systematic error, for Cuts\,II, was estimated as
\begin{equation}
  \left( \frac{\delta P}{P} \right)^\text{syst}_{pp\rightarrow Xp} = +0.15\pm0.15\,\%.
\end{equation}

\subsection{pA scattering}

The $p\text{A}$ scattering gives the dominant contribution to the recoil proton background which is assumed to be Si strip $z_\text{str}$ coordinate independent.
In units of  $k_\text{str}$, the strip number in a detector, the background rate can be approximated by 
\begin{equation}
\begin{aligned}[c]
  \eta(T_R,z_\text{str}) =\:& dN_\text{bgr}^{(p\text{A})}/d\sqrt{T_R}\\
  \propto\:& 1+c_1(T_R)k_\text{str}+c_2(T_R)k_\text{str}^2
\end{aligned}
\end{equation}
where $c_{1,2}$ are expected to be small. Since both RHIC beams, {\em blue} and {\em yellow}, contribute to the background in each Si strip, $c_1$ must be additionally suppressed.

\begin{figure*}[t]
  \begin{center}
    \includegraphics[width=0.32\textwidth]{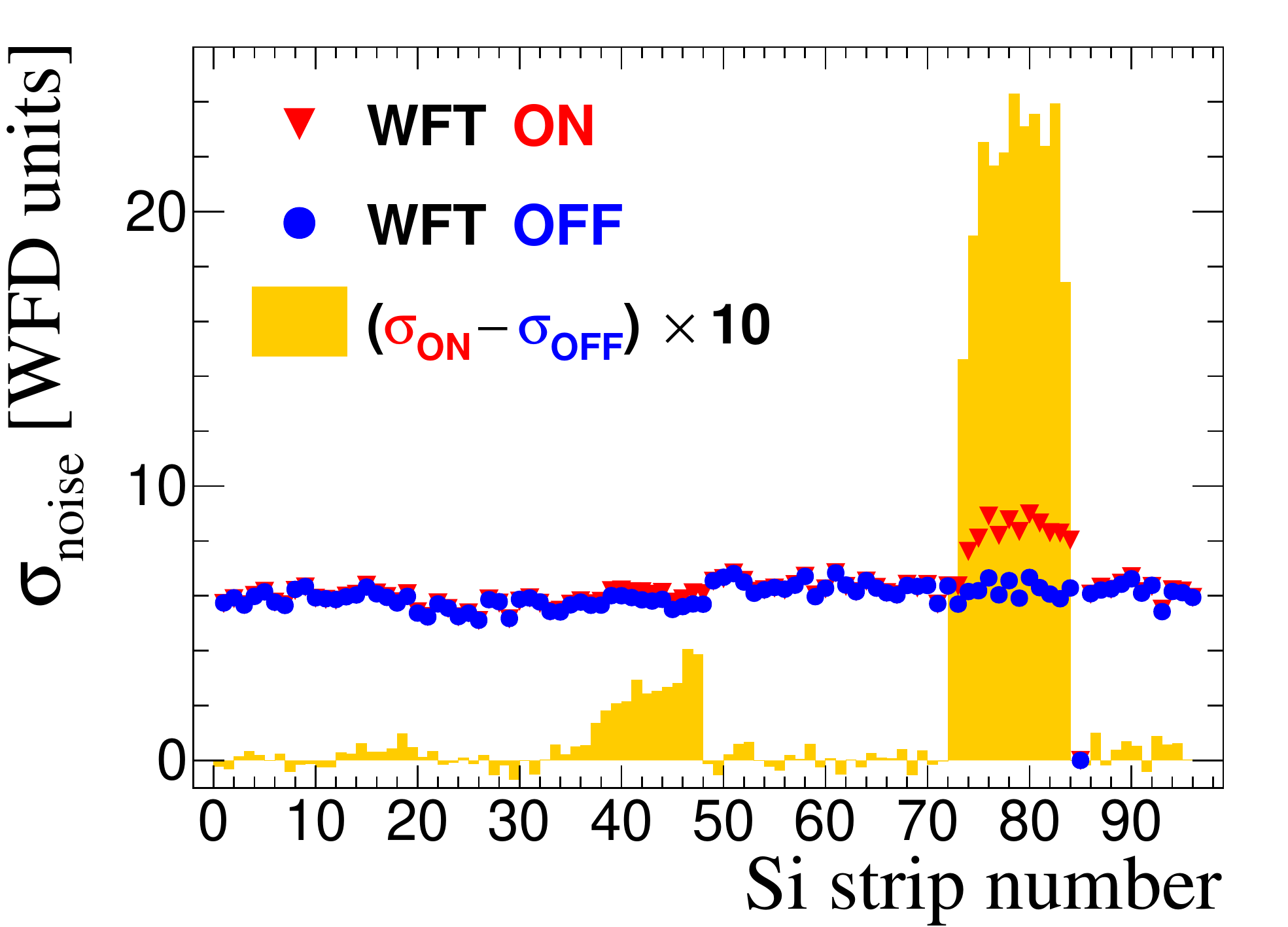} 
    \includegraphics[width=0.32\textwidth]{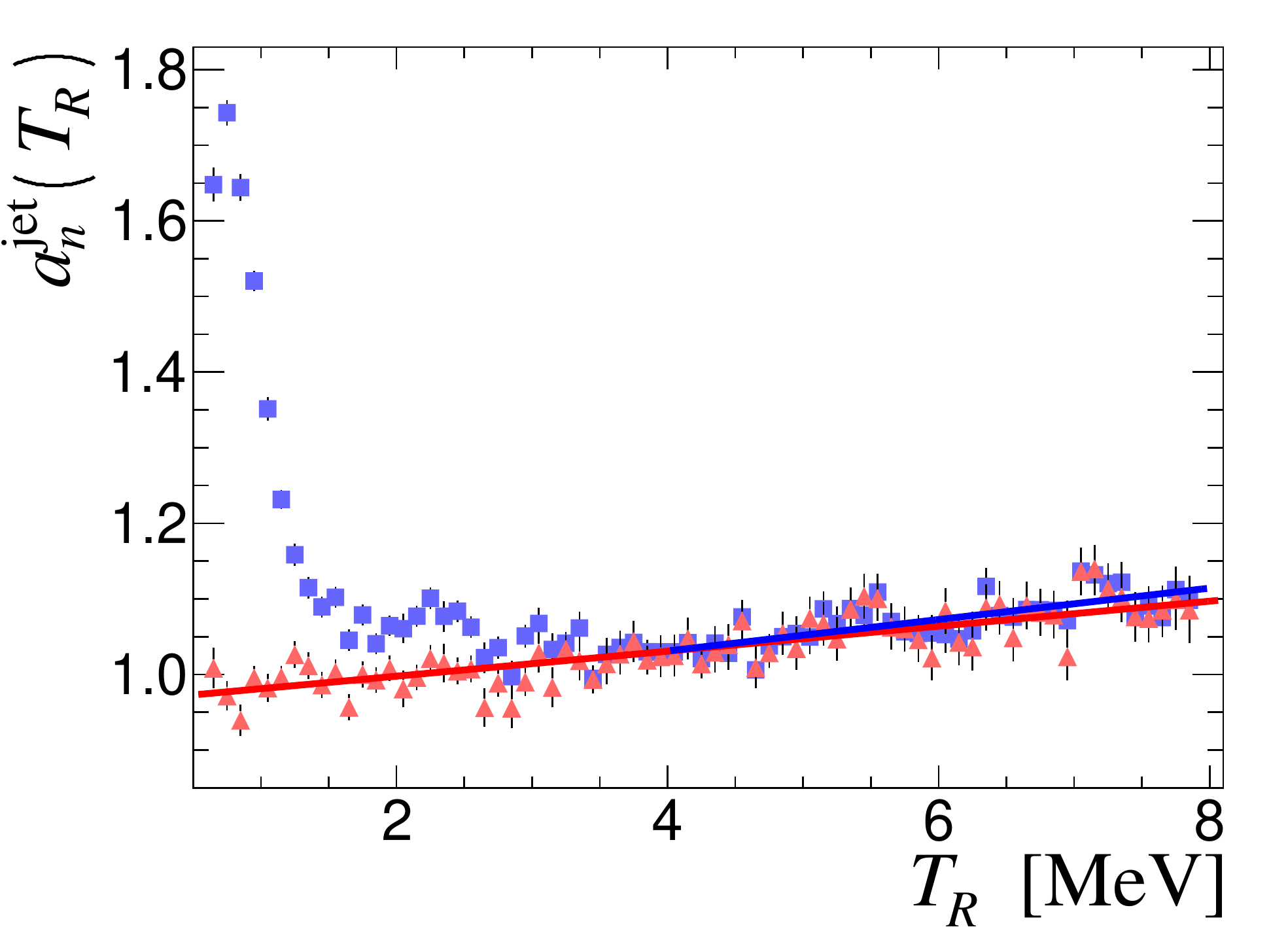} 
    \includegraphics[width=0.32\textwidth]{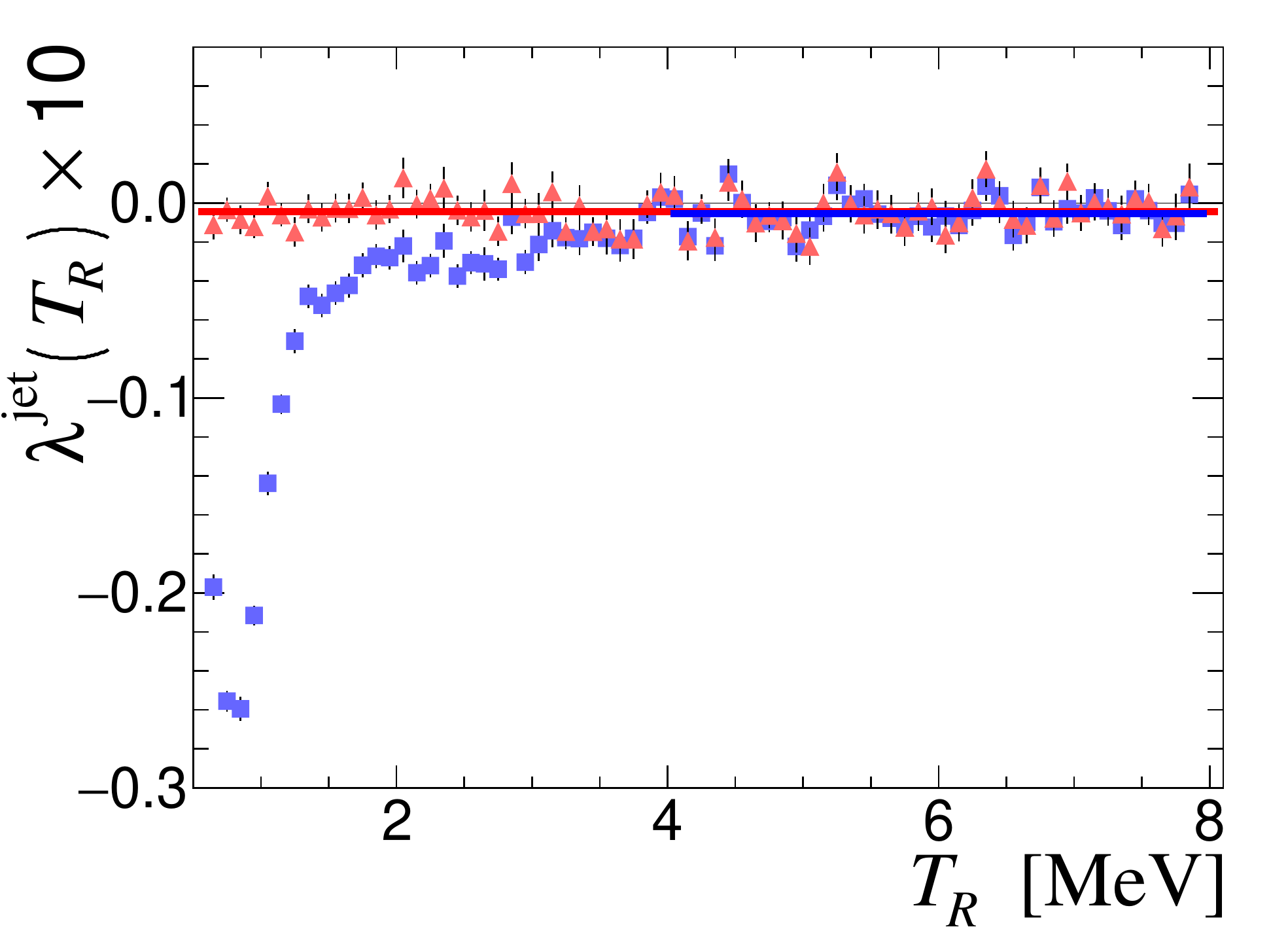} 
    \caption{\label{fig:Noise}
      Left: Effective electronic noise for the jet spin up and down in Run\,15. The difference scaled by factor 10 is shown by the histogram. For the measured jet spin (center) and intensity (right) asymmetries,
      {\color{blue}\scriptsize$\blacksquare$} points are for the all {\em blue} Si detectors and 
      {$\color{red}\blacktriangle$} are for the analysis with no strips 72--83 data used.}
  \end{center}
\end{figure*}

To evaluate a possible non-uniformity of $\eta(z_\text{str})$ we compare $dN/d\sqrt{T_R}$ distributions in two strips, $k_\text{str}$=1 $(z_\text{str}\!\approx\!11\,\text{mm})$ and 10 $(45\,\text{mm})$. The distributions, shown in Fig.\,\ref{fig:pA_bgr} were averaged over all 8 Si detectors to eliminate discrepancy due to the left/right dependent backgrounds. To avoid corruption by the inelastic $\mathit{pp}$ events, we have used $100\,\text{GeV}$ data. For that beam energy, the inelastic background can be detected only in strip 12. For the $\sqrt{T_R}$ range $1.2\text{--}1.7\,\text{MeV}^{1/2}$, where the contribution from the beam scattering of the jet protons is strongly suppressed, the $\eta_1(T_R)$ and $\eta_{10}(T_R)$ backgrounds are consistent at the 0.01\% level relative to elastic peak rate. Thus, systematic errors in $p\text{A}$ background subtraction can be neglected. However, since the obtained result is slightly dependent on the cuts used and because the entire Cuts\,II $T_R$ range was not tested, we used a more conservative estimate 
\begin{equation}
  \left( \frac{\sigma_P}{P}\right)_\text{pA}^\text{syst} < 0.2\%
\end{equation}
based on the observed fluctuations of the background lines. The specified systematic error is more due to the background subtraction method uncertainties rather than due to the $p\text{A}$ background itself. 

\subsection{Noise correlation with the jet spin}

It was found that the WFT induces a $14\,\text{MHz}$ ringing in some detectors. In Run\,15, this effective noise (see Fig.\,\ref{fig:Noise}) in {\em blue} upper left detector (strips 72--83) resulted in a dependence of the event selection efficiency on the  jet spin state. Therefore, for {\em blue} beam, the asymmetry $\delta\omega_L$ (\ref{eq:omega})  appeared to be significantly non-zero and, in accordance with Eqs.\,(\ref{eq:aN_syst}),\,(\ref{eq:Lambda_syst}), the correlated changes of the measured $a_n^\text{jet}(T_R)$ and $\lambda^\text{jet}(T_R)$ were observed. Removing strips 72--83 from the analysis fixed the distributions. For the beam spin asymmetries, the issue was not detected.

In RHIC Run\,17, no evidence of the WFT related systematic error was found. The upper limit, mostly defined by the statistical uncertainties in the studied distributions, was established as
\begin{equation}
  \left(\frac{\sigma_P}{P}\right)_\text{WFT}^\text{syst} < 0.2\%.
\end{equation}

\subsection{Systematic uncertainties summary}

\begin{table}[t]
  \begin{center}
    \begin{tabular}{ l | c | c}
      Source & $\delta P/P$~[\%] & $\sigma_P/P$~[\%] \\
      \hline \hline
      A.\quad Long term stability       &         & $0.1$  \\
      B.\quad Jet polarization          &         & $0.1$  \\
      C.\quad Jet H$_2$                    & $+0.06$ & $0.06$ \\
      D.\quad Beam gas H$_2$                   &         & $\lesssim0.1$ \\
      E.\quad $p\mathrm{A}$             &         & $\lesssim0.2$ \\
      F.\quad $pp\rightarrow Xp$        & $+0.15$ & $0.15$ \\
      G.\quad WFT                       &         & $\lesssim0.2$ \\
      \hline
      ~~Total                     & $+0.21$ & $\lesssim0.37$ \\
    \end{tabular}
  \end{center}
  \caption{Systematic uncertainties summary for the event selection Cuts II.
   Some systematic uncertainties were canceled in the beam/jet asymmetry ratio and, therefore, were omitted here.
   } 
\label{tab:Syst}
\end{table}

The systematic uncertainties in the beam polarization measurements are summarized in Table \ref{tab:Syst}. For event selection Cuts\,II, the total correction to compensate a systematic error bias was evaluated as
\begin{equation}
   \tilde{\delta}_\text{corr} = -0.21\pm0.37\,\%
\end{equation}
According to Eq.\,(\ref{eq:ANeff}) and using $\left\langle P_\text{jet}\right\rangle=0.957$, the effective Run\,17 {\em blue} and {\em yellow} beam spin analyzing powers are equal to
\begin{align}
  \left(A_N^\text{eff}\right)_\text{B} &=
  3.749 \pm 0.013_\text{stat} \pm 0.014_\text{syst}\,\%,
  \label{eq:ANeffB}\\
  \left(A_N^\text{eff}\right)_\text{Y} &=
  3.739 \pm 0.012_\text{stat} \pm 0.014_\text{syst}\,\%,
  \label{eq:ANeffY}
\end{align}
respectively. The effective systematic error in the HJET beam polarization measurements can be approximated by
\begin{equation}
  \sigma_P^\text{syst}/P \lesssim 0.5\%
\end{equation}
The provided value includes statistical uncertainties in evaluation of the Run\,17 average analyzing powers (\ref{eq:ANeffB}),\,(\ref{eq:ANeffY}).

\section{Single spin analyzing power measurement}
\begin{figure*}[t]
      \includegraphics[width=0.5\textwidth]{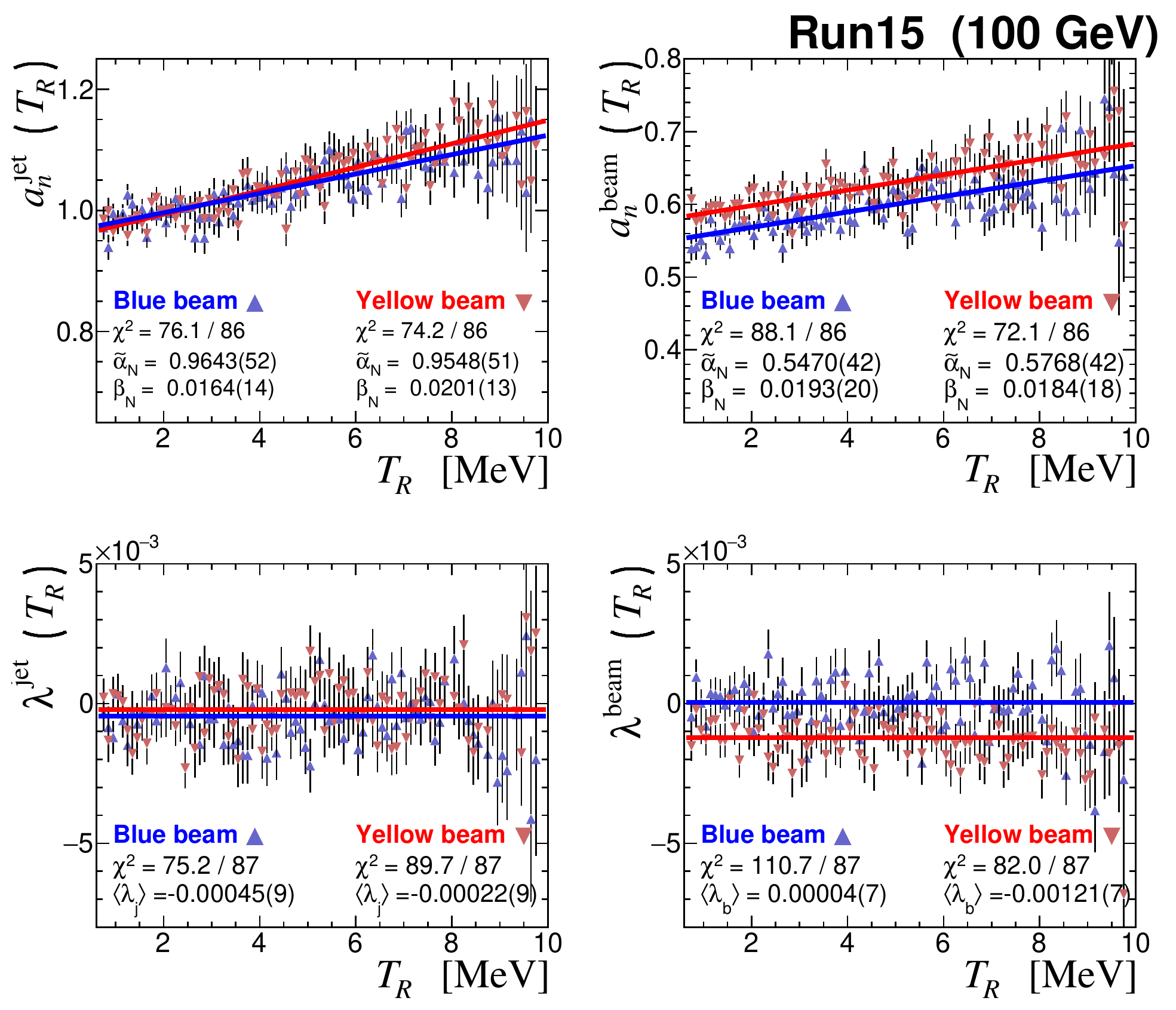} 
      \includegraphics[width=0.5\textwidth]{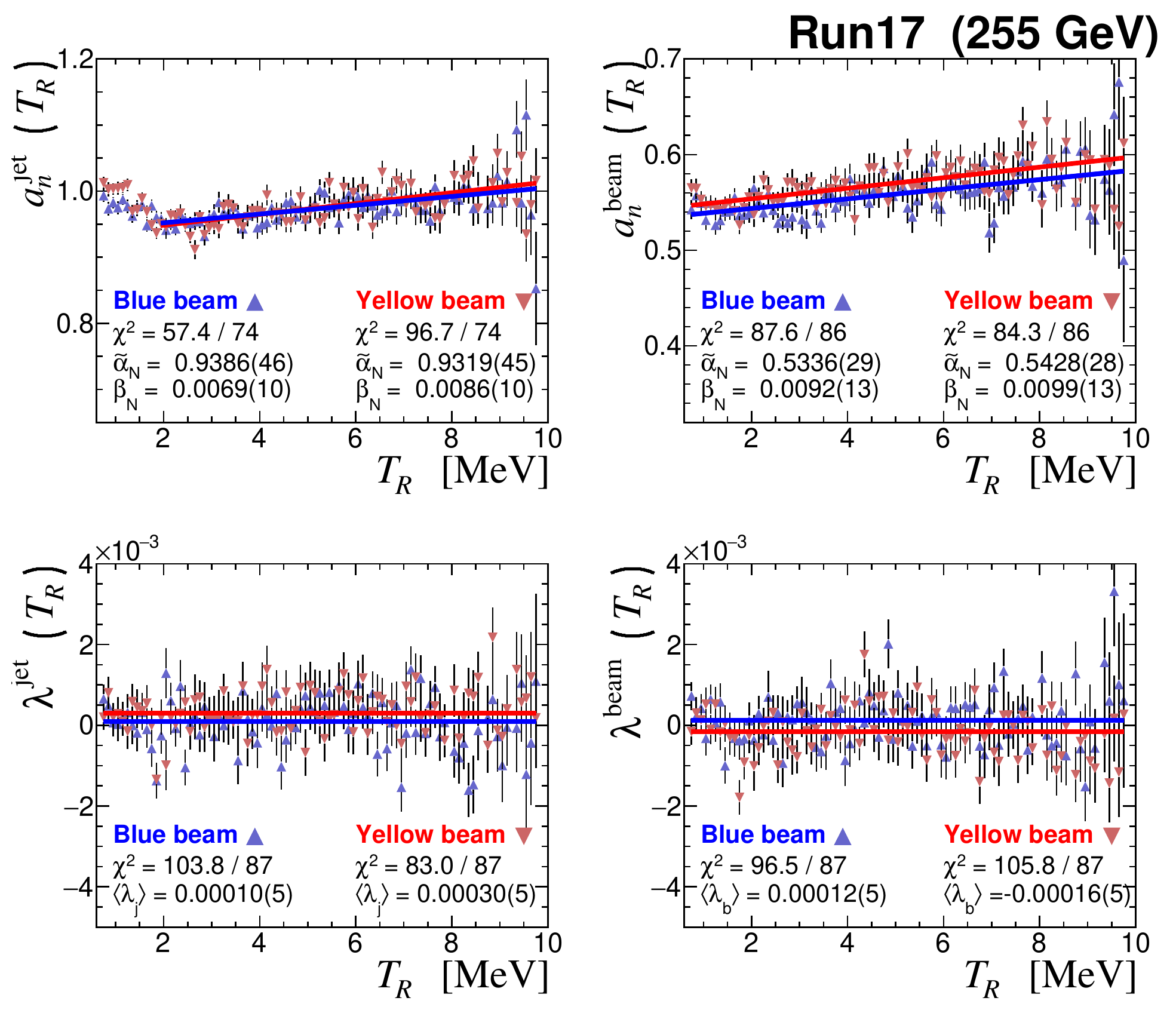}
      \caption{ Measured asymmetries in RHIC Run15 (100\,GeV) and Run\,17 (255 GeV). Blue/red polymarker colors are used for
        {\em blue / yellow} RHIC beams. 
        The fit energy range is $1.9<T_R<9.9\ \text{MeV}$ for the 255 GeV $a_n^\text{j}(T_R)$ and $0.7<T_R<9.9\ \text{MeV}$ for all other graphs. The fit parameter  $\tilde{\alpha}_\text{N}$ is defined as $\tilde{\alpha}_\text{N}=\langle P\rangle \alpha_\text{N}$.
}
      \label{fig:Asym}
\end{figure*}

Low systematic uncertainties in the HJET spin correlation measurements gave a possibility of high precision determination\,\cite{bib:HJET19} of the elastic $\mathit{pp}$ single spin analyzing power
\begin{equation}
  A_\text{N}(t) = a_\text{N}^\text{jet}(T_R)/P_\text{jet}
\end{equation}
at two proton beam energies, 100 and 255\,GeV. To fit $A_\text{N}(t)$, it is convenient to use $\alpha_\text{N}$ and $\beta_\text{N}$ based  parametrization (\ref{eq:AN_norm}). However, since the main goal of this analysis was experimental isolation of the hadronic spin flip amplitude $r_5$\,\cite{bib:BKLST}, the fit in Ref.\,\cite{bib:HJET19} was done in terms of $\text{Re}\,r_5$ and $\text{Im}\,r_5$ using the theoretically expected expressions for  $\alpha_\text{N}(r_5)$ and $\beta_\text{N}(r_5)$\,\cite{bib:AN_corr}.

In this study, the quadratic sum of uncertainties B, C, and F in Table\,\ref{tab:Syst} of 0.25\% was attributed to the effective systematic error in the value of the jet polarization, and the mean value of $P_\text{jet}$ was corrected by 0.06\%. In the combined Regge fit of the 100 and 255 GeV data\,\cite{bib:HJET09}, this systematic uncertainty was assumed to be strictly correlated for the both energies. Similarly, the quadratic sum of uncertainties A and G was interpreted as systematic error in $P_\text{jet}$, uncorrelated for the 100 and 255\,GeV.

The systematic corrections due to the data contamination by the inelastic $\mathit{pp}$ events were evaluated using the same method as in section \ref{sec:Inelastic}. The dependence of measured $r_5$ on the $\delta\sqrt{T_R}$ event selection cut was studied.

In the beam polarization measurement \cite{bib:PSTP2017}, some systematic uncertainties common to $a_\text{N}^\text{j}$ and  $a_\text{N}^\text{b}$ were effectively canceled. However, these uncertainties should be thoroughly re-assessed in the experimental evaluation of the single spin analyzing power:\\
\noindent{--\quad}%
Compared to section \ref{sec:H2BeamGas}, a more accurate analysis of the magnetic field corrections to the evaluated beam gas $\text{H}_2$ background is needed. The residual background, both for {\em forward} and {\em backward} beams, was simulated and corrections to the background subtractions were applied. The simulation included possible non-uniformity of the beam gas density due to pumping and used a more accurate description of the effective collimator shown in Fig.\,\ref{fig:Collimators}. For adjustment, we compared the calculated and measured difference in the left/right background rates (see Fig.\,\ref{fig:bgrLR}).\\
\noindent{--\quad}%
  The vertical size of the detectors effectively alter the jet polarization $P_\text{jet}^\text{eff}=P_\text{jet}\langle\sin\varphi\rangle\approx0.997\,P_\text{jet}$.\\
  \noindent{--\quad}%
  Systematic uncertainties in the energy calibration of the detectors, can alter the result of the measurement of $r_5$ or, equivalently, $\alpha_\text{N}$ and $\beta_\text{N}$.
  To estimate these systematic errors, we used the calibration uncertainties given in Eq.\,(\ref{eq:systCalib}) considering them strictly correlated for 100 and 255 GeV measurements.

  The effective, i.e. after systematic corrections, jet polarization used in data analysis was $P_\text{jet}^\text{eff}\!=0.954$ (100 GeV) and $P_\text{text}^\text{eff}\!=0.953$ (255 GeV). The small difference is due to change of the holding field magnet currents.  

\begin{figure}[t]
\begin{center}
\includegraphics[width=0.8\columnwidth]{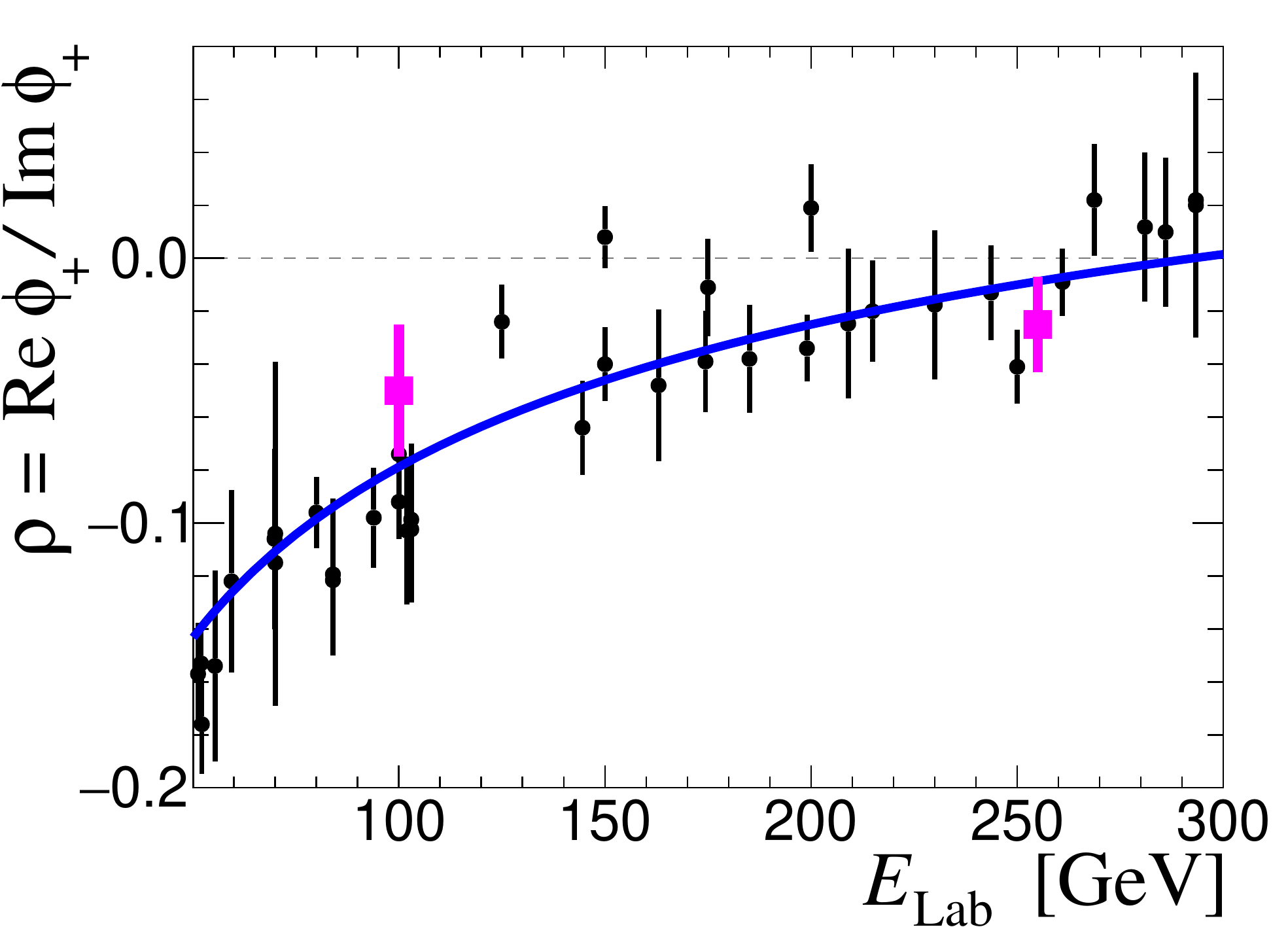} 
\end{center}
\caption{\label{fig:rho} 
  Comparison of the elastic $\mathit{pp}$ forward imaginary to real ratio $\rho$ determined in ({\scriptsize$\color{magenta}\blacksquare$}) the HJET analyzing power fit  \cite{bib:HJET19} and ($\bullet$) the unpolarized measurements  \cite{bib:PDG}.
}
\end{figure}
\begin{figure}[t]
\begin{center}
\includegraphics[width=0.8\columnwidth]{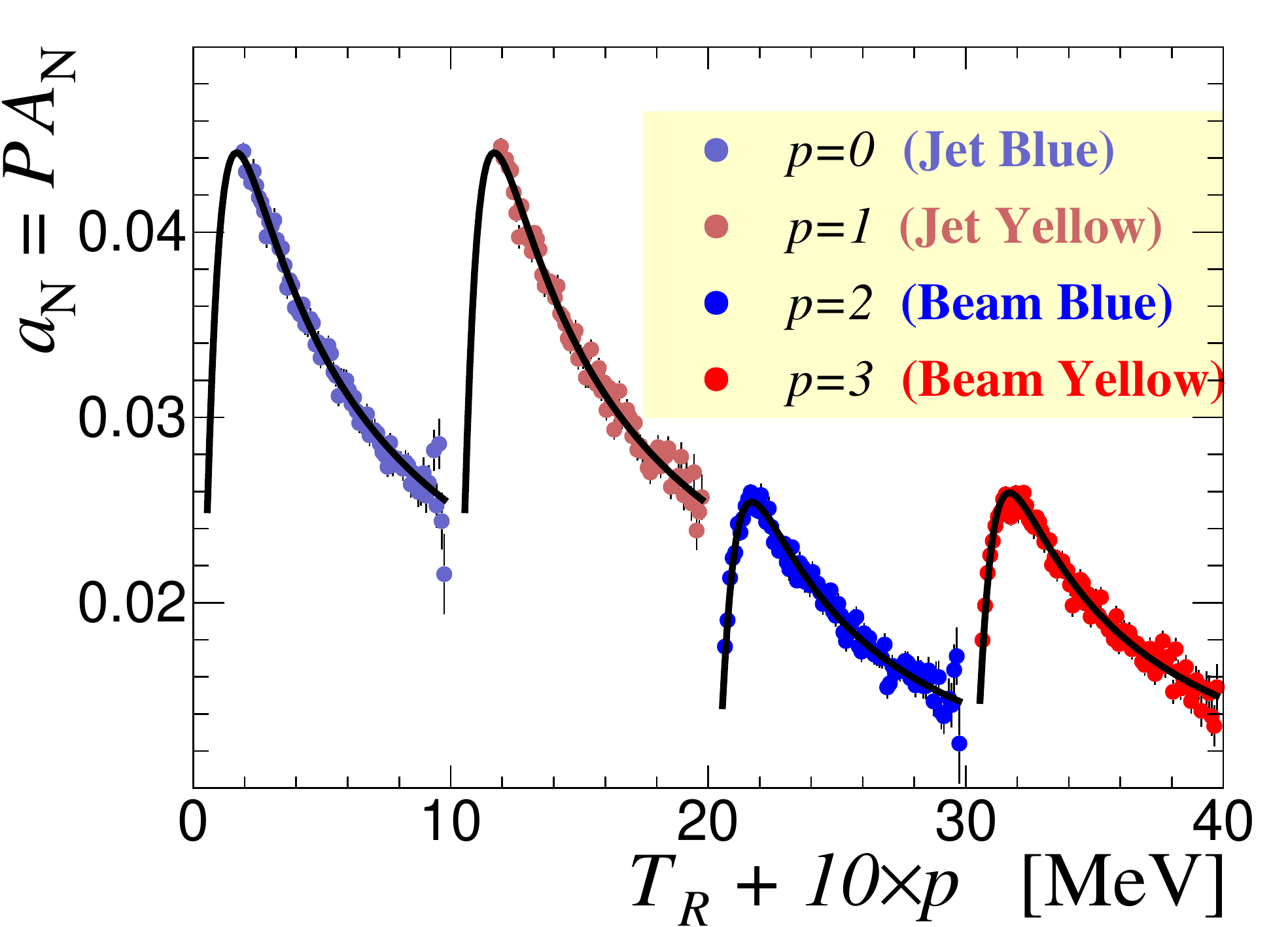} 
\end{center}
\caption{\label{fig:fitAN}
  The combined single spin analyzing power fit in Run\,17 (255\,GeV). All the displayed data was concurrently fit using four free parameters, $\alpha_\text{N}$, $\beta_\text{N}$, and Run average values of the {\em blue} and {\em yellow} beam polarizations.
  }
\end{figure}
\begin{figure}[t]
  \begin{center}
    \includegraphics[width=0.49\columnwidth]{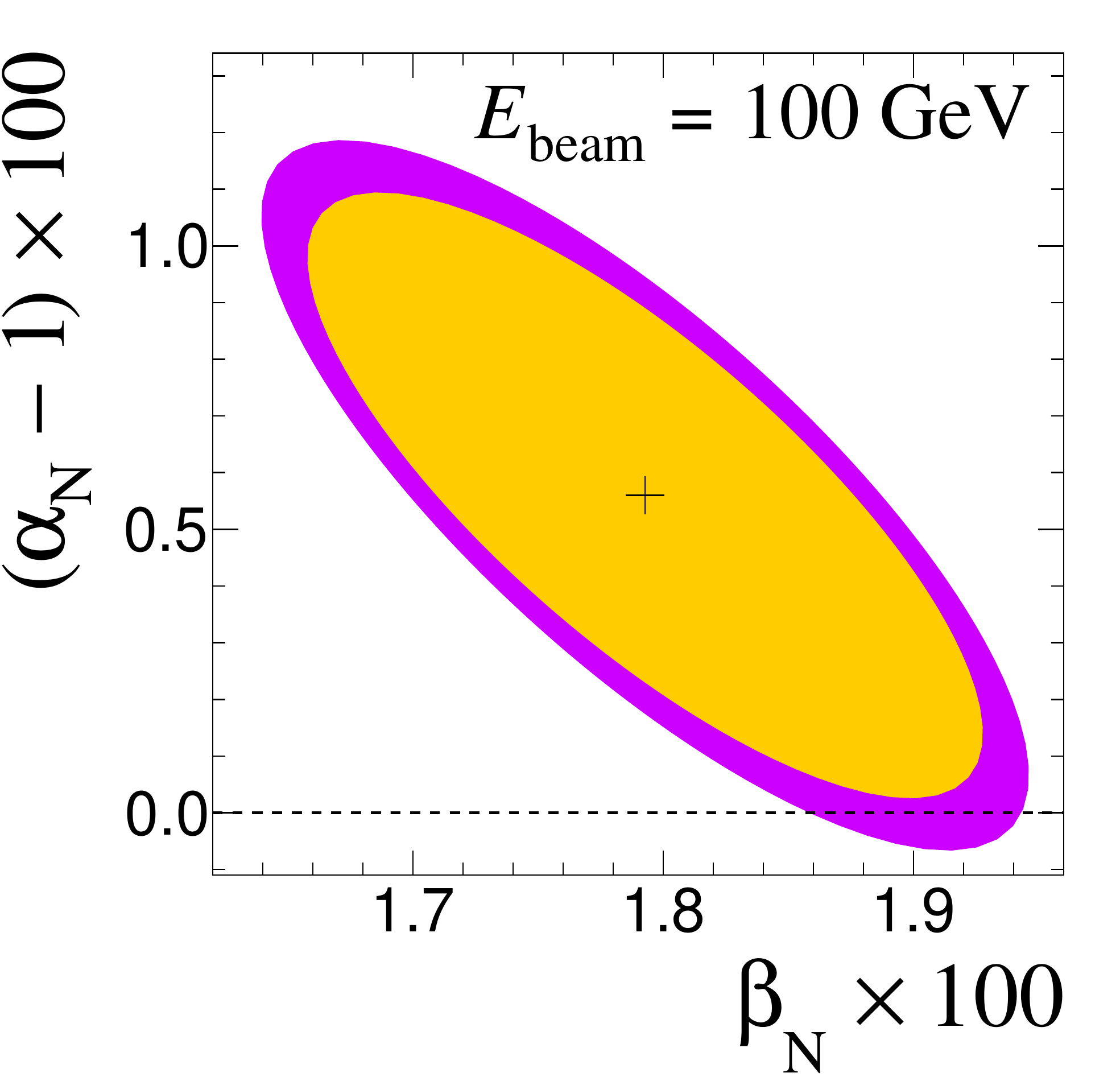} \hfill
    \includegraphics[width=0.49\columnwidth]{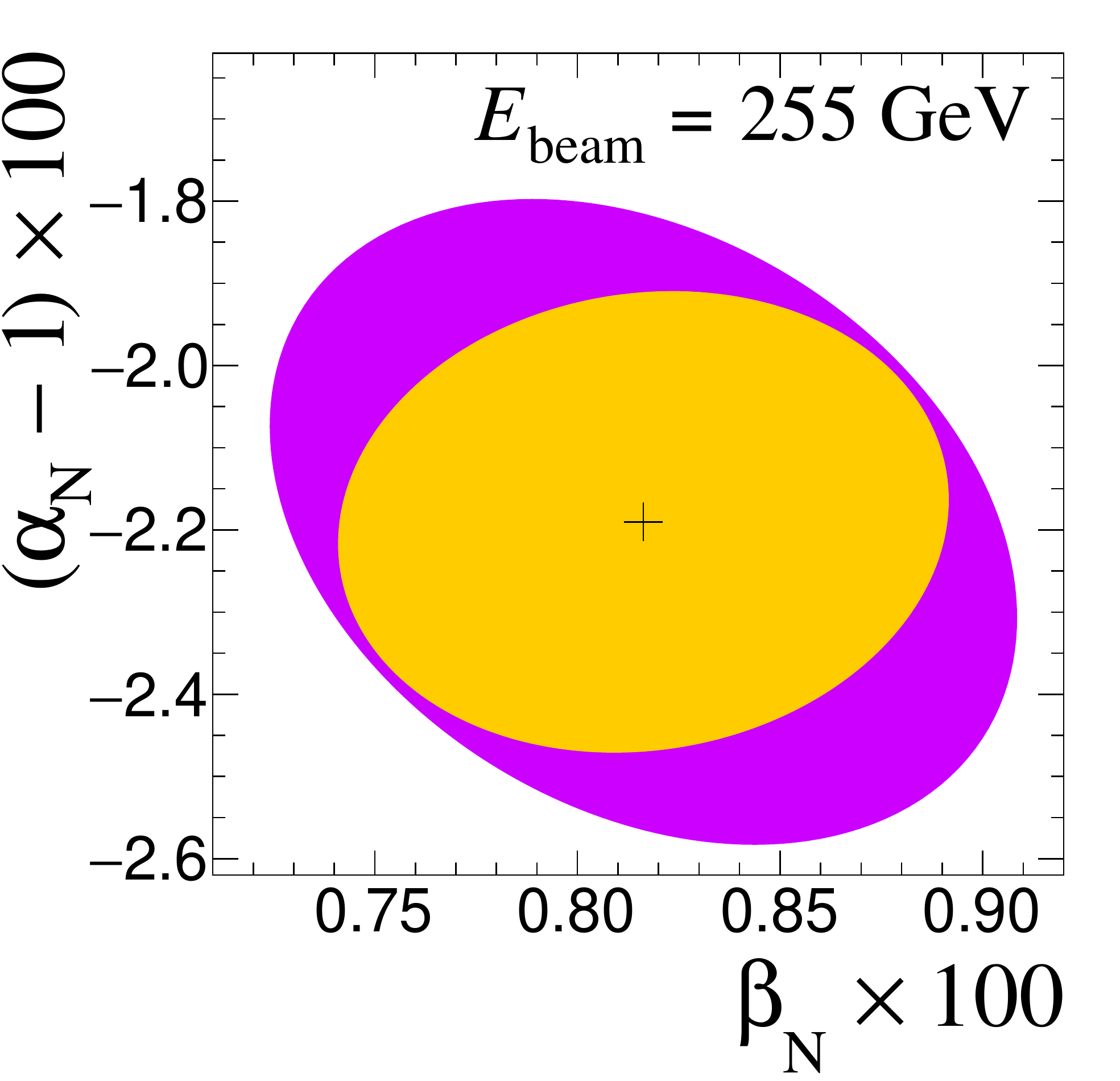}
    \caption{\label{fig:r5}
      $1\text{-}\sigma$ correlation contours, in terms of $\alpha_\text{N}$ and $\beta_\text{N}$, for the HJET measured single spin--flip hadronic amplitudes $r_5$. Orange and violet ellipses are for systematic and total (stat.+syst.) uncertainties, respectively.}    
  \end{center}
\end{figure}

\begin{figure*}[t]
  \begin{minipage}[t]{0.64\textwidth}
  \begin{center}
    \includegraphics[width=0.49\textwidth]{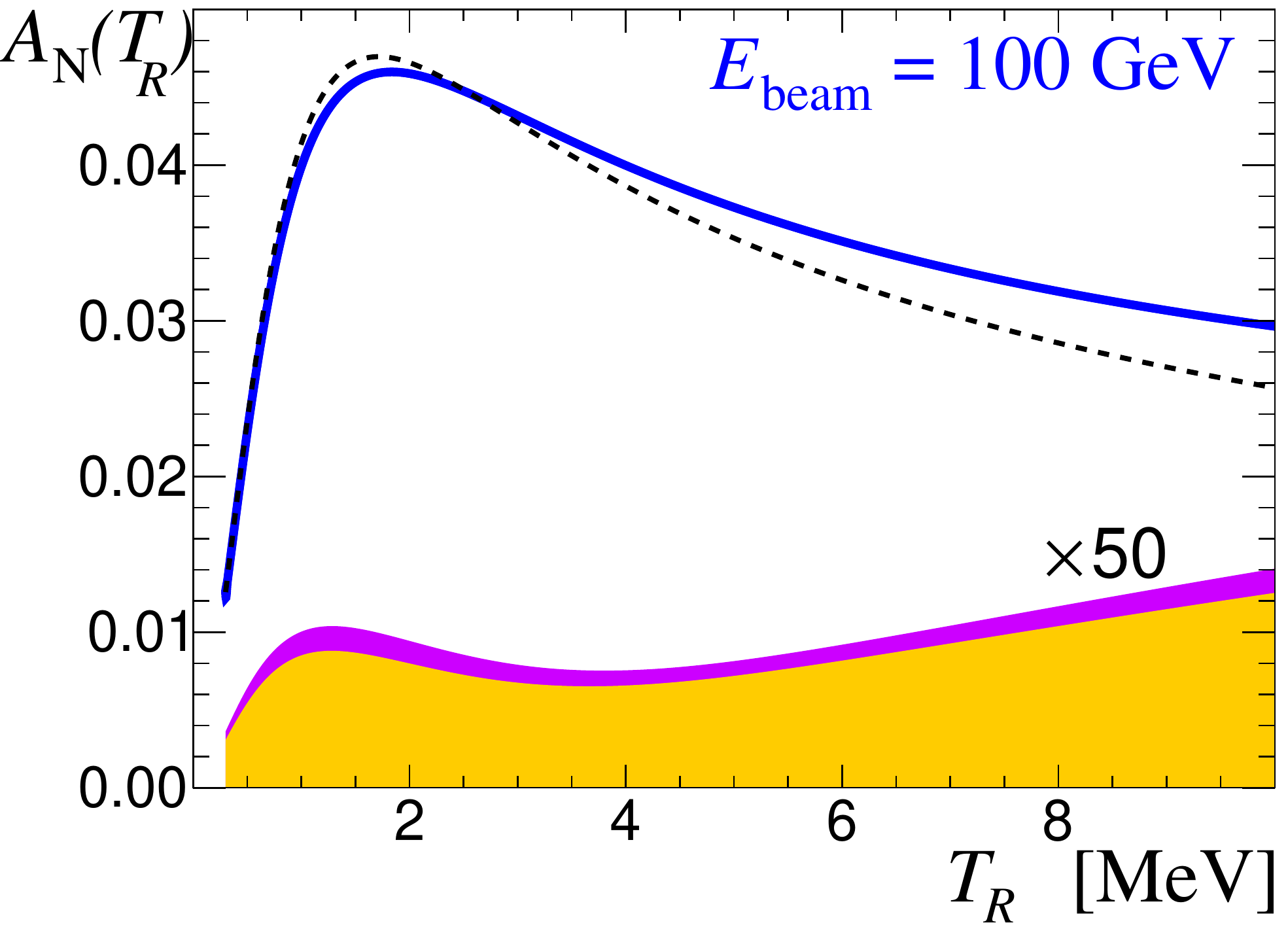} 
    \includegraphics[width=0.49\textwidth]{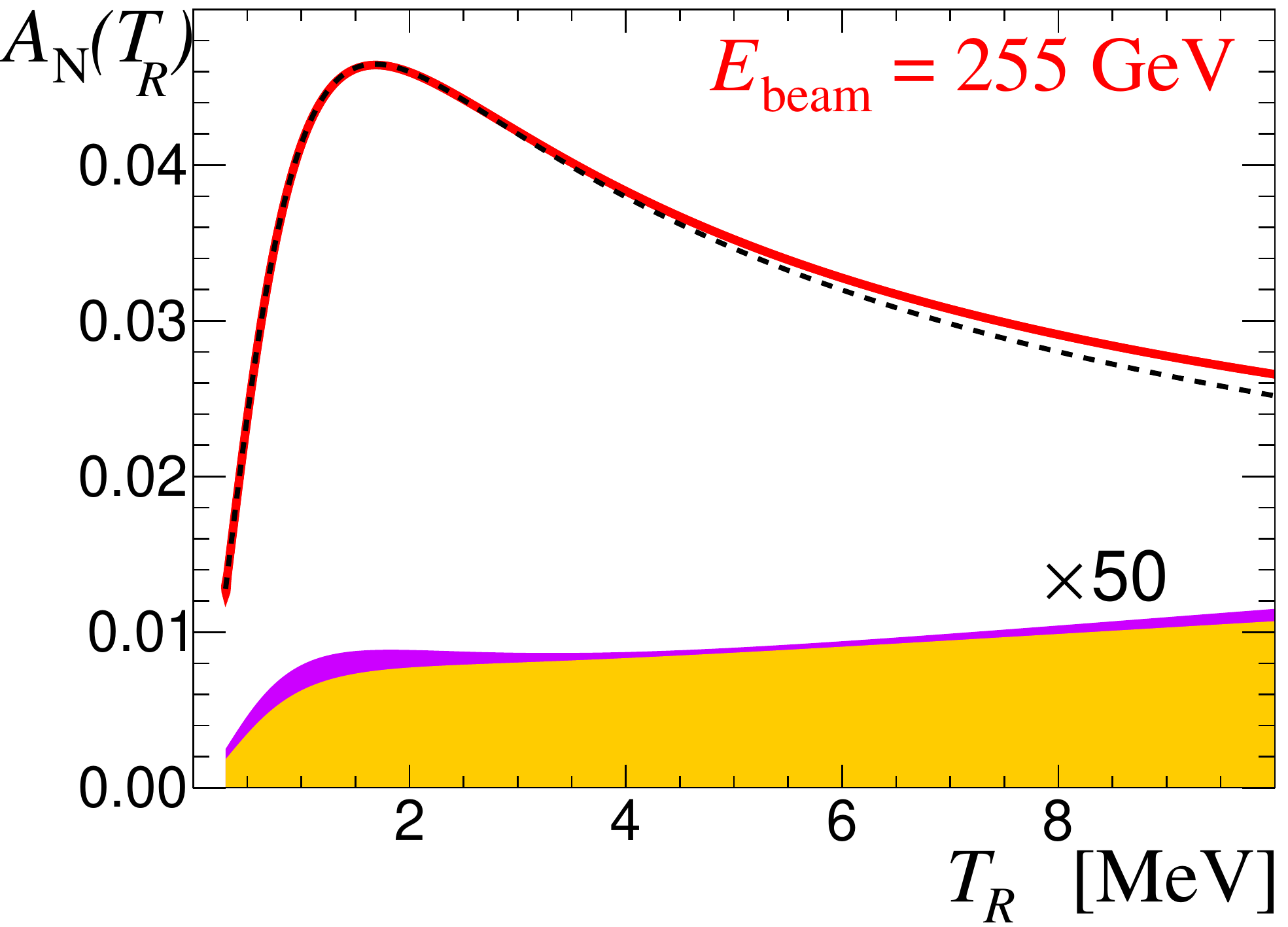}
  \end{center}
  \caption{\label{fig:AN}
    Elastic $\mathit{pp}$ single spin analyzing powers measured at HJET\,\cite{bib:HJET19}. Filled areas are 50-fold systematic and total (stat+syst) errors in the measurements. Dashed lines specifies the theoretical predictions given by Eq.\,(\ref{eq:AN_KL}).
  }
  \end{minipage}
  \hfill
  \begin{minipage}[t]{0.32\textwidth}
    \begin{center}
      \includegraphics[width=0.98\textwidth]{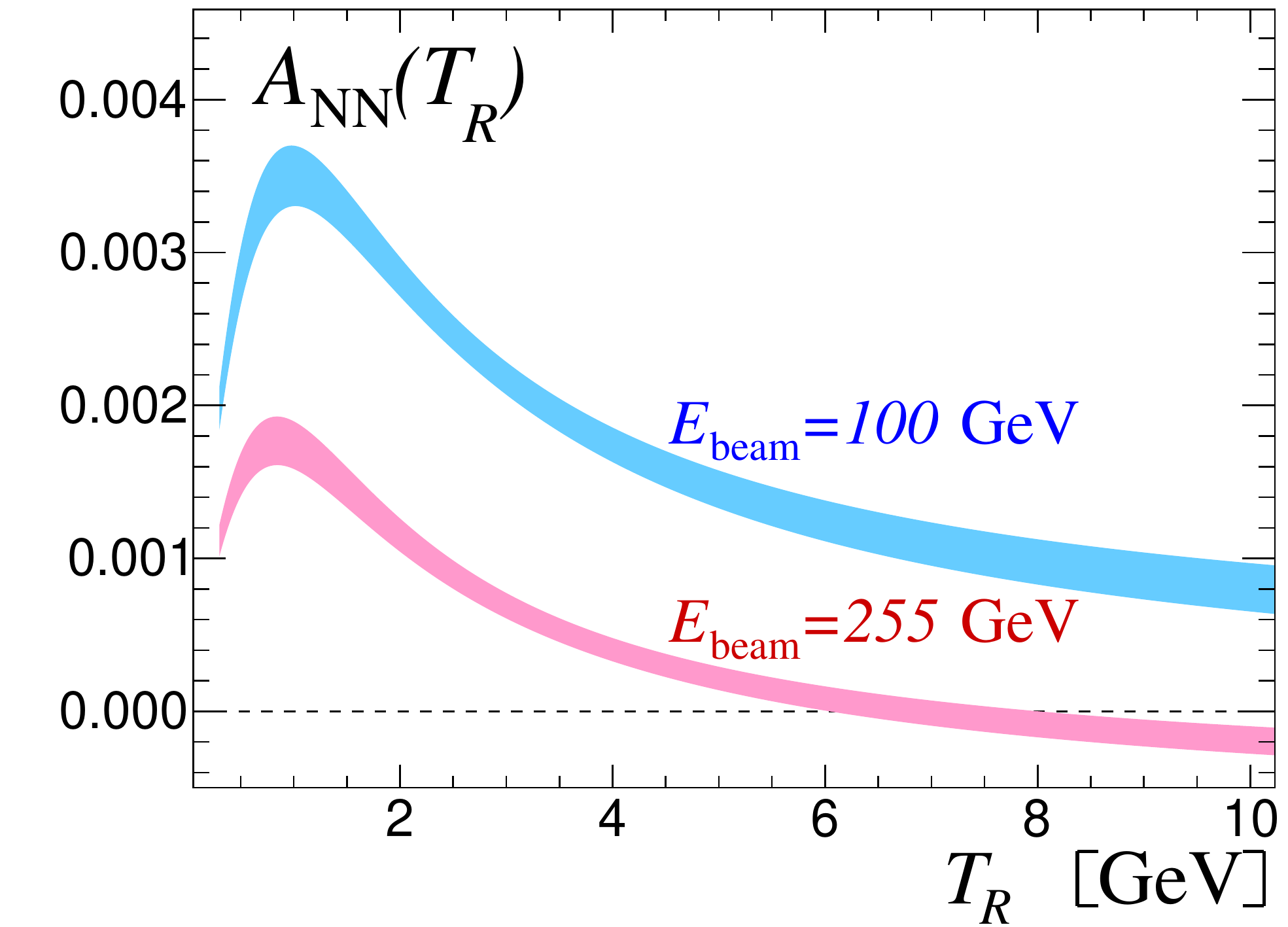} 
    \end{center}
    \caption{\label{fig:ANN}
      Elastic $\mathit{pp}$ double spin analyzing powers measured at HJET\,\cite{bib:HJET19}. Filled areas indicate $\pm1\sigma$ statistical errors. 
    }
  \end{minipage}
\end{figure*}

The event selection cuts
\begin{equation}
  \begin{aligned}
    \text{100\,GeV:}\qquad
    & |\delta t|<5.9\,\text{ns},\\
    & |\delta\sqrt{T}|<0.36\,\text{MeV}^{1/2},\\
    \text{255\,GeV:}\qquad
    & |\delta t|<7.1\,\text{ns},\\
    & -0.18<\delta\sqrt{T}<0.36\,\text{MeV}^{1/2},\\
  \end{aligned}
\label{eq:CutsAN}
\end{equation}
were optimized using procedures similar to that illustrated in Fig.\,\ref{fig:MMcut}. The normalized single spin, $a_\text{N}(T_R)/A_\text{N}(T_R)$ (\ref{eq:a_n(T_R)}), and intensity, $\lambda(T_R)$ (\ref{eq:N}),  asymmetries determined with these cuts are shown in Fig.\,\ref{fig:Asym}.

For the 100\,GeV beam, all dependencies  agree with a systematic error free expectation. For example, four statistically independent measurements of the slope $\beta_\text{N}$ are well consistent $\langle\beta_\text{N}\rangle=0.0185\pm0.0008\ (\chi^2/\text{ndf}=3.93/3)$.

For 255\,GeV, our assumption that $a_n^\text{j}(T_R)$ is a linear function of $T_R$ is strongly violated at low energies, $T_R<1.9\ \text{MeV}$, but there are no issues with other distributions. 
Such a signature attributes  the problem to an overestimated $\text{H}_2$ background subtraction.
For low $T_R$, the accuracy of the $\text{H}_2$ background evaluation may be affected by the recoil proton tracking in the magnetic field and, for the 255 GeV beam, by the data contamination from inelastic $pp\to Xp$ events. 
However, as we currently have no proven model to describe the phenomenon, we have eliminated
the 255\,GeV low energy ($T_R<1.9\ \text{MeV}$) jet asymmetry $a_\text{N}^\text{j}$ results from the analysis. With such a cut, the $\beta_\text{N}$ test gives  $\langle\beta_\text{N}\rangle=0.0084\pm0.0006\ (\chi^2/\text{ndf}=3.99/3)$.

An incorrect value of the forward elastic real-to-imaginary ratio $\rho$ used in the calculation of $A_\text{N}^{(0)}(t)$ [Eq.\,(\ref{eq:AN_norm})] may result in a false non-linearity of Eq.\,(\ref{eq:AN_norm}). In the fits with $\rho$ being a free parameter, we obtained $\rho=-0.050\pm0.025$ (100 GeV) and $\rho=-0.028\pm0.018$ (255 GeV), the values which agree (Fig.\,\ref{fig:rho}) with unpolarized $\mathit{pp}$ data to within one standard deviation. So, this test does not indicate any statistically significant discrepancy with the theoretical expectation (\ref{eq:AN_norm}).

The described fit of $\rho$ provided an important contribution to the upper limit (\ref{eq:systCalib}) for the systematic uncertainties in the energy calibration.

To determine the hadronic spin--flip amplitude $r_5$ (separately for 100 and 255\,GeV), we concurrently fit all four (the jet and beam spin for both RHIC beams) measured asymmetries, $a_\text{N}^\text{j,b}(t)=P_\text{j,b}A_\text{N}(t,r_5)$ as illustrated in Fig.\,\ref{fig:fitAN}.

The detailed description of the fit results is given in Ref.\,\cite{bib:HJET19}. The correlation contours for $\alpha_\text{N}\!\approx\!1\!-\!(2/\varkappa)\,\text{Im}\,r_5$ and $\beta_\text{N}\!\approx\!-(2/\varkappa)\,\text{Re}\,r_5$ are shown in Fig.\,\ref{fig:r5}. The measured analyzing power dependence on $T_R$, including the experimental uncertainties, are shown in Fig.\,\ref{fig:AN}. It should be noted that the drown analyzing powers can be interpreted\,\cite{bib:HJET19} as model independent, i.e. insensitive to the actual  theoretical parametrization of $A_\text{N}(t)$ used in the fit. Instructions to calculate $A_\text{N}(t)$ including extrapolations to other energies can be found in Ref.\,\cite{bib:HJET19}.

\section{Double spin analyzing power measurement}

The jet spin correlated systematic uncertainties are canceled
in the ratio, $a_\text{NN}(T_R)/a_\text{N}^\text{j}(T_R)$, of the measured asymmetries. This statement was verified by comparing the ratio for data with and without background subtraction. Therefore, for experimental determination of the double spin analyzing power $A_\text{NN}(t)$, it is convenient to use the following relation:
\begin{equation}
  A_\text{NN}(t\!=\!-2m_pT_R) = \frac{A_\text{N}^2(T_R,r_5)}{a_\text{N}^\text{b}(T_R)}
  \times \frac{a_\text{NN}(T_R)}{a_\text{N}^\text{j}(T_R)}
  \label{eq:ANN_meas}
\end{equation}
Since $A_\text{N}(T_R,r_5)$ is well known from the single spin asymmetry study and since there are no issues with systematic error in the experimental determination of $a_\text{N}^\text{b}(T_R)$, the experimental uncertainties of such defined $A_\text{NN}(t)$  are strongly dominated by the statistical errors of the measured $a_\text{NN}(T_R)$.   

Double spin analyzing power measured at HJET\,\cite{bib:HJET19} is shown in Fig.\,\ref{fig:ANN}.

\section{HJET in Heavy Ion beam \label{sec:IonBeam}}

\begin{figure}[t]
\begin{center}
\includegraphics[width=0.8\columnwidth]{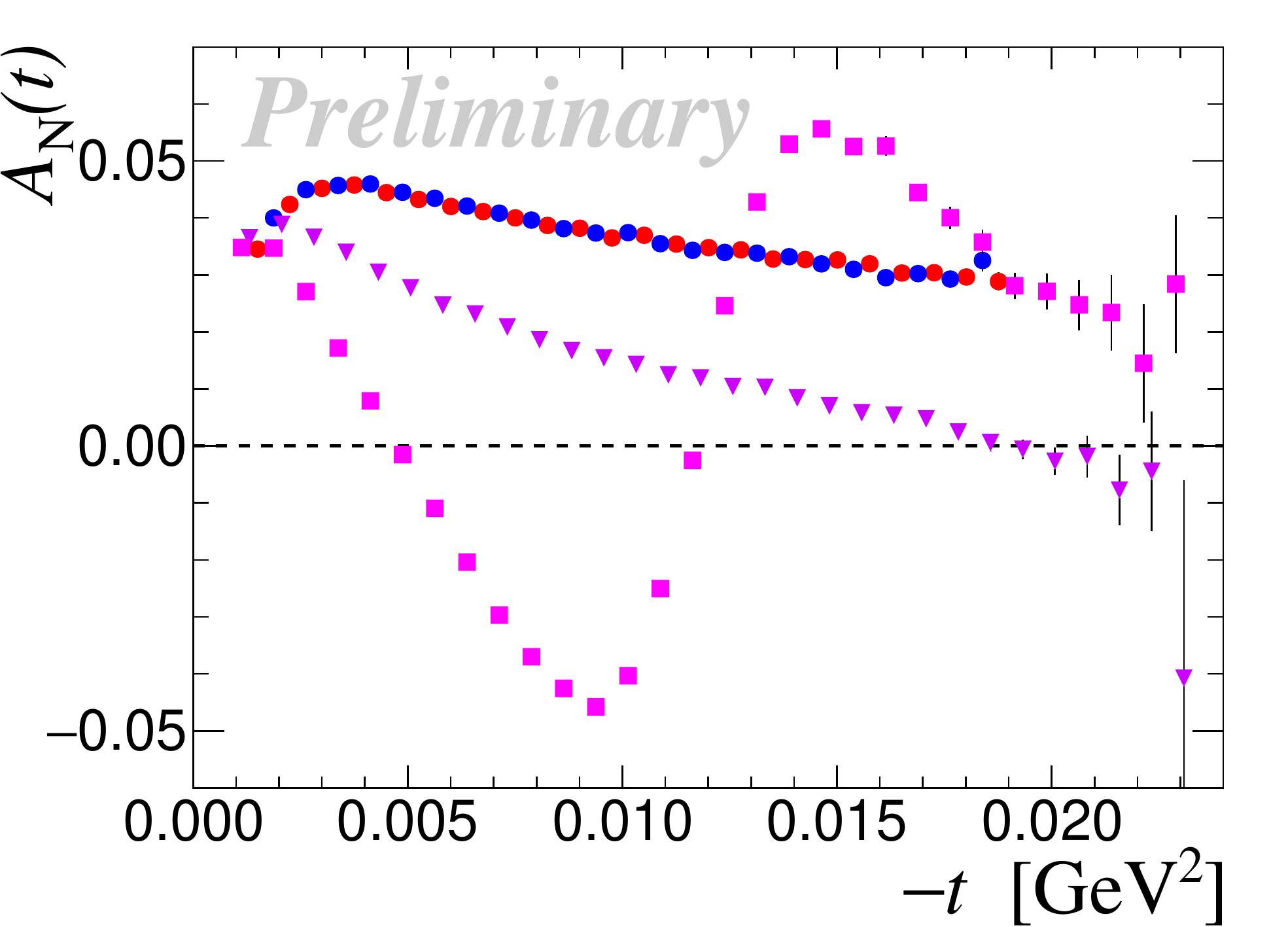} 
\end{center}
\caption{\label{fig:AN2015}
  Preliminary results for $p^\uparrow p$ ({\large $\color{blue}\bullet$} and {\large $\color{red}\bullet$} for blue and yellow beams, respectively), $p^\uparrow\text{Al}$ ({$\color{violet}\blacktriangledown$}), and $p^\uparrow\text{Au}$ ({\scriptsize$\color{magenta}\blacksquare$}) analyzing powers measured with 100\,GeV/n beams in RHIC Run15.
}
\end{figure}
\begin{figure}[t]
\begin{center}
\includegraphics[width=0.77\columnwidth]{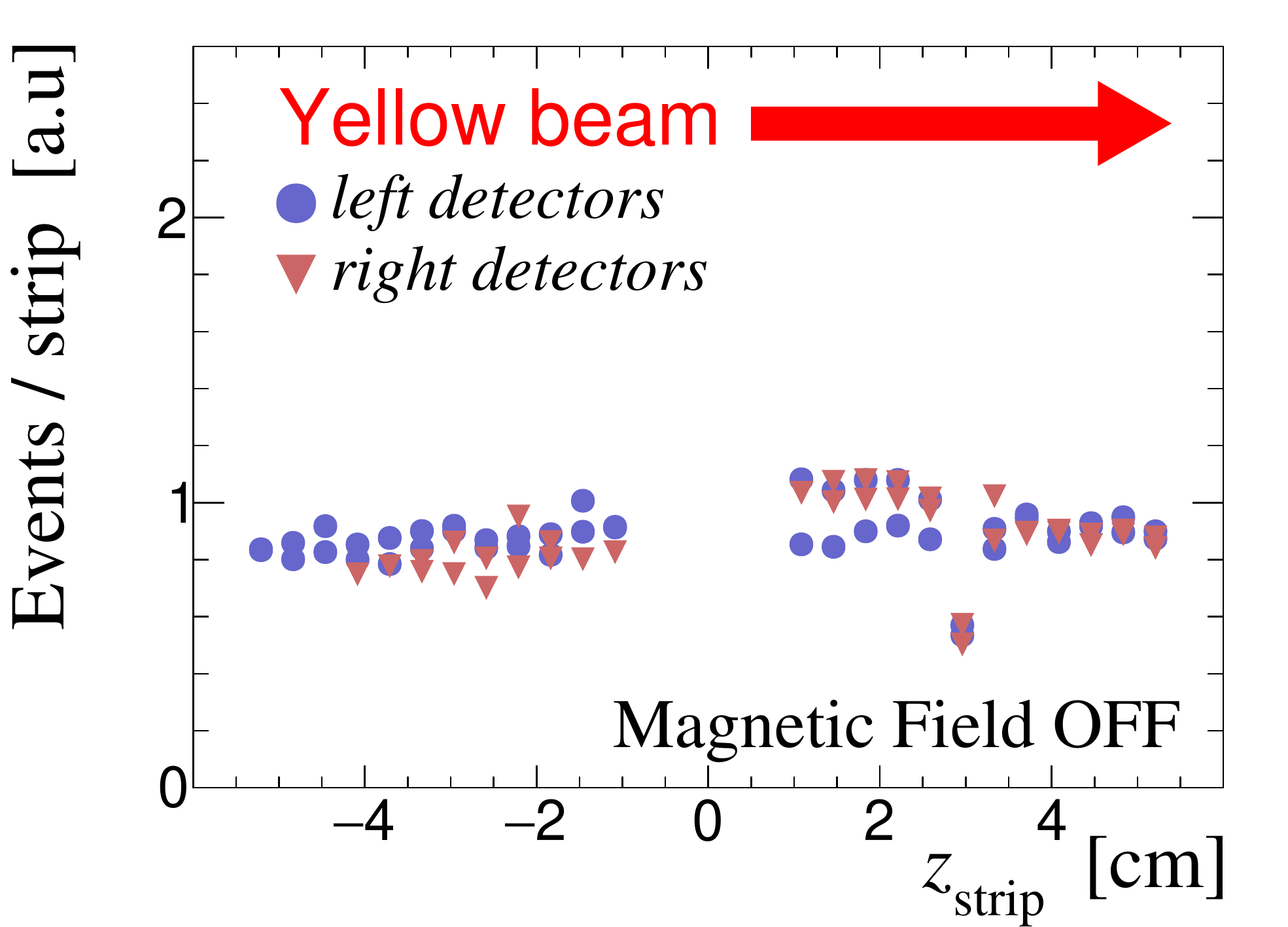}
\includegraphics[width=0.77\columnwidth]{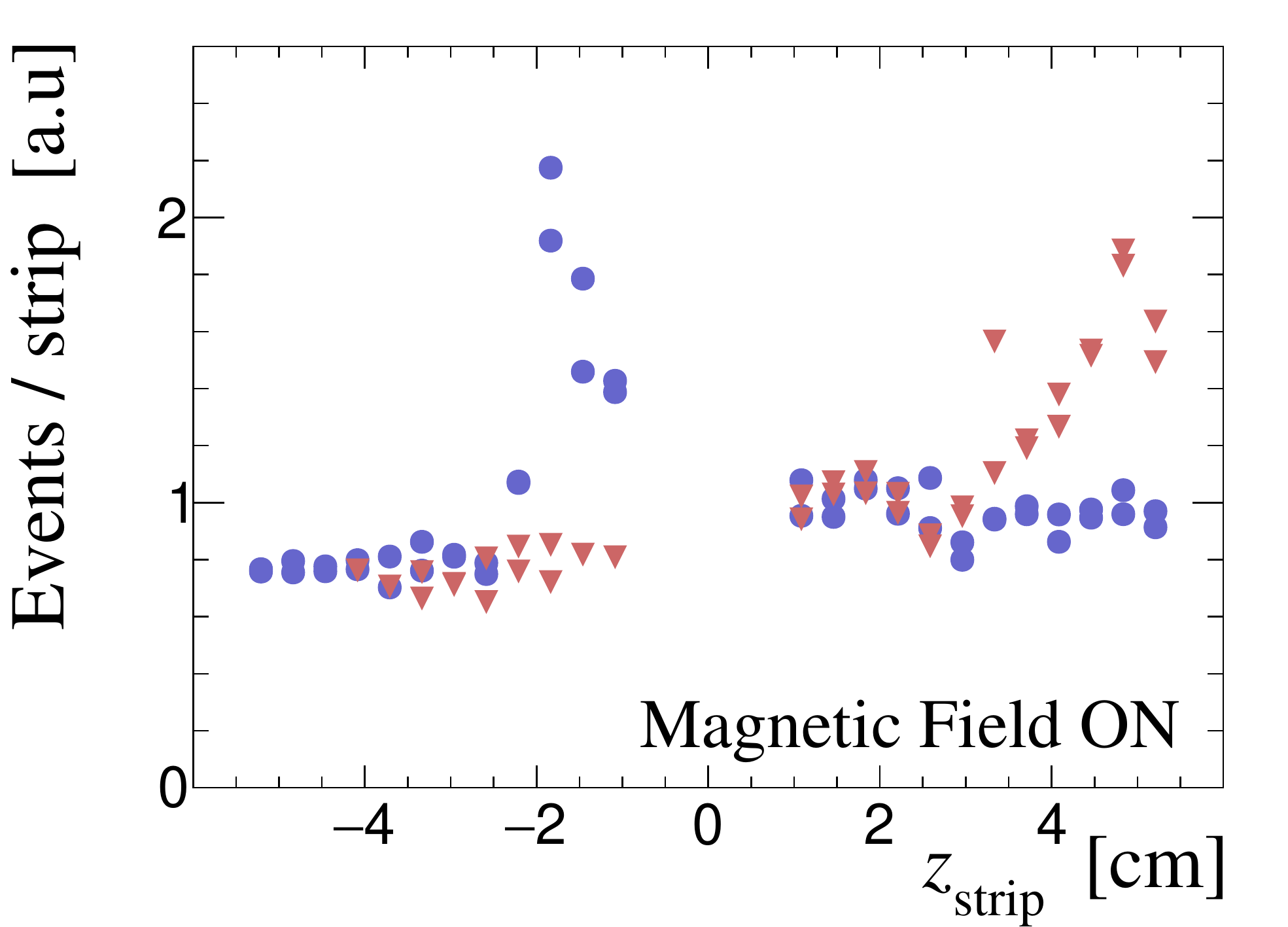}
\end{center}
\caption{\label{fig:AuY}
  Recoil proton $z$-coordinate distribution for $T_R\!=\!1.0\pm0.1\,\text{MeV}$ with the holding field magnet On and Off. The measurements were done with {\em yellow} Gold beam and $\text{H}_2$ injected to Chamber\,7.
}
\end{figure}

The HJET performance in a heavy ion RHIC beam is very similar to that of in the proton beam. The observed correlation between the recoil angle and recoil energy is well consistent with Eq.\,(\ref{eq:kappaA}) suggesting a strong dominance of the elastic and/or nucleus excitation $\mathit{pp\to{}p}\mathrm{A}^*$ scattering. Therefore, the $p^\uparrow\text{A}$ single spin analyzing power can be studied in the same way as it was done to measure elastic $p^\uparrow p$ $A_\text{N}$ even though we cannot experimentally separate the elastic and $\mathrm{A}^*$ events at the low $t$. Some preliminary results (statistical errors only), obtained in 2015 are displayed in Fig.\,\ref{fig:AN2015}. The data obtained with 100\,GeV/nucleon $d$, Al, Ru, Zr, and Au beams as well as energy scans for Deuterium (9.8, 19.5, 31.2, and 101\,GeV/n) and Gold (3.85, 4.59, 5.76, 9.8, 13.2, 19.5, 27.2, 31.2, and 100\,GeV/n) are being analyzed.

For the Gold beam, the low energy ($T_R\!\sim\!1\,\text{MeV}$) recoil proton rate is about the {\em prompts} rate. This makes the Gold beam a convenient tool to study some systematic uncertainties. For example, in RHIC Run\,20, we studied the magnetic field correction to the low energy recoil $dN/dz$ distribution. For that, $\text{H}_2$ was injected to Chamber\,7 and, using the single ({\em yellow}) Gold beam, we measured the recoil proton rate in the Si strips with the holding field magnet switched on and off. The results for $T_R\!=\!1.0\!\pm\!0.1\,\text{MeV}$ are shown in Fig.\,\ref{fig:AuY}. These measurements provided experimental confirmation of the assumptions discussed in sections \ref{sec:mfTracking} and \ref{sec:H2BeamGas}.

\section{The HJET based polarimeters at EIC}

The proton beam energies and intensities proposed for EIC\,\cite{bib:eRHIC} are similar to those at RHIC, which allows for a straightforward transfer of  our experience with HJET to the hadronic polarimetry at EIC. However, high bunch frequency (10\,ns bunch spacing) at EIC may be a challenge due to mixing signals from the differently polarized beam bunches.

To evaluate the HJET performance in these conditions, we emulated the beam polarization measurement data at EIC using the real data acquired in RHIC Run\,17\,\cite{bib:PSTP2019}. These generated events were processed using the regular HJET analysis described in this paper. The proton beam polarization systematic error is not changed much if recoil protons with energy cut $T_R\!>\!2\,\text{MeV}$ are selected\,\cite{bib:PSTP2019}. A possibility to lower this threshold to $\lesssim\!1\,\text{MeV}$ requires more studies including the optimization of the waveform shape event selection cuts shown in Fig.\,\ref{fig:nA_Cut}.

Employing double layer Si strip detectors to veto high intensity {\em prompts} is also being investigated.

\section{Summary and outlook}

Since it was commissioned in 2004 the HJET has exhibited stable and reliable performance. The sources of systematic errors are well controlled, which allowed us to measure the high energy proton beam polarization with little systematic uncertainty $\sigma_P^\text{syst}/P\lesssim0.5\%$.

The accuracy achieved in determining $A_\text{N}(t)$\,\cite{bib:HJET19} allows one to use a higher density unpolarized hydrogen jet target in a high precision absolute polarimeter, e.g., at EIC. 

The unpolarized H-jet thickness of a $10^{14}\,\text{H}_2/\text{cm}^2$ (with the jet profile similar to the HJET one) can be produced by the cluster-jet technique, which was developed for the PANDA experiment at GSI\,\cite{bib:PANDA}. Considering the higher EIC beam intensity and a possibility to increase the recoil proton detectors solid angle, one can expect 1000 times higher elastic $\mathit{pp}$ event rate compared to that in the RHIC HJET measurements.

For such a polarimeter, only a 3 min. exposure is needed to determine the beam polarization with a statistical accuracy of $\sigma_P^\text{stat}\sim1\%$, while the systematic uncertainty can be kept as low as $\delta^\text{syst}P/P\!\lesssim\!1\%$. Therefore, an unpolarized jet can also provide accurate measurement of the polarization decay time and, with additional detectors at different azimuthal angles, the beam's spin tilt.

\section{Acknowledgments}

Many results of this work were presented at RHIC Spin Group meetings. The authors are thankful to 
I.~Alekseev, E.~Aschenauer, A.~Bazilevsky, K.~O.~Eyser, H.~Huang, W.~B.~Schmidke, and D.~Svirida for numerous useful discussions and stimulating comments. We acknowledge support from the Office of Nuclear Physics in the Office of Science of the US Department of Energy and the RIKEN BNL Research Center.

\appendix
\section{Experimental determination of the spin correlated asymmetries\label{app:SqrtFormula}}

For evaluation of the spin-correlated asymmetries, an experiment schematically shown in Fig.\,\ref{fig:HjetView} is equivalent to counting of recoil protons in right/left ($RL$) symmetric detectors depending on the beam ($\uparrow\downarrow$) and target ($+-$) spins. Following Eq.\,(\ref{eq:dsdt}) one can expect
\begin{align}
  N^{\uparrow+}_R = N_0&
  \left(1+a_\text{N}^\text{j}+a_\text{N}^\text{b}+a_\text{NN}\right)\times
  \nonumber\\ &
  \left(1\!+\!\lambda_j\right)
  \left(1\!+\!\lambda_\text{b}\right)
  \left(1\!+\!\epsilon\right)
  \left(1\!+\!b_\text{NN}\right),
  \label{eq:N++R} \\
  N^{\downarrow+}_R = N_0&
  \left(1+a_\text{N}^\text{j}-a_\text{N}^\text{b}-a_\text{NN}\right)\times
  \nonumber\\ &
  \left(1\!+\!\lambda_j\right)
  \left(1\!-\!\lambda_\text{b}\right)
  \left(1\!+\!\epsilon\right)
  \left(1\!-\!b_\text{NN}\right),
  \label{eq:N-+R} \\
  N^{\uparrow-}_R = N_0&
  \left(1-a_\text{N}^\text{j}+a_\text{N}^\text{b}-a_\text{NN}\right)\times
  \nonumber\\ &
  \left(1\!-\!\lambda_j\right)
  \left(1\!+\!\lambda_\text{b}\right)
  \left(1\!+\!\epsilon\right)
  \left(1\!-\!b_\text{NN}\right),
  \label{eq:N+-R} \\
  N^{\downarrow-}_R = N_0&
  \left(1-a_\text{N}^\text{j}-a_\text{N}^\text{b}+a_\text{NN}\right)\times
  \nonumber\\ &
  \left(1\!-\!\lambda_j\right)
  \left(1\!-\!\lambda_\text{b}\right)
  \left(1\!+\!\epsilon\right)
  \left(1\!+\!b_\text{NN}\right),
  \label{eq:N--R} \\
  N^{\uparrow+}_L = N_0&
  \left(1-a_\text{N}^\text{j}-a_\text{N}^\text{b}+a_\text{NN}\right)\times
  \nonumber\\ &
  \left(1\!+\!\lambda_j\right)
  \left(1\!+\!\lambda_\text{b}\right)
  \left(1\!-\!\epsilon\right)
  \left(1\!-\!b_\text{NN}\right),
  \label{eq:N++L} \\
  N^{\downarrow+}_L = N_0&
  \left(1-a_\text{N}^\text{j}+a_\text{N}^\text{b}-a_\text{NN}\right)
  \nonumber\\ &
  \left(1\!+\!\lambda_j\right)
  \left(1\!-\!\lambda_\text{b}\right)
  \left(1\!-\!\epsilon\right)
  \left(1\!+\!b_\text{NN}\right),
  \label{eq:N-+L} \\
  N^{\uparrow-}_L = N_0&
  \left(1+a_\text{N}^\text{j}-a_\text{N}^\text{b}-a_\text{NN}\right)\times
  \nonumber\\ &
  \left(1\!-\!\lambda_j\right)
  \left(1\!+\!\lambda_\text{b}\right)
  \left(1\!-\!\epsilon\right)
  \left(1\!+\!b_\text{NN}\right),
  \label{eq:N+-L} \\
%
  N^{\downarrow-}_L = N_0&
  \left(1+a_\text{N}^\text{j}+a_\text{N}^\text{b}+a_\text{NN}\right)\times
  \nonumber\\ &
  \left(1\!-\!\lambda_j\right)
  \left(1\!-\!\lambda_\text{b}\right)
  \left(1\!-\!\epsilon\right)
  \left(1\!-\!b_\text{NN}\right)
  \label{eq:N--L} 
\end{align} 
where $a_\text{N}^\text{j}$, $a_\text{N}^\text{b}$, and $a_\text{NN}$ are the spin correlated asymmetries defined in Eq.\,(\ref{eq:asym}), $\lambda_\text{j,b}$ are the target and beam intensity asymmetries, $\epsilon$ is the $RL$ detector acceptance asymmetry, and $N_0$ is a normalization factor.

We deliberately added the variable $b_\text{NN}$ to this system of eight equations. The parameter $b_\text{NN}$ can be considered as right/left asymmetric analogue of the double spin asymmetry $a_\text{NN}$. According to (\ref{eq:dsdt}), $b_\text{NN}\!=\!0$. However, the experimental evaluation of $b_\text{NN}$ may serve as an indicator of some systematic uncertainties. For example, a non-zero value of $b_\text{NN}$ can be found if both, $A_\text{NN}\!\neq\!A_\text{SS}$ and $\cos{\varphi_L}\neq \cos{\varphi_R}$.

If all of the asymmetries are small, Eqs.\,(\ref{eq:N++R})--(\ref{eq:N--L}) can be readily linearized, indicating that the above defined asymmetries are orthogonal to each other and are similarly normalized. Consequently, the statistical uncertainties of the experimental determination of the asymmetries, commonly denoting as $a_i$, are mutually uncorrelated
\begin{equation}
  \big\langle\delta^\text{stat}a_i\,\delta^\text{stat}a_j\big\rangle = \delta_{ij}/N_\text{tot}
\end{equation}
where $\delta_{i,j}$ is the Kronecker delta and $N_\text{tot}$ is the total statistics in the measurement.

Denoting
\begin{equation}
  F(A,B) = (A-B)/(A+B)
\end{equation}
the exact solution of Eqs.\,(\ref{eq:N++R})--(\ref{eq:N--L}) can be given as
\begin{align}
  a_\text{N\phantom{N}}^j &=F\big(
  \sqrt{ N_R^{\uparrow+}N_L^{\downarrow-}}\!+\!
  \sqrt{N_R^{\downarrow+}N_L^{\uparrow-}},
  \nonumber \\ & \phantom{=F\big(}
  \sqrt{ N_R^{\uparrow-}N_L^{\downarrow+}}\!+\!
  \sqrt{N_R^{\downarrow-}N_L^{\uparrow+}}
  \big), \label{eq:a_Nj}\\
  a_\text{N\phantom{N}}^b &= F\big(
  \sqrt{ N_R^{\uparrow+}N_L^{\downarrow-}}\!+\!%
  \sqrt{ N_R^{\uparrow-}N_L^{\downarrow+}},
  \nonumber \\ & \phantom{=F\big(}
  \sqrt{N_R^{\downarrow+}N_L^{\uparrow-}}\!+\!
  \sqrt{N_R^{\downarrow-}N_L^{\uparrow+}}
  \big), \label{eq:a_Nb}\\
  a_\text{NN} &= F\big(
  \sqrt{ N_R^{\uparrow+}N_L^{\downarrow-}}\!+\!%
  \sqrt{N_R^{\downarrow-}N_L^{\uparrow+}},
  \nonumber \\ & \phantom{=F\big(}
  \sqrt{N_R^{\downarrow+}N_L^{\uparrow-}}\!+\!
  \sqrt{ N_R^{\uparrow-}N_L^{\downarrow+}}
  \big), \label{eq:a_NN}\\
  \lambda^j_\text{\phantom{NN}} &= F\big(
  \sqrt[4]{ N_R^{\uparrow+}N_R^{\downarrow+}N_L^{\uparrow+}N_L^{\downarrow+}},
  \nonumber \\ &\phantom{=F\big(}
  \sqrt[4]{ N_R^{\uparrow-}N_R^{\downarrow-}N_L^{\uparrow-}N_L^{\downarrow-}}
  \big),\label{eq:a_Lj}\\
  \lambda^b_\text{\phantom{NN}} &= F\big(
  \sqrt[4]{ N_R^{\uparrow+}N_R^{\uparrow-}N_L^{\uparrow+}N_L^{\uparrow-}},
  \nonumber \\ &\phantom{=F\big(}
  \sqrt[4]{ N_R^{\downarrow+}N_R^{\downarrow-}N_L^{\downarrow+}N_L^{\downarrow-}}
  \big), \label{eq:a_Lb} \\
  \epsilon_\text{\phantom{NN}} &= F\big(
  \sqrt[4]{ N_R^{\uparrow+}N_R^{\downarrow+}N_R^{\uparrow-}N_R^{\downarrow-}},
  \nonumber \\ &\phantom{=F\big(}
  \sqrt[4]{ N_L^{\uparrow-}N_L^{\downarrow-}N_L^{\uparrow-}N_L^{\downarrow-}}
  \big), \label{eq:epsilon} \\
  b_\text{NN} &= F\big(
  \sqrt[4]{ N_R^{\uparrow+}N_R^{\downarrow-}N_L^{\uparrow+}N_L^{\downarrow-}},
  \nonumber \\ &\phantom{=F\big(}
  \sqrt[4]{ N_R^{\uparrow-}N_R^{\downarrow+}N_L^{\uparrow-}N_L^{\downarrow+}}
  \big). \label{eq:b_NN}
\end{align}
Such experimentally determined asymmetries are systematic error free as long as Eqs.\,(\ref{eq:N++R})--(\ref{eq:N--L}) are valid. The second order effects, which may result in the possible systematic uncertainties in the measured asymmetries are discussed in Section\,\ref{sec:SystErr}.

Eqs.\,(\ref{eq:N++R})--(\ref{eq:N--L}) were derived assuming the HJET scheme of measurements in which both spin up and down beam bunches see the same target (the jet) density. More commonly, e.g. for scattering of the polarized proton beams at colliders, two asymmetries $\lambda^j$ and $\lambda^b$ should be replaced by four luminosity corrections $\lambda^{ (\uparrow\downarrow)(+-)}$ constrained by the equality
\begin{equation}
  \lambda^{\uparrow+} + \lambda^{\uparrow-} +    \lambda^{\downarrow+} +    \lambda^{\downarrow-} = 0.
\end{equation}
In terms of the linearized equations, $\lambda^{ (\uparrow\downarrow)(+-)}$ can, generally, be expanded via linear combinations of $\lambda^j$, $\lambda^b$, and $a_\text{NN}$. Thus, to avoid the consequent systematic error in measurement of double spin asymmetry $a_\text{NN}$, the effective dependence of luminosity on the beam spins,  $\lambda^{ (\uparrow\downarrow)(+-)}$, should be externally evaluated\,\cite{bib:Svirida}.

For the unpolarized target (similarly for the unpolarized beam), the  above solutions can be reduced to well-known and widely used expressions
\begin{eqnarray}
  a_\text{N}   &=& F\big(\sqrt{N^\uparrow_RN^\downarrow_L},\sqrt{N^\downarrow_RN^\uparrow_L}\big),
  \label{eq:a_N} \\
  \lambda     &=& F\big(\sqrt{N^\uparrow_RN^\uparrow_L},\sqrt{N^\downarrow_RN^\downarrow_L}\big),
  \label{eq:a_L} \\
  \epsilon &=& F\big(\sqrt{N^\uparrow_RN^\downarrow_R},\sqrt{N^\downarrow_LN^\uparrow_L}\big)
  \label{eq:a_acc} 
\end{eqnarray}
by omitting the target polarization indexes $+-$ in Eqs.\,(\ref{eq:a_Nb}), (\ref{eq:a_Lb}), and (\ref{eq:epsilon}), respectively. At HJET, an unpolarized target measurements can be reasonably approximated by combining data with both jet asymmetries, e.g. $N_R^{\uparrow+}\!+\!N_R^{\uparrow-}\!\to\!N_R^{\uparrow}$. This is why, the earlier analysis of HJET data\,\cite{bib:HJET06,bib:HJET09} was based on Eqs.\,(\ref{eq:a_N})--(\ref{eq:a_acc}).

If the intensity asymmetry $\lambda$ is somehow known, the single spin asymmetry $a_\text{N}$ can be measured independently in left or right detectors
\begin{equation}
  a_\text{N}
  =F\left(\frac{N_R^\uparrow}{1\!+\!\lambda},\frac{N_R^\downarrow}{1\!-\!\lambda}\right)
  =F\left(\frac{N_L^\downarrow}{1\!-\!\lambda},\frac{N_L^\uparrow}{1\!+\!\lambda}\right)
\end{equation}
For example, $\lambda$ can be determined, following Eqs.\,(\ref{eq:a_N})--(\ref{eq:a_acc}), in the part of the recoil proton energy range and, than, the obtained value can be used  to determine $a_N$ for energies at which left and right detector acceptances are significantly different. Obviously, the method can be straightforwardly expanded for a  more general case (\ref{eq:N++R})--(\ref{eq:N--L})


\begin{thebibliography}{99}

 \bibitem{bib:KL}
  B.~Z.~Kopeliovich and L.~I.~Lapidus,
  On the necessity of polarization experiments in colliding $\mathit{pp}$ and $\mathit{\bar{p}p}$ beams,
  Yad, Fiz. 19 (1974) 218--223 [Sov. J. Nucl. Phys. 19 (1974) 114];
  JINR-P2-72-34 [CERN-Trans-73-7, \url{https://cds.cern.ch/record/396833}].


\bibitem{bib:jetConcept}
  N.~Akchurin {\em et al.},
  Polarimetry for high energy polarized proton colliders with a jet target,
  in:
  C.W.~de~Jager, T.j.~Ketel, P.J.~Mulders, J.E.J.~Oberski, and M.~Oskam-Tamboezer (Eds.),
  Proceedings of the 12th International Symposium on High-Energy Spin Physics (Spin\,96), Amsterdam, The Netherlands, 1996, 
 World Scientific, pp. 810--812.

\bibitem{bib:Bunce}
  G.~Bunce,
  Proton Polarimetry at RHIC,
  in: A. De Roeck, D. Barber, G. Radel (Eds.),
  Polarized protons at high-energies - accelerator challenges and physics op-
  portunities. Proceedings, Workshop, Hamburg, Germany, May 17-20, 1999,
  pp. 283--288.


\bibitem{bib:BreitRabi} 
  C.~Baumgarten {\it et al.} [HERMES Target Group],
  An atomic beam polarimeter to measure the nuclear polarization in the HERMES gaseous polarized hydrogen and deuterium target,
  Nucl.\ Instrum.\ Methods A 482 (2002) 606-618.
  \href{http://dx.doi.org/10.1016/S0168-9002(01)01738-7}%
       {doi:10.1016/S0168-9002(01)01738-7}.


\bibitem{bib:SpinProgram}
  G.~Bunce, N.~Saito, J.~Soffer, and W.~Vogelsang,
  Prospects for spin physics at RHIC,
  Ann.\ Rev.\ Nucl.\ Part.\ Sci.\ 50 (2000) 525--575.
  \href{https://arxiv.org/abs/hep-ph/0007218}{arXiv:hep-ph/0007218},
  \href{http://dx.doi.org/10.1146/annurev.nucl.50.1.525}%
       {doi:10.1146/annurev.nucl.50.1.525}.


\bibitem{bib:ABS}
  A.~Zelenski {\em et al.},
  Absolute polarized H-jet polarimeter development for RHIC,
  Nucl. Instrum. Methods A 536 (2005) 248--254.
  \href{http://dx.doi.org/10.1016/j.nima.2004.08.080}%
       {doi:10.1016/j.nima.2004.08.080}.

 \bibitem{bib:pC} 
  H.~Huang and K.~Kurita,
  Fiddling carbon strings with polarized proton beams,
  AIP Conf.\ Proc.\  868 (1) (2006) 3--21.
  \href{http://dx.doi.org/10.1063/1.2401392}%
       {doi:10.1063/1.2401392}.

\bibitem{bib:HJET06}
  H.~Okada {\em et al.},
  Measurement of the analyzing power in $\mathit{pp}$ elastic scattering in the peak CNI region at RHIC,
  Phys.\ Lett.\ B 638 (2006) 450--454.
  \href{https://arxiv.org/abs/nucl-ex/0502022}{arXiv:nucl-ex/0502022},
  \href{http://dx.doi.org/10.1016/j.physletb.2006.06.008}%
       {doi:10.1016/j.physletb.2006.06.008}.

\bibitem{bib:HJET09}
  I.~G.~Alekseev {\em et al.},
  Measurements of single and double spin asymmetry in pp elastic scattering in the CNI region with a polarized atomic hydrogen gas jet target,
  Phys.\ Rev.\ D 79 (2009) 094014.
  \href{http://dx.doi.org/10.1103/PhysRevD.79.094014}%
       {doi:10.1103/PhysRevD.79.094014}.

  \bibitem{bib:PSTP2015} 
  A.~Poblaguev,
  New DAQ for the HJET polarimeter at RHIC,
  PoS PSTP2015 (2015) 032.
  \href{http://dx.doi.org/10.22323/1.243.0032}%
  {doi:10.22323/1.243.0032}.

\bibitem{bib:Run15}
  V.~Schoefer {\em et al.},
  RHIC Polarized Proton--Proton Operation at 100 GeV in Run 15,
  in: Proc. of 6th International Particle Accelerator Conference (IPAC'15), Richmond, VA, USA, May 3-8, 2015, no. 6 in International Particle Accelerator Conference, JACoW, Geneva, Switzerland, 2015, pp. 2384--2386. 
  \href{http://dx.doi.org/10.18429/JACoW-IPAC2015-TUPWI060}%
  {doi.org/10.18429/JACoW-IPAC2015-TUPWI060}.

\bibitem{bib:Run17}
  V.~Ranjbar {\em et al.},
  RHIC Polarized Proton Operation for 2017,
  in: Proc. of 8th International Particle Accelerator Conference (IPAC'17), Copenhagen, Denmark, 14-19 May, 2017, no. 8 in JACoW, Geneva, Switzerland, 2017, pp. 2188--2190. 
  \href{http://dx.doi.org/10.18429/JACoW-IPAC2017-TUPVA050}%
       {doi:10.18429/JACoW-IPAC2017-TUPVA050}

\bibitem{bib:HJET19}
  A.~A.~Poblaguev {\it et al.},
  Precision Small Scattering Angle Measurements of Elastic Proton--Proton Single and Double Spin Analyzing Powers at the RHIC Hydrogen Jet Polarimeter,
  Phys.\ Rev.\ Lett.\  123 (2019) 162001.
  \href{https://arxiv.org/abs/1909.11135}{arXiv:1909.11135},
  \href{http://dx.doi.org/10.1103/PhysRevLett.123.162001}%
  {doi:10.1103/PhysRevLett.123.162001}.
    
\bibitem{bib:RHIC} 
  I.~Alekseev {\it et al.},
  Polarized proton collider at RHIC,
  Nucl.\ Instrum.\ Methods\ A 499 (2003) 392--414.
  \href{http://dx.doi.org/10.1016/S0168-9002(02)01946-0}%
       {doi:10.1016/S0168-9002(02)01946-0}.
       
\bibitem{bib:OPPIS}
  A.~Zelenski, G.~Atoian, D.~Raparia, J.~Ritter and D.~Steski,
  The RHIC polarized H${}^-$ ion source,
  Rev. Sci. Instrum. 87 (2015) 02B705.
  \href{http://dx.doi.org/10.1063/1.4932392}%
       {doi:10.1063/1.4932392}.
  A.~Zelenski, G.~Atoian, D.~Raparia, J.~Ritter, A.~Kolmogorov and V.~Davydenko,
  High-intensity polarized H${}^-$ ion source for the RHIC SPIN physics,
  AIP Conf. Proc. 1869 (2017) 030015.
  \href{http://dx.doi.org/10.1063/1.4995735}%
       {doi.org/10.1063/1.4995735}.
       
  \bibitem{bib:FADC250}
    H. Dong {\em et al.},
    Integrated tests of a high speed VXS switch card and 250 MSPS flash ADCs,
    in: 2007 IEEE Nuclear Science Symposium Conference Record, Vol.1, 2007, pp. 831--833.
    \href{http://dx.doi.org/10.1109/NSSMIC.2007.4436457}
    {doi:10.1109/NSSMIC.2007.4436457}.
  
\bibitem{bib:Convention}
  J.~Ashkin {\em et al.},
  Convention for Spin Parameters in High-Energy Scattering Experiments,
  AIP Conf. Proc. 42 (1978) 142--146.
  \href{http://dx.doi.org/10.1063/1.31281}%
  {doi:10.1063/1.31281};
  E. Leader,
  Spin in Particle Physics, Cambridge Monographs on Particle Physics, Nuclear Physics and Cosmology, Cambridge University Press, 2001, p.~119.
  \href{http://dx.doi.org/10.1017/CBO9780511524455}%
  {doi:10.1017/CBO9780511524455}.
  
\bibitem{bib:BKLST}
  N.~H.~Buttimore, B.~Z.~Kopeliovich, E.~Leader, J.~Soffer, and T.~L.~Trueman,
  The spin dependence of high-energy proton scattering,
  Phys.\ Rev.\ D 59 (1999) 114010.
  \href{https://arxiv.org/abs/hep-ph/9901339}{arXiv:hep-ph/9901339},  
  \href{http://dx.doi.org/10.1103/PhysRevD.59.114010}%
       {doi:10.1103/PhysRevD.59.114010}.
       

\bibitem{bib:BGL}
  N.~H.~Buttimore, E.~Gotsman, and E.~Leader,
  Spin Dependent Phenomena Induced By Electromagnetic Hadronic Interference At High-energies,
  Phys.\ Rev.\ D 18 (1978) 694--716.
  \href{http://dx.doi.org/10.1103/PhysRevD.18.694}%
       {doi:10.1103/PhysRevD.18.694};
       Erratum: 35 (1987) 407.
  \href{http://dx.doi.org/10.1103/PhysRevD.35.407}%
       {doi:10.1103/PhysRevD.35.407}.
       
\bibitem{bib:PDG}
  M. Tanabashi {\em et al.} (Particle Data Group),
  Review of Particle Physics,
  {Phys. Rev. D 98 030001 (2018) 030001.
    \href{http://dx.doi.org/10.1103/PhysRevD.98.030001}
         {doi:10.1103/PhysRevD.98.030001}.

\bibitem{bib:AbsorptiveCorr}
  B.~Z.~Kopeliovich and M.~Krelina,
  Probing the Pomeron spin--flip with Coulomb-nuclear interference,
  \href{https://arxiv.org/abs/1910.04799}{arXiv:1910.04799}.

\bibitem{bib:AN_corr}
   A.~A.~Poblaguev,
   Corrections to the Elastic Proton--Proton Analyzing Power Parametrization at High Energies,
   Phys.\ Rev.\ D 100 (2019) 116017.
   \href{https://arxiv.org/abs/1910.02563}{arXiv:1910.02563},
   \href{http://dx.doi.org/10.1103/PhysRevD.100.116017}%
        {doi:10.1103/PhysRevD.100.116017}.

\bibitem{bib:StoppingPower}
  Berger, M.J., Coursey, J.S., Zucker, M.A., and Chang, J. (2005), ESTAR, PSTAR, and ASTAR: Computer Programs for Calculating Stopping-Power and Range Tables for Electrons, Protons, and Helium Ions (version 1.2.3). [Online] Available: \href{http://physics.nist.gov/Star}{http://physics.nist.gov/Star [2020, January 12]}. National Institute of Standards and Technology, Gaithersburg, MD.

\bibitem{bib:Am141}
  M.~S.~Basunia,
  Nuclear Data Sheets for A = 237*,
  Nucl.\ Data Sheets 107 (2006) 2323.
  \href{http://dx.doi.org/10.1016/j.nds.2006.07.001}%
  {doi:10.1016/j.nds.2006.07.001}.

\bibitem{bib:SiSimulation}
  A.~Poblaguev, 
  Waveform dependence on signal amplitude in the RHIC H-Jet polarimeter,
  BNL-104-366-2014-IR; C-A/AP/505; \url{https://www.bnl.gov/isd/documents/85162.pdf}
  (unpublished).
  
\bibitem{bib:GeomCalib}
  A.~Poblaguev, 
  A precise in situ calibration of the RHIC H-Jet polarimeter,
  BNL-104-363-2014-IR; C-A/AP/504; \url{https://www.bnl.gov/isd/documents/85159.pdf}
  (unpublished).

\bibitem{bib:H2}
  A.~Zelenski {\em et al.},
  Atomic beam studies in the RHIC H-jet polarimeter,  
  in: F.~Bradamante, A.~Bressan, A.~Martin, K.~Aulenbacher (Eds.),
  Proc. of the 16th International Spin Physics Symposium and Workshop on Polarized Electron Sources and Polarimeters (Spin\,2004), Trieste, Italy, 2004, 
  World Scientific, 2005, pp. 761--764.

\bibitem{bib:PSTP2017}
  A.~Poblaguev {\em et al.},
  The HJET polarimeter in RHIC Run 2017,
  PoS PSTP2017 (2018) 022.
  \href{http://dx.doi.org/10.22323/1.324.0022}%
  {doi:10.22323/1.324.0022}.

\bibitem{bib:PANDA}
  A.~Taschner, E.~Kohler, H.~W.~Ortjohann and A.~Khoukaz,
  High density cluster jet target for storage ring experiments,
  Nucl.\ Instrum.\ Methods\ A 660 (2011) 22.
   \href{https://arxiv.org/abs/1108.2653}{arXiv:1108.2653},  
  \href{http://dx.doi.org/10.1016/j.nima.2011.09.024}%
       {doi.org/10.1016/j.nima.2011.09.024}.

\bibitem{bib:eRHIC}
   C. Montag \emph{et al.},
   eRHIC Electron Ring Design Status,
   in: Proc. 10th Int. Particle Accelerator Conf. (IPAC'19)}, Melbourne, Australia, 19--24 May 2019, no 10 International Particle Accelerator Conference,
  JACoW Publishing, Geneva, Switzerland, 2019, pp. 794--796,
   \href{http://dx.doi.org/10.18429/JACoW-IPAC2019-MOPRB093}%
   {doi:10.18429/JACoW-IPAC2019-MOPRB093}.
     
\bibitem{bib:PSTP2019}
  A.A.~Poblaguev, A.~Zelenski, and G.~Atoian,
  The prospects on the absolute proton beam polarimetry at EIC,
  in: Proc. of 18th International Workshop on Polarized Sources, Targets, and Polarimetry (PSTP2019), Knoxville, Tennesee, 2019,
  PoS PSTP2019 (2020) 007.

   
\bibitem{bib:Svirida} 
  D.~Svirida,
  Transverse Spin Asymmetries in the CNI Region of Elastic Proton--Proton Scattering at $\sqrt s=$\,200\,GeV,
  Int.\ J.\ Mod.\ Phys.\ Conf.\ Ser. 40 (2016) 1660056.
  \href{http://dx.doi.org/10.1142/S2010194516600569}%
       {doi.org/10.1142/S2010194516600569}.

\end{thebibliography}
\end{document}